\documentclass[
aps,
prd,
twocolumn,
superscriptaddress,
floatfix,
nobibnotes,
nofootinbib,
]
{revtex4-2}
\usepackage[english]{babel}
\usepackage{caption}
\usepackage{subcaption}
\usepackage[]{hyperref}
\usepackage[utf8]{inputenc}
\usepackage{amsmath,amssymb,amsthm,amstext}
\usepackage{mathtools}
\usepackage[nameinlink,capitalise]{cleveref}
\usepackage{siunitx}
\usepackage{braket}
\usepackage{verbatim}
\usepackage{multirow}
\usepackage{dsfont}
\usepackage{siunitx}
\usepackage{float}
\usepackage{graphicx}
\usepackage{changebar}
\usepackage{graphicx}
\usepackage{dcolumn}
\usepackage{bm}
\usepackage{orcidlink}
\usepackage{xcolor}
\usepackage{slashed}
\usepackage{esvect}
\usepackage{crossreftools}
\usepackage{hyperref}
\usepackage{enumitem}
\usepackage[normalem]{ulem}
\usepackage{makecell}

\definecolor{bubbles}{rgb}{0.91, 1.0, 1.0}
\definecolor{brightcerulean}{rgb}{0.11, 0.67, 0.84}
\definecolor{mbcol}{rgb}{1, 0, 1}
\definecolor{umcol}{RGB}{0, 128, 0}
\definecolor{bjcol}{RGB}{128, 6, 6}
\usepackage[colorinlistoftodos]{todonotes}
\presetkeys{todonotes}{size=\footnotesize}{}
\usepackage{soul}
\DeclareMathOperator\arctanh{arctanh}
\newcounter{prop}[section]
\renewcommand*{\theprop}{\thesection.\arabic{prop}}

\hypersetup{
	colorlinks=true,
	citecolor=blue,
	citebordercolor=red,
	linktoc=all,
	linkcolor=blue,
	urlcolor=blue
}
\AddToHook{cmd/appendix/before}{
    \renewcommand*{\thesubsection}{\Roman{subsection}}
    \crefalias{section}{appendix}
    \crefalias{subsection}{appendix}
}

\crefformat{equation}{#2Eq.~(#1)#3}
\crefformat{appendix}{#2App.~#1#3}
\crefformat{section}{#2Sec.~#1#3}
\crefformat{table}{#2Tab.~#1#3}
\crefformat{figure}{#2Fig.~#1#3}

\crefmultiformat{section}{#2Secs.~#1#3}{ and~#2#1#3}{, #2#1#3}{ and~#2#1#3}
\crefmultiformat{appendix}{#2Apps.~#1#3}{ and~#2#1#3}{, #2#1#3}{ and~#2#1#3}

\crefrangeformat{equation}{#3Eqs.~(#1)#4 to #5(#2)#6}

\Crefformat{figure}{#2Fig.~#1#3}

\renewcommand{\eqref}[1]{\cref{#1}}

\newcommand{\vcfpi}{\tilde{f}_\pi}

\newcommand{\vcmsigma}{\tilde{m}_\sigma^2}
\newcommand{\vcmDelta}{\tilde{m}_\Delta^2}
\newcommand{\vclambdamix}{\tilde{\lambda}_\text{mix}}
\newcommand{\vclambdaDelta}{\tilde{\lambda}_\Delta}
\newcommand{\vcZDelta}{\tilde{Z}_\Delta}

\newcommand{\MeV}{\;\text{MeV}}
\newcommand{\GeV}{\;\text{GeV}}
\newcommand{\Tr}{\;\text{Tr}}
\newcommand{\tp}{\mathsf{T}}
\newcommand{\sqvec}[1]{\vec{#1}^{\,2}}

\newcommand{\bDeltaZero}{\bar\Delta_0}
\newcommand{\Delgap}{\Delta_{\text{gap}}}
\newcommand{\DelgapZero}{\Delta_{\text{gap},0}}
\newcommand{\bDelgap}{\bar\Delta_{\text{gap}}}
\newcommand{\bDelgapZero}{\bar\Delta_{\text{gap},0}}
\newcommand{\epsvac}{\epsilon_{q,\text{vac}}}
\newcommand{\sigvac}{\vcfpi}
\newcommand{\mqvac}{m_{q,\text{vac}}}
\newcommand{\lrvac}[1]{\left. \left( #1 \right) \right|_{\text{vac}}}

\newcommand{\onefig}{0.49\textwidth}

\newcommand{\Darmstadt}{
    Technische Universität Darmstadt, % 
    Fachbereich Physik, % 
    Institut für Kernphysik, %
    Theoriezentrum, %
    Schlossgartenstr. 2, %
    D-64289 Darmstadt, %
    Germany %
}

\newcommand{\JLU}{%
	Institut f\"{u}r Theoretische Physik, %
	Justus-Liebig-Universit\"{a}t Gie\ss{}en, %
	35392 Gie\ss{}en, %
	Germany%
}
\newcommand{\HFHFGiessen}{%
	Helmholtz Forschungsakademie Hessen f\"{u}r FAIR (HFHF), %
	GSI Helmholtzzentrum f\"{u}r Schwerionenforschung, %
	Campus Gie\ss{}en, %
	35392 Gie\ss{}en, %
	Germany%
}
\newcommand{\HFHFDarmstadt}{
    Helmholtz Forschungsakademie Hessen für FAIR (HFHF),
    GSI Helmholtzzentrum für Schwerionenforschung,
    Campus Darmstadt, D-64289 Darmstadt, Germany
}

\graphicspath{
{./}
{./plots/}
}

\begin{document}

\title{Renormalizing the Quark-Meson-Diquark Model}

\author{Hosein Gholami \orcidlink{0009-0003-3194-926X}}
\email[]{mohammadhossein.gholami@tu-darmstadt.de}
\affiliation{\Darmstadt}
\author{Lennart Kurth \orcidlink{0009-0005-2723-6405}}
\email[]{lennart.kurth@stud.tu-darmstadt.de}
\affiliation{\Darmstadt}
\author{Ugo Mire \orcidlink{0009-0009-2345-691X}}
\email[]{ugo.louis.tryphon.mire@physik.uni-giessen.de}
\affiliation{\JLU}
\author{Michael Buballa \orcidlink{0000-0003-3747-6865}}
\email[]{michael.buballa@tu-darmstadt.de}
\affiliation{\Darmstadt}
\affiliation{\HFHFDarmstadt}
\author{Bernd-Jochen Schaefer \orcidlink{0000-0003-0659-2679}}
\email[]{bernd-jochen.schaefer@uni-giessen.de}
\affiliation{\JLU}
\affiliation{\HFHFGiessen}

\begin{abstract}
  We present a comprehensive study of the two-flavor
  Quark--Meson--Diquark (QMD) model by comparing a renormalization
  approach with a renormalization-group (RG) consistent mean-field
  formulation based on the functional renormalization group (FRG). The
  renormalized QMD model allows analytical investigations of key
  quantities such as the zero-temperature diquark gap and the critical
  temperature for color superconductivity, ultimately reproducing the
  exact BCS relation in the high-density limit. We carry out the same
  analysis for different schemes of RG-consistent QMD models. We show
  that the RG-consistent approach yields a phase diagram and
  thermodynamic properties qualitatively similar to those of the
  renormalized model, provided both are embedded within a unified
  scheme that ensures consistent vacuum properties.  In particular,
  both treatments recover the Stefan--Boltzmann limit at high
  densities. On the other hand, whether the BCS relation for the
  critical temperature is satisfied depends on the details of the
  RG-consistent setup.  Our results highlight the relevance of
  renormalization and RG-consistent methods for accurately capturing
  the thermodynamics of QMD and related effective models with diquark
  degrees of freedom.
\end{abstract}

\maketitle

\section{Introduction}

Quantum Chromodynamics (QCD) is the fundamental theory of the strong
interaction among quarks and gluons. In the low-energy regime,
however, QCD becomes inherently non-perturbative, making direct
analytical calculations extremely challenging. Over the years, a
variety of non-perturbative tools have been developed to study QCD in
this regime. Notably, lattice QCD simulations \cite{Aarts:2023vsf} and
continuum functional methods (such as Dyson--Schwinger equations
\cite{Fischer:2018sdj} and the functional renormalization group
\cite{Fu:2022gou}) have provided important insights into the QCD phase
structure.  Despite these advances, first-principles calculations
remain challenging, particularly for the thermodynamics of QCD at
finite density, motivating the use of effective models to capture its
essential low-energy physics.

Effective models of QCD are designed to respect its symmetries and
replicate the dynamics of the full theory, enabling the study of
phenomena such as spontaneous chiral symmetry breaking and
hadronization within a more manageable framework.  Two prominent
examples are the Nambu--Jona-Lasinio (NJL) and the quark--meson model,
also known as the linear sigma model coupled to quarks. These models
dynamically generate constituent quark masses through chiral
condensates and successfully capture many qualitative features of
low-energy QCD. They have been widely employed to investigate the QCD
phase diagram and its transition properties~\cite{ Buballa:2003qv,
  Schaefer:2004en,Fukushima:2011}.

At high baryon density color-superconducting (CSC) phases are expected
to emerge, i.e., phases where quarks are paired in so-called diquark
condensates \cite{Alford:2008}.  While the quark-meson model and
earlier versions of the NJL model account for quark-antiquark
condensates only, CSC phases can straightforwardly be described within
the NJL model by adding the appropriate quark-quark interactions
\cite{Berges:1998rc,Schwarz:1999dj,Buballa:2003qv}.  Similarly, a
natural extension of the quark-meson model to enable the modeling of
CSC is the inclusion of diquark degrees of freedom.  This leads to the
so-called Quark--Meson--Diquark (QMD)
model~\cite{Braun:2018svj,Andersen:2024qus,Andersen:2025ezj}, first
introduced in the context of two-color
QCD~\cite{Kogut:1999iv,Andersen:2010vu,Strodthoff:2011tz}, whose
effective Lagrangian incorporates not only mesonic fields (such as the
sigma ($\sigma$) and pions $(\vec{\pi})$, associated with the chiral
symmetry) but also explicit complex-valued diquark fields ($\Delta$,
$\Delta^{*}$) representing correlated quark--quark and
antiquark-antiquark pairs. The diquark fields{\tt } can condense in
the color-antitriplet channel, providing a description of the
color-superconducting phase of quark matter~\cite{Alford:2008}.

Despite their successes, NJL-type models are non-renormalizable. As a
result, a momentum cutoff (or another regularization scheme) must be
introduced to handle ultraviolet (UV) divergences, and physical
predictions can, in general, depend on the choice of the
regularization scheme. In particular, a naïve cutoff regularization
can induce significant artifacts, making certain observables
unphysically sensitive to the regulator~\cite{Farias:2005cr}.

The quark--meson model, which can be regarded as a bosonized NJL model with
Yukawa couplings, is renormalizable. However, when treated with a
simple cutoff and without proper renormalization, it may exhibit
spurious regulator dependencies, see, e.g.,~\cite{Skokov:2010sf}.
Over the years, various techniques have been developed to mitigate
these issues and ensure proper renormalization-group (RG) behavior in
effective models.

One strategy to eliminate cutoff artifacts is renormalization --
explicitly incorporating a dependence of the bare model parameters
(such as couplings and masses) on the regulator scale so that all
divergent contributions are absorbed into redefinitions of these
parameters. In practice, this involves the introduction of
counterterms to ensure that physical observables remain finite and
independent of the UV cutoff eventually. As a (perturbatively)
renormalizable theory, the QMD model permits a fully renormalized
formulation in which all UV divergences are absorbed into the bare
parameters of the Lagrangian (at least on the mean-field level, where
bosonic fluctuations are neglected).  This issue was recently
discussed in Ref.~\cite{Gholami:2024diy} and explicitly demonstrated
using dimensional regularization in Ref.~\cite{Andersen:2024qus} -- a
scheme often favored for preserving symmetries, particularly Lorentz
invariance. At finite temperature and chemical potential, however,
Lorentz symmetry is explicitly broken, reducing the advantages of this
scheme.  In the present work, we instead employ a technically simpler
three-momentum cutoff, which eventually yields more tractable
expressions.

An alternative and powerful approach is provided by the functional
renormalization group (FRG) formalism, pioneered by
Wetterich~\cite{Wetterich:1992yh}, which allows for an RG-consistent
formulation of the model. In this FRG-based treatment, a
scale-dependent effective action is introduced, with the requirement
that physical quantities in the infrared (IR) of the flow remain
independent of the chosen regulator cutoff. Such FRG-improved schemes
have been applied to low-energy QCD models, see
e.g.~\cite{Herbst:2013ufa,Braun:2018svj,Gholami:2024diy}.  In
particular, these RG-consistent formulations carefully handle
additional divergences that arise at finite chemical potential (often
referred to as medium-induced divergences) by adjusting the flow of
couplings and potentials accordingly. The QMD model can be analyzed
using an RG-consistent mean-field methodology, ensuring that cutoff
artifacts are canceled out by flow corrections.  In
Ref.~\cite{Gholami:2024diy} this method was applied to the
three-flavor NJL model, and it was found that the RG consistent
treatment does not only remove obvious cutoff artifacts, like
decreasing diquark condensates with increasing chemical potentials,
but also less expected ones concerning the ordering of the
color-superconducting phases \cite{Iida_2004}. It should be noted,
however, that the NJL model is non-renormalizable and, as a
consequence, the cutoff cannot be chosen freely but is the result of
the parameter fit in vacuum, e.g., Ref.~\cite{Rehberg:1995kh}. The RG
consistent treatment applied in Ref.~\cite{Gholami:2024diy} then
removes cutoff artifacts related to medium scales (temperatures or
chemical potentials) of the order of the cutoff. But the results still
depend to some extent on the initial cutoff scale of the vacuum
fit. Related to this, there is some scheme dependence in the NJL model
due to the treatment of the medium-induced divergences. Different
subtraction schemes lead to variations in the results, including
changes in the predicted phase diagram.

In the QMD model, the initial UV scale can be chosen arbitrarily
large. For sufficiently large values the results of the ``RG
consistent treatment'' then become fully RG invariant, i.e.,
independent of the initially chosen UV scale and identical to the
results of the fully renormalized model. In this paper, however, we
deliberately choose a relatively small value of the initial UV scale,
similar to a typical NJL cutoff, and then compare the results obtained
with different schemes of the RG consistent treatment with the
``exact'' ones of the renormalized formulation. In this way, the QMD
model serves as a controlled testing ground for directly contrasting
these schemes, thereby giving a hint on their quality also for other
models.

In this paper, we thus present a comprehensive comparison between the
renormalized QMD model and an RG-consistent QMD formulation at the
mean-field level. First, we implement a straightforward
renormalization procedure for the QMD model using a three-dimensional
momentum cutoff regularization, yielding a renormalized theory with no
residual dependence on the cutoff scale. Second, we formulate an
RG-consistent mean-field approximation within the FRG framework:
starting from the microscopic QMD action, we integrate out quantum
fluctuations down to the infrared scale while maintaining RG
invariance of the effective potential.

For a meaningful comparison, we fix the model parameters in both
approaches to the same physical vacuum observables -- such as the pion
mass, decay constant, and vanishing diquark gap in vacuum -- ensuring that
they describe an identical vacuum state. We then analyze how the
predictions of these two treatments diverge away from the vacuum and
in particular at finite chemical potential and temperature.

An important advantage of implementing an explicit renormalization
procedure (or an RG-consistent approximation) is that it enables us to
derive analytical results, providing benchmarks for the model’s
predictions under extreme conditions.  In particular, we demonstrate
that in the high-density limit, both approaches recover the expected
BCS relation between the zero-temperature diquark gap, $\Delgap(T=0)$,
and the critical temperature, $T_c$, of the superconducting
phase. Furthermore, we verify that as the quark chemical potential
$\mu$ increases, both treatments approach the free-quark
Stefan--Boltzmann limit for thermodynamic quantities, ensuring
consistency with standard thermodynamic expectations.  On the other
hand, we find that, in the QMD model, the diquark gap asymptotically
approaches a constant at large $\mu$, whereas in full QCD, the pairing
gap is expected to diverge in the limit $\mu\to\infty$ despite the
weakening of the attractive interaction at high
density~\cite{Son:1998uk,Schafer:1999jg}.  Nonetheless, the ability to
obtain analytic high-density results within the QMD framework remains
valuable, as it allows us to pinpoint the limitations of the model and
identify the missing physics necessary for a more accurate description
of dense quark-matter.

Our approach is based on the mean-field approximation (MFA). However
we note that, several alternative methods are being developed to
access the intermediate-density regime beyond the MFA. In particular,
these include approaches based on Dyson-Schwinger equations
\cite{Nickel:2006kc, Nickel:2006vf, Nickel:2008ef, Marhauser:2006hy,
  Muller:2013pya, Muller:2016fdr}, the FRG
\cite{Strodthoff:2011tz,Khan:2015puu,Braun:2021uua} and holographic
QCD \cite{CruzRojas:2025fzs}.

The remainder of this paper is organized as follows. In
~\cref{sec:qmd} we introduce the QMD model and its field content,
describing the quark, meson, and diquark sectors and introduce the
standard regularized mean-field approximation. In
\cref{sec:renormalization}, we present the renormalization scheme for
the QMD model, demonstrating how divergent vacuum contributions can be
absorbed into the couplings and discuss the determination of the model
parameters from vacuum observables. \cref{sec:rg-consistency} is
devoted to the FRG approach and the implementation of the
RG-consistent mean-field approximation, with special attention to the
treatment of medium-dependent divergences at finite chemical
potential. In \cref{sec:analytical} we present analytical results for
the different approximations of QMD model considered in this work. In
particular, we derive analytically the asymptotic behavior of the
diquark condensate and the BCS relation. In \cref{sec:results} we
compare numerical results for the renormalized and RG-consistent
approaches, examining the resulting phase diagram and thermodynamic
properties such as the diquark gap, critical temperatures, and
pressure. Finally, we summarize our findings in \cref{sec:conc},
highlighting the agreements and differences between the two treatments
and offering some outlook for future investigations.

\section{Quark-meson-diquark Model}\label{sec:qmd}

In this section, we introduce the $N_f=2$ quark-meson-diquark model
which serves as the foundation of our analysis. This model describes
the interaction between the quark fields $q$ and $\bar q$ and the
effective degrees of freedom associated with the $\sigma$ and
$\vec\pi$ mesons, as well as the diquark fields $\Delta$ and
$\Delta^*$. The action of the model in the Euclidean space at
finite temperature $\beta=1/T$ is given by
\begin{widetext}
\begin{align}
        S[\bar{q},q,\phi,  {\Delta}, {\Delta}^{*}] = &
        \int_0^\beta\!\! dx_4 \int\!\! d^3x \bigg\{
        \bar{q}\left[
            Z_q\left(\slashed{\partial} - \mu \gamma_4\right)
            + g_\phi \left(
            \sigma + i \gamma_5 \vec{\pi} \cdot \vec{\tau}
            \right)
        \right] q
        + \frac{1}{2} g_\Delta \left(
        \Delta_a \bar{q} \gamma_5 \tau_2 i\epsilon_a C \bar{q}^\tp
        - \Delta_a^* q^\tp C \gamma_5 \tau_2 i\epsilon_a q
        \right)                          \nonumber \\ &
        + \frac{Z_\phi}{2} (\partial_\mu \phi) (\partial_\mu \phi)
        + Z_\Delta (\partial_\mu + 2\mu\delta_{\mu 4})\Delta_a^*
        (\partial_\mu - 2\mu\delta_{\mu 4})\Delta_a
        + U(\phi^2, |\Delta|^2) - c\sigma
        \bigg\} \; .
        \label{eq:minimal_qmdm}
\end{align}
\end{widetext}

Here, the quark fields $q$ and $\bar q$ are understood as vectors in
color, flavor and Dirac space. The mesonic fields are grouped into the
$O(4)$-symmetric chiral field $\phi^\tp = (\sigma, \vec{\pi}^\tp)$
which provides a convenient basis for constructing the chirally
invariant quantity $\phi^2 = \sigma^2 +\sqvec{\pi}$. The
complex-valued diquark fields $\Delta$ and $\Delta^*$ are understood
as vectors in color space and their components $\Delta_a$ and
$\Delta^{*}_a$ carry a color index $a = 1,2,3$.

Throughout this work, we generally omit field indices for readability,
except for the diquark fields, where the color indices are always
shown explicitly. Additionally, we define the diquark invariant as
$|\Delta|^2 = |\Delta_1|^2 + |\Delta_2|^2 + |\Delta_3|^2$.
The Euclidean $\gamma$-matrices are defined such that
$\{\gamma_\mu,\gamma_\nu\}=2\delta_{\mu\nu}$ and the charge
conjugation matrix reads $C=\gamma_2\gamma_4$. We also introduced the
fully antisymmetric tensor in color space
$(\epsilon_a)_{bc}=\epsilon_{abc}$ and the three Pauli matrices
$\tau_i$ acting in flavor space.

In \cref{eq:minimal_qmdm}, the kinetic terms of the quark, meson and
diquark field are included with their associated wave function
renormalizations $Z_q$, $Z_\phi$ and $Z_\Delta$. In particular, note
the structure of the diquark kinetic term, which explicitly couples to
the chemical potential $\mu$ because the diquark field carries a
nonzero $U(1)_B$ charge.

We collect all bosonic interaction terms in a potential
$U(\phi^2, |\Delta|^2)$, which depends only on the chiral invariant
$\phi^2$ and the diquark invariant $|\Delta|^2$. The most general form
of this potential, which includes all relevant and marginal couplings,
is given by
\begin{equation}\label{eq:potential_min}
    \begin{split}
         & U(\phi^2, |\Delta|^2) =
        \frac{1}{2} m^2_\phi \phi^2
        + \frac{1}{4} \lambda_\phi \phi^4 \\ & \quad
        + m^2_\Delta |\Delta|^2
        + \lambda_\Delta |\Delta|^4
        + \frac{1}{2} \lambda_{\text{mix}} \phi^2 |\Delta|^2  \; .
    \end{split}
\end{equation}
In practical applications, it is common to omit some of the couplings
due to the lack of physical observables sensitive to diquark pairing,
which limits constraints on these couplings. A typical and simple
approximation sets $\lambda_\Delta = \lambda_{\text{mix}} =
0$. However, as we will show in \cref{sec:renormalization}, all of
these couplings are essential for renormalization.  We revisit the
issue of missing constraints in \cref{sec:parameter_fixing}.

\subsection*{Regularized mean-field approximation}

A common approach for studying the model's thermodynamic properties is
the mean-field approximation, where the bosonic fields are assumed to
take constant field configurations:
$\phi^\tp = (\sigma, \vec{0}^{\,\tp})$ and
$\Delta_a = \Delta \delta_{a3}$ with $\sigma \in \mathbb{R}$ and
$\Delta \in \mathbb{R}$. Note that in a slight abuse of notation we
use the same symbol for the homogeneous field configuration and for
the space-time dependent field. Then, neglecting all bosonic
fluctuations in the path integral, the effective thermodynamic
potential of the model can be computed as

\begin{equation}
  \label{eq:regularized_potential}
  \begin{aligned}
    \Omega_{\text{reg}}^{\text{eff}} (\sigma,\Delta) = &
    U(\sigma^2,\Delta^2) - c\sigma - 4 Z_\Delta \mu^2 \Delta^2 \\ 
    & + L_{\Lambda}(m_q, \Delta_{\text{gap}}) \; ,
  \end{aligned}
\end{equation}
where
\begin{equation}
  \label{eq:Omega_reg_therm}
  \begin{aligned}
    & L_{\Lambda}(m_q, \Delta_{\text{gap}}) = -2 N_f \int_{|\vec{p}|<\Lambda}
    \bigg\{ E_q^+ + E_q^- + \epsilon_q                        \\
    & \quad + 2T \ln\left( 1 + e^{- E_q^+ / T} \right)
    + 2T \ln\left( 1 + e^{- E_q^- / T} \right)                \\
    & \quad + T \ln\left( 1 + e^{- \epsilon_q^+ / T} \right) + T
    \ln\left( 1 + e^{- \epsilon_q^- / T} \right) \bigg\} \; ,
  \end{aligned}
\end{equation}
stands for the quark loop contribution with the dispersion relations
\begin{equation}
  \label{eq:3}
  E_q^\pm = \sqrt{{\epsilon_q^\pm}^2 + \Delta_{\text{gap}}^2} \; ,
\end{equation}
and 
\begin{equation}
  \label{eq:4}
  \epsilon_q^\pm = \epsilon_q \pm \mu
  \quad \text{with } \quad
  \epsilon_q = \sqrt{\sqvec{p} + m_q^2} \; .
\end{equation}
The quark mass $m_q$ and the Fermi-surface gap $\Delta_{\text{gap}}$
are related to the homogeneous field configurations through
\begin{equation}
  \label{eq:5}
  m_q = g_\phi \sigma
  \quad \text{and} \quad
  \Delta_{\text{gap}} = g_\Delta \Delta \; ,
\end{equation}
and correspond to the physical implications of the chiral and diquark
condensate on the quark sector. In particular, the dependence of
$L_\Lambda$ on $m_q$ and $\Delgap$, and not $\sigma$ or $\Delta$
individually, highlights that it originates from the quark physics.

In general, we use a short-hand notation for momentum integrals, where
the integral index specifies any possible
cutoffs
\begin{equation}
\label{eq:2}
\int_{\vec{p}} \equiv \int \!\!\frac{d^3p}{(2\pi)^3} \; ,
\quad \text{and} \quad
\int_{|\vec{p}|<\Lambda} \equiv \int \!\!\frac{d^3p}{(2\pi)^3} \theta(\Lambda^2 - \sqvec{p}) \; .
\end{equation}
Thus, the effective potential \cref{eq:regularized_potential} consists
of the tree-level bosonic potential with the explicit symmetry
breaking term, a quadratic diquark contribution from the diquark
kinetic term that still depends on the chemical potential, and a quark
loop contribution. The divergent quark loop is regularized with a
three-momentum cutoff, $|\vec{p}|<\Lambda$\footnote{In
  non-superconducting models, it is common to only regularize the
  divergent vacuum contribution, leading to reduced cutoff
  artifacts. However, for superconducting models the dispersion
  relation $E_q^\pm$ depends on the chemical potential and a
  separation into a divergent vacuum contribution and a convergent
  medium contribution is not possible. For details, see
  \cite{Gholami:2024diy}.}. We refer to
\cref{eq:regularized_potential} as the \textit{regularized} mean-field
approximation (regMFA) and use the subscript $\text{reg}$ accordingly.

Note that in obtaining \cref{eq:regularized_potential} we implicitly
chose to set $Z_q=1$. At the mean-field level, the quark wave function
renormalization does not receive any loop contribution and only
contributes through a rescaling of the Yukawa couplings $g_\phi$ and
$g_\Delta$. As such any dependency on $Z_q$ can be safely removed by a
rescaling of $g_\phi$ and $g_\Delta$, which is fully equivalent to
choosing $Z_q=1$ in \cref{eq:minimal_qmdm}. This is in contrast to the
meson and diquark wave function renormalizations which receive a loop
contribution. As we only consider homogeneous condensates, the meson
wave function $Z_\phi$ does not contribute to the effective potential
and can safely be ignored. However, the diquark wave function
renormalization $Z_\Delta$ contributes to the effective potential and
must, without further approximations, be considered.

The three-momentum cutoff regularization scheme is often used in the
literature due to its simplicity and the fact that at finite $T$ and
$\mu$ the Lorentz symmetry is already explicitly broken. 
Only the $T=0$ contribution of the quark loop diverges, as
can be seen from the asymptotic expansion
\begin{equation}
    \label{eq:vacuum_divergences}
    \begin{split}
         & L_{\Lambda}(m_q, \Delta_{\text{gap}})
        \underset{\Lambda \to \infty}{\simeq}{}
        -\frac{3 }{2 \pi ^2}{\Lambda}^4
        - \frac{ \left(3 m_q^2 + 2 \Delta^2_{\text{gap}}\right)}{2 \pi ^2}{\Lambda} ^2 \\
         & +\left(\frac{3 m_q^4}{4 \pi ^2}
        + \frac{m_q^2 \Delta_{\text{gap}}^2}{\pi^2}
        + \frac{\Delta_{\text{gap}}^4}{2 \pi ^2}
        - \mu ^2\frac{2 \Delta_{\text{gap}}^2 }{\pi ^2}\right)\ln \Lambda \; .
    \end{split}
\end{equation}
Note in particular the presence of a $\mu^2$ divergence, which we call
a \textit{medium divergence}.

The regularized effective potential \cref{eq:regularized_potential}
suffers from cutoff artifacts due to the absence of high-momentum
modes $|\vec{p}| > \Lambda$. A straightforward yet naïve solution is
to choose $\Lambda$ much larger than any physical scale in the system.
However, this procedure presents two difficulties.

The first difficulty is technical: the vacuum and medium contributions
to the effective potential diverge as $\Lambda\to\infty$ (see
\cref{eq:vacuum_divergences}). Due to these divergences, fixing the
bare parameters as $\Lambda\to\infty$ becomes a fine-tuning
problem. The renormalization of the model is the procedure that
resolves this fine-tuning issue.

The second difficulty is conceptual: in non-renormalizable models like
the NJL model, the cutoff $\Lambda$ is an intrinsic parameter -
similar to the bare couplings - and is chosen to reproduce vacuum
phenomenology. As a result, it may be impossible to eliminate all
cutoff artifacts by simply tuning $\Lambda$. RG-consistency offers a
way to addresss this issue, enabling the removal of cutoff artifacts
in both renormalizable and non-renormalizable models
\cite{Braun:2018svj,Gholami:2024diy}. We are thus faced with two
approaches that aim to achieve the same goal. It is therefore natural
to compare them directly.

In the following \cref{sec:renormalization}, we present a
straightforward renormalization procedure to eliminate the cutoff
dependence. Then, \cref{sec:rg-consistency} briefly reviews the
RG-consistency approach introduced in Ref.~\cite{Braun:2018svj} and
discusses how it handles medium divergences. Finally, in
\cref{sec:analytical,sec:results}, we compare the predictions of the
two methods using both analytical and numerical results.

\section{Renormalized Approach}
\label{sec:renormalization}

In the renormalized approach, the bare parameters are assumed to
depend on the cutoff scale $\Lambda$ such that the effective potential
remains finite as $\Lambda \to \infty$. To detail this, we derive in
\cref{sec:divergences_in_qmd} the asymptotic behavior of the
couplings required for finiteness of the potential. In
\cref{sec:parameter_fixing_and_renormalization}, we present an
explicit strategy for implementing the renormalized mean-field
approximation.

\subsection{Asymptotic expansion}
\label{sec:divergences_in_qmd}

We begin by noting that all divergent contributions to the effective
potential take the form shown in \cref{eq:vacuum_divergences}. By
rewriting this expression in terms of the condensates and comparing it
to the various bare couplings in $U(\phi^2,|\Delta|^2)$, we find that
all purely bosonic bare couplings are required to absorb the
divergences - except for the medium divergence, which can be handled
via the wave function renormalization of the diquarks.

Promoting all couplings and the diquark wave function
  renormalization in the regularized potential
  $\Omega_{\text{reg}}^{\text{eff}}$, \cref{eq:regularized_potential},
  to functions of the cutoff $\Lambda$, the renormalized effective
  potential $\Omega_{\text{ren}}^{\text{eff}}$ is eventually obtained
  by taking the limit $\Lambda \to \infty$.  An asymptotic expansion
  of the renormalized potential together with
  \cref{eq:vacuum_divergences} then yields
\begin{align}  \label{eq:Omega_asympt_ren}
         & \Omega_{\text{ren}}^{\text{eff}} (\sigma,\Delta)
        \underset{\Lambda\to\infty}{\simeq}
        \frac{1}{2}\bigg[
            \lambda_{\text{mix}}(\Lambda)
            + \frac{2g^2_\phi g^2_\Delta}{\pi^2} \ln \Lambda
        \bigg] \sigma^2 \Delta^2                         \nonumber  \\ &
        +\frac{1}{2} \bigg[
            m^2_\phi(\Lambda)
            - \frac{3g_\phi^2}{\pi^2}{\Lambda}^2
            \bigg] \sigma^2
        + \frac{1}{4} \bigg[
            \lambda_\phi(\Lambda)
            + \frac{3g_\phi^4}{\pi^2}\ln\Lambda
        \bigg] \sigma^4                                   \nonumber \\ &
        + \bigg[
            m^2_\Delta(\Lambda) - \frac{g_\Delta^2}{\pi^2} {\Lambda}^2
            \bigg] \Delta^2
        + \bigg[
            \lambda_\Delta(\Lambda)
            + \frac{g^4_\Delta}{2\pi^2} \ln\Lambda
        \bigg] \Delta^4                                  \nonumber  \\ &
        - 4 \bigg[
            Z_\Delta(\Lambda) + \frac{g_\Delta^2}{2\pi^2} \ln\Lambda
            \bigg] \mu^2 \Delta^2
        - \frac{3}{2\pi^2} {\Lambda}^4    
        \; .
\end{align}
From
\cref{eq:Omega_asympt_ren}, we conclude that maintaining a finite
effective potential as $\Lambda \to \infty$ requires the bare
couplings to scale with $\Lambda$ as follows:
\begin{align}
    \label{eq:34}
    m_\phi^2(\Lambda)           & = \frac{3g_\phi^2}{\pi^2}\Lambda^2 + \mathcal{O}(1) \; ,               \\
    m_\Delta^2(\Lambda)         & = \frac{g_\Delta^2}{\pi^2}\Lambda^2 + \mathcal{O}(1) \; ,              \\
    Z_\Delta(\Lambda)           & = -\frac{g_\Delta^2}{2\pi^2}\ln \Lambda + \mathcal{O}(1) \; ,         \\
    \lambda_\phi(\Lambda)       & = -\frac{3g_\phi^4}{\pi^2}\ln \Lambda + \mathcal{O}(1) \; ,           \\
    \lambda_\text{mix}(\Lambda) & = -\frac{2g_\phi^2g_\Delta^2}{\pi^2}\ln \Lambda + \mathcal{O}(1) \; , \\
    \label{eq:39}
    \lambda_\Delta(\Lambda)     & = -\frac{g_\Delta^4}{2\pi^2}\ln \Lambda +\mathcal{O}(1) \; .
\end{align}
The remaining parameters $g_\phi$, $g_\Delta$, and $c$ stay finite, as
they do not receive loop corrections at mean-field level, and can
therefore be treated as cutoff-independent.

\subsection{Vacuum matching scheme}
\label{sec:parameter_fixing_and_renormalization}

\begin{table}[t]
\centering
\bgroup
\def\arraystretch{2.2}
\begin{tabular}{l  c}
\textbf{Vacuum parameters} & \textbf{Physical interpretation} 
\\ \hline\hline
$\vcfpi = \langle\sigma\rangle_{\text{vac}}$ 
& pion decay constant \\
$\vcmsigma = \displaystyle \lrvac{\partial_\sigma^2\,\Omega^\text{eff}}$
& sigma mass \\
$\vcmDelta = \displaystyle \frac{1}{2}\lrvac{\partial_\Delta^2\,\Omega^\text{eff}}$
& diquark mass \\
$\vclambdamix = \displaystyle \frac{1}{2}\lrvac{\partial_\sigma^2\,\partial_\Delta^2\,\Omega^\text{eff}}$
& $\Delta$--$\sigma$ scattering amplitude \\
$\vclambdaDelta = \displaystyle \frac{1}{24}\lrvac{\partial_\Delta^4\,\Omega^\text{eff}}$
& $\Delta$--$\Delta$ scattering amplitude \\
$\vcZDelta = \displaystyle -\frac{1}{16}\lrvac{\partial_\mu^2\,\partial_\Delta^2\,\Omega^\text{eff}}$
& \makecell{diquark wave function\\ renormalization} \\
\end{tabular}
    \egroup
    \caption{Vacuum parameters and their definitions in terms of the
      effective action, along with their physical interpretation (in
      vacuum). The subscript $"{\text{vac}}"$ denotes evaluation at
      $\sigma=\sigvac$ and $\Delta=0$ in vacuum.
      }
    \label{tab:observables_description}
\end{table}

We now outline a concrete strategy to fix the model parameters. In
general, determining all couplings and their cutoff dependence
requires a sufficient number of constraints, typically obtained by
fitting a suitable set of vacuum parameters.  If the analysis is
restricted to the homogeneous phase diagram and momentum-dependent
observables (such as the pole masses) are not of interest, the mesonic
wave function renormalization $Z_\phi$ does not enter the calculation.
In contrast, the diquark wave function renormalization $Z_\Delta$ must
be taken into account, as it appears explicitly in the effective
potential through its coupling to the chemical potential, as
previously noted.

The importance of wave function renormalization for the
renormalizability of such effective models at finite density was
recently emphasized in
Refs.~\cite{Gholami:2024diy,Brandt:2025tkg}. Particularly in
Ref.~\cite{Brandt:2025tkg} and in the context of the two-flavor
quark-meson model at finite isospin density, it was shown within an
RG-invariant mean-field approximation that, once a single RG-invariant
scale is matched to lattice data, the resulting equation of state and
phase diagram agree quantitatively with modern lattice simulations
over a broad range of $\mu_I$.

As already mentioned, $g_\phi$, $g_\Delta$ and $c$ receive no loop
contributions in MFA. They can therefore be fixed directly, without
requiring corresponding vacuum parameters. This leaves six couplings
from \crefrange{eq:34}{eq:39} that must be determined. To illustrate
the procedure, we fix them using a set of \emph{vacuum parameters} --
a process we refer to as the \emph{vacuum matching scheme}. The chosen
set of parameters is
\begin{equation}\label{eq:obs}
    \mathcal{P}_{\text{vac}} = \{
        \vcfpi , 
        \vcmsigma , 
        \vcmDelta , 
        \vcZDelta ,
        \vclambdamix , 
        \vclambdaDelta
    \} \; .
\end{equation}
In \cref{tab:observables_description}, we list the vacuum parameters,
their definitions in terms of the effective action, and their physical
interpretations.  It is worth noting that this choice of parameters is
not unique -- alternative sets of vacuum parameters could be used. In
particular, no parameter is directly associated with $\lambda_\phi$,
which will instead be fixed through a combination of $\vcfpi$ and
$\vcmsigma$, as discussed in \cref{app:ren_pot_derivation}.

Note that the diquark wave function renormalization, \(Z_\Delta\), is
associated with single particle states and, in this regard, is not
directly linked to any physical observable. One can always fix
\(Z_\Delta\) to an arbitrary constant at a fixed scale, which in
turn rescales the fields and consequently modifies the values of
other model couplings. Here, we choose to fix this parameter in
vacuum. In contrast, the remaining parameters can be related to
physical observables involving multi-particle states that are, for
example, realized through certain scattering or decay processes. To
this end, we define a set of vacuum parameters that can be
associated with these physical quantities.

The pion decay constant is directly related to the chiral condensate
in vacuum\footnote{This identity assumes that the residue of the pion
  propagator at its pole is normalized to one. This condition can
  always be achieved by choosing an appropriate $\tilde Z_\phi$, which
  we have implicitly done here.  In contrast, in
  Refs.~\cite{Carignano:2014jla, Carignano:2016jnw}, the choice
  $Z_\phi=1$ was made. The relation between the pion decay constant
  and the chiral condensate was then given by
  $f_\pi = \sigma/\sqrt{Z_\pi}$, where $Z_\pi^2$ is the residue of the
  dressed pion propagator at the pion pole.}
   \begin{equation} \sigvac=\langle \sigma  \rangle_{\text{vac}}\ ,
\end{equation}
which is determined as the solution of the gap equation evaluated for
vanishing diquark condensate
\begin{equation}
  \label{eq:6}
  \frac{\partial\Omega^{\text{eff}}_{\text{ren}} (\sigma, \Delta=0)}{\partial \sigma} = 0 \; .
\end{equation}
For $\Delta=0$ we can use  \cref{eq:regularized_potential} and  obtain
\begin{equation} \label{eq:sigma_gap_ren}
  m_\phi^2\sigvac 
  + \lambda_\phi \sigvac^3 
  + \frac{\partial L_\Lambda (m_q, \Delta_{\text{gap}}=0)}{\partial \sigma } - c = 0 \; .
\end{equation}
With the definitions of the remaining vacuum parameters, one can
derive a linear system of equations that allows one to express the
bare parameters in terms of vacuum quantities. Substituting these
expressions into the effective potential and subsequently taking the
limit $\Lambda \to \infty$ yields a renormalized form of the
effective potential. The details of this procedure are presented in
\cref{app:ren_pot_derivation}. The renormalized potential takes the
following form
\begin{align}
\label{eq:ren_potential}
    \Omega^{\text{eff}}_{\text{ren}}(\sigma, \Delta) 
    = {}& U_{\text{ren}}(\sigma^2, \Delta^2) - c\sigma
    - 4 \vcZDelta \Delta^2 \mu^2 \nonumber\\ &
     + \Omega_{\text{ct}}(\sigma,\Delta)
    + L_{\text{ren}}(m_q,\Delta_\text{gap}) \; ,
\end{align}
where $U_{\text{ren}}$ is the renormalized potential contribution, now
expressed in terms of the vacuum parameters as
\begin{align}
    & U_{\text{ren}}(\sigma^2, \Delta^2) = 
    \vclambdaDelta\Delta^4  
    + \Bigl(-\frac{\vcmsigma}{4}+\frac{3c}{4\sigvac}\Bigr)\sigma^2 \nonumber\\ & \quad 
    + \Bigl(-\frac{c}{8\sigvac^3}+\frac{\vcmsigma}{8\sigvac^2}\Bigr)\sigma^4 
    + \frac{1}{2} \vclambdamix \sigma^2\Delta^2 \nonumber\\ & \quad
    + \Bigl(
        \vcmDelta 
        - \frac{1}{2} \vclambdamix\sigvac^2
    \Bigr)\Delta^2 \; ,
\end{align}
and $L_{\text{ren}}(m_q,\Delta_\text{gap})$ represents the renormalized loop
contribution
\begin{widetext}
\begin{align}
    & L_{\text{ren}}(m_q, \Delta_{\text{gap}}) = 
    2N_f \int_{\vec{p}} \bigg\{ 3p 
    + \frac{\Delgap^2\mu^2}{\epsvac^3}
    - \frac{\Delgap^4}{4\epsvac^3} 
    + \frac{\Delgap^2}{2\epsvac^5}p^2\left( 2p^2 + 5\mqvac^2 \right)  
    - \frac{\Delgap^2 m_q^2}{2\epsvac^5}\left( p^2 - 2\mqvac^2 \right) \nonumber \\ & \quad
    + \frac{3 m_q^2}{4\epsvac^3}\left( 2p^2 + 3 \mqvac^2 \right)  
    - \frac{3 m_q^4}{8\epsvac^3} 
    - E_q^+ - E_q^- - \epsilon_q \nonumber \\ & \quad
    - 2T \ln\left( 1 + e^{- E_q^+ / T} \right)
    - 2T \ln\left( 1 + e^{- E_q^- / T} \right)                
    - T \ln\left( 1 + e^{- \epsilon_q^+ / T} \right)
    - T \ln\left( 1 + e^{- \epsilon_q^- / T} \right)
    \bigg \} \; .
\label{eq:Lren}
\end{align}
\end{widetext}
It includes both the standard loop contribution and the additional
counterterm contribution arising from the bare parameters. This
results in a finite expression for the effective potential in the
limit $\Lambda \to \infty$.

Using this, we can define the gap equations
\begin{subequations}  \label{eq:gap_eq}
\begin{align}
    \left.
    \frac{\partial \Omega^\text{eff}_\text{ren}(\sigma, \Delta; T, \mu)}{\partial\sigma}
    \right|_{{\bar\sigma, \bar\Delta}}
    = 0 \; , \label{eq:gap_eq_sigma} \\
    \left.
    \frac{\partial \Omega^\text{eff}_\text{ren}(\sigma,\Delta; T, \mu)}{\partial\Delta}
    \right|_{{\bar\sigma, \bar\Delta}}
    = 0 \; .
    \label{eq:gap_eq_Delta}
\end{align}
\end{subequations}
By solving these two equations simultaneously, we get the physical gap
solutions $\bar\sigma(T,\mu)$, and $\bar\Delta(T,\mu)$, also denoted
as the condensates in the following and the thermodynamic potential is
then obtained by inserting these solutions back into the effective
potential. What is now missing are the values of the vacuum parameters
we have introduced. We will come back to them in
\cref{sec:parameter_fixing}.

Lastly, we note that one could, if desired, absorb the wave function
renormalization factor $\vcZDelta$ via the field redefinition
\(\Delta \!\to\! \Delta \vcZDelta^{-1/2}\). One can simultaneously
rescale the diquark Yukawa coupling as
\(g_\Delta \!\to\! g_\Delta \vcZDelta^{1/2}\), the diquark vacuum
curvature mass as \(\vcmDelta \!\to\! \vcmDelta\vcZDelta \), the
meson-diquark vacuum scattering coupling as
\(\vclambdamix \!\to\! \vclambdamix\vcZDelta \) and the diquark vacuum
scattering coupling as
\(\vclambdaDelta \!\to\! \vclambdaDelta\vcZDelta^2 \).  In that
representation the RG-invariant combinations
\(\Delgap = g_\Delta \Delta\), \(\vcmDelta \Delta^2\),
\(\vclambdamix \Delta^2\phi^2\) and \(\tilde\lambda_\Delta \Delta^4\)
remain unchanged. Proceeding similarly for the mesons, one can write
the effective action \eqref{eq:ren_potential} in an RG invariant way.
However, in this work and for clarity we keep the original
normalization, with \(\vcZDelta\) explicit.  Conversely, ignoring the
\(\vcZDelta\) factor when quoting the vacuum parameters, e.g. the
vacuum curvature masses in \cref{tab:observables_description}, would
mis-identify the intended parameter set.  Since this work enforces a
common vacuum baseline across all approximations, we retain
\(\vcZDelta\) explicitly in every step, akin to the canonical choice
of \(\vcZDelta=1\).

\section{RG-consistent Mean-field Approximation}
\label{sec:rg-consistency}

We now adopt an alternative perspective on the mean-field
approximation, which is based on the functional renormalization group
(FRG). First, we demonstrate how the FRG framework reproduces the
regMFA. We then use this approach to construct RG-consistent (RGC)
approximations, with a particular focus on the treatment of medium
divergences. This leads to alternative forms of the effective
potential, which we compare with the renormalized approach in
\cref{sec:analytical,sec:results}.

\subsection{FRG and MFA}
\label{sec:frg-mfa}

The FRG is rooted in the Wilsonian formulation of the renormalization
group in quantum field theory. The Wilsonian coarse-graining procedure
is captured by a functional differential equation for an effective
action. In this work, we employ the Wetterich equation
\cite{Wetterich:1992yh}, where the central object is the
scale-dependent one-particle irreducible effective action $\Gamma_k$
with the infrared (IR) cutoff scale $k$. For a purely fermionic
theory, the Wetterich equation takes the form
\begin{equation}
  \label{eq:WetterichEq}
  \partial_t \Gamma_k = - \frac{1}{2} \text{Tr} \left[
    \left(\Gamma^{(2)}_k + R_k\right)^{-1} \partial_t R_k
  \right] \; ,
\end{equation}
where $\partial_t = k \partial_k$ denotes the derivative with respect
to the dimensionless scale $t = \log(k/\Lambda)$ with the UV cutoff
scale $\Lambda$. In \eqref{eq:WetterichEq}, $\Gamma^{(2)}_k$ denotes
the full two-point function (i.e., the inverse full propagator) for
the fermion fields. With the superfield
$\Psi^{\tp} = (q^{\tp}, \bar{q})$, it is defined in momentum space as
\begin{equation}
  \label{eq:two_point_def}
  \Gamma^{(2)}_k =
  \frac{\delta^2  \Gamma_k}{\delta\Psi^{\tp} (p) \delta \Psi(p)}\ .
\end{equation}
The trace $\text{Tr}$ in \eqref{eq:WetterichEq} runs over all relevant
spaces, including momentum, color, flavor, and Dirac spaces on which
the fermionic fields are defined. The regulator function $R_k (p)$
implements the Wilsonian coarse-graining procedure within the
path-integral formulation by acting as a scale-dependent mass term.
In this work, we employ the sharp three-dimensional regulator function
for fermions \cite{Pawlowski:2015mlf}, given by
\begin{equation}
  \label{eq:3momentum_regulator}
  R_k(p) = i \slashed{\vec{p}} \, \left(
    \frac{1}{\theta(\sqvec{p} - k^2)} - 1
  \right) \; ,
\end{equation}
which renders momentum modes with $p<k$ infinitely massive, while
vanishing for $p>k$. This choice is particularly convenient, as it
leads to an expression for the effective potential that coincides with
the result obtained in the regMFA using a three-momentum cutoff, as
shown in \cref{eq:regularized_potential}.

Solving the Wetterich equation \eqref{eq:WetterichEq} with the initial
condition $\Gamma_{\Lambda}$ at the UV scale $k=\Lambda$ yields the
full effective action $\Gamma = \Gamma_{k=0}$ in the infrared
limit. In general, $\Gamma_k$ admits an infinite expansion in terms of
effective operators, rendering the exact treatment of the effective
action intractable without suitable truncations.

To recover the MFA results, we employ the following ansatz for the
effective action of the quark-meson-diquark model in Euclidean space
\begin{align}
    \label{eq:mfa_truncation}
  & \qquad \Gamma_k  = \int_0^\beta\!\! dx_4 \int\!\! d^3x \bigg\{
    \bar{q}\left[
    \slashed{\partial} - \mu \gamma_4
    + g_\phi
    \sigma
    \right] q \\
  &
    + \frac{1}{2} g_\Delta \Delta \left(
    \bar{q} \gamma_5 \tau_2 i\epsilon_3 C \bar{q}^\tp
    - q^\tp C \gamma_5 \tau_2 i\epsilon_3 q
    \right) + \Omega_k^{\text{eff}}(\sigma,\Delta) \bigg\} \; , \nonumber 
\end{align}
where $\sigma$ and $\Delta$ denote the homogeneous background fields
associated with the scalar $\sigma$ and the diquark $\Delta_3$
channels, respectively. The scale-dependent effective potential
$\Omega_k^{\text{eff}}(\sigma, \Delta)$ is treated as a general
function of $\sigma$ and $\Delta$. In the infrared limit, it
corresponds to the thermodynamic effective potential
\begin{equation}
  \label{eq:Omega_eff_definition}
  \Omega^{\text{eff}}(\sigma,\Delta)
  \equiv \frac{\Gamma_{k=0}(\sigma,\Delta)}{\beta V}
  = \Omega_{k=0}^{\text{eff}}(\sigma, \Delta) \; ,
\end{equation}
where $V$ denotes the three-dimensional spatial volume.  At the UV
scale $k=\Lambda$, we initialize the flow with the polynomial form
\begin{equation}
\label{eq:OmegaeffUV}
    \Omega_{k=\Lambda}^{\text{eff}}(\sigma,\Delta) =
    U(\sigma^2, \Delta^2) - 4Z_\Delta\mu^2\Delta^2 - c\sigma \; ,
\end{equation}
which is consistent with the classical action of the
quark-meson-diquark model in \cref{eq:minimal_qmdm}.

Next, we observe that for the truncation \cref{eq:mfa_truncation}, the
inverse two-point function is independent of the RG scale $k$ and
coincides with the second derivative of the classical action given in 
\cref{eq:minimal_qmdm},
\begin{equation}
  \Gamma^{(2)}_k = S^{(2)} \; ,
\end{equation}
where $S^{(2)}$ is defined analogously to
\cref{eq:two_point_def}. This identification lies at the heart of the
mean-field approximation and allows the Wetterich equation to be
rewritten as
\begin{align}\label{eq:mfa_wetterich}
    \partial_k \Gamma_k ={}&-\frac{1}{2} \Tr \left[
        \left( S^{(2)} + R_k \right)^{-1} \partial_k R_k
    \right]
    \nonumber \\
    ={}&-\frac{1}{2} \partial_k \Tr  \left[
        \ln \left( S^{(2)} + R_k \right)
    \right] \; .
\end{align}
With the truncation \cref{eq:mfa_truncation}, we obtain the flow
equation for the effective potential
\begin{equation}
    \label{eq:potential_flow}
    \partial_k \Omega_k^{\text{eff}}(\sigma,\Delta) = \partial_k f_k(\sigma,\Delta) \; ,
\end{equation}
where we introduce the shorthand notation
\begin{equation}
  \label{eq:definition_fk}
  f_k = - \frac{1}{\beta V} \frac{1}{2} \Tr \left[
    \ln \left(S^{(2)} + R_k\right) \right] \; .
\end{equation}
Evaluating the trace over flavor, color and Dirac indices, and
performing the Matsubara sums in \cref{eq:definition_fk}, one obtains
\begin{equation} \label{eq:f_equal_L}
  f_k(\sigma,\Delta) = - L_k(m_q,\Delta_\text{gap}) \; ,
\end{equation}
where $L_k(m_q,\Delta_\text{gap})$ is defined in \cref{eq:Omega_reg_therm}. 

To derive the infrared mean-field potential, we integrate
\cref{eq:potential_flow} from $k=\Lambda$ to $k=0$   and recover the
regularized effective potential
\eqref{eq:regularized_potential}
\begin{equation}
    \label{eq:meanfield_frg_potentail}
    \Omega^{\text{eff}}_{\text{reg}}(\sigma,\Delta)
    = \Omega_{\Lambda}^{\text{eff}}(\sigma,\Delta)
    + L_{\Lambda}(m_q, \Delta_{\text{gap}}) \; ,
\end{equation}
where we have used $L_{k=0}(m_q,\Delta_\text{gap})=0$.

Lastly, we remark that the mean-field flow \eqref{eq:mfa_wetterich}
with the sharp momentum regulator \cref{eq:3momentum_regulator} can be
understood as a mapping that directly relates the RG scale $k$ to the
UV cutoff scale $\Lambda$. With this in mind, one can alternatively
derive the scale dependence of the couplings, the
  $\beta$-functions, directly from the Wetterich equation by solving
their associated flow equations. This procedure reproduces the scale
dependency of the couplings in \crefrange{eq:34}{eq:39} and is
explicitly demonstrated in \cref{app:coupling_flow}.

In the following, we extend this analysis by utilizing the FRG flow to
eliminate cutoff artifacts, building on the concept of RG-consistency
\cite{Braun:2018svj}. In particular, we focus on the treatment of
medium divergences within this framework.

\subsection{RG-consistency of the QMD model}

The concept of RG-consistency was introduced in Ref.~\cite{Braun:2018svj}
and has recently been discussed in the context of the $N_f=3$ NJL
model in Ref.~\cite{Gholami:2024diy}. It requires that the full quantum
effective action $\Gamma$ remains  independent of the explicit cutoff
dependence of $\Gamma_{\Lambda}$,
\begin{equation}
    \Lambda \frac{\mathrm{d}\Gamma}{\mathrm{d}\Lambda} = 0 \; .
\end{equation}
In practice, however, this property is not automatically guaranteed if
external scales of the theory, such as the temperature or the quark
chemical potential, approach the cutoff scale $\Lambda$.

This issue can be resolved using the FRG.  The underlying idea is
  most transparently illustrated in the context of the NJL model
  \cite{Gholami:2024diy}.  Suppose the initial conditions are
  originally specified by the effective potential
  $\Omega_{\Lambda'}^\text{eff}(\sigma,\Delta)$ at some fixed scale
  $\Lambda'$, for instance obtained via a vacuum fit at that cutoff.
  By formally integrating the flow in vacuum upwards to a higher scale
  $k=\Lambda > \Lambda'$, one can construct a modified initial
  condition $\Omega_{\Lambda}^{\text{eff}}(\sigma,\Delta)$, that
  reproduces the same infrared effective potential in vacuum.
However, flowing downward from $k=\Lambda$ to $k=0$ yields an
RG-consistent effective potential
$\Omega_{k=0}^{\text{eff}}(\sigma,\Delta)$ even in the presence of
finite external scales, provided that $\Lambda$ is much larger than
these scales (e.g., $\Lambda \gg T,\mu$).

For the QMD model, the general strategy remains the same.
  However, a small but important difference arises: the effective
  potential at the initial UV scale, denoted henceforth as $\Lambda'$,
  includes an explicitly $\mu$-dependent term, see
  \eqref{eq:OmegaeffUV}.  This term, which is absent in the NJL
model, originates from the kinetic energy of the diquarks and is
essential for the renormalizability of the QMD model.  In the
following, we therefore explicitly separate it from the pure vacuum
initial condition by writing
\begin{equation}
\label{eq:Ominimu}
\Omega_{\Lambda'}^\text{eff}(\sigma,\Delta;\mu)
=
\Omega_{\Lambda'}^\text{eff}(\sigma,\Delta)
- 4Z_\Delta\mu^2\Delta^2, 
\end{equation}
where 
\begin{equation}
\Omega_{\Lambda'}^\text{eff}(\sigma,\Delta)
\equiv
\Omega_{\Lambda'}^\text{eff}(\sigma,\Delta;\mu=0)
=
    U(\sigma^2, \Delta^2) - c\sigma 
\end{equation}
is a $T$- and $\mu$-independent starting point, similar to the UV
initial condition in the NJL model.

From this, the RG-consistent initial effective potential at $k=\Lambda$,
$\Omega_{\text{rgc},\Lambda}^{\text{eff}}$, can be constructed by
integrating the mean-field flow, \eqref{eq:potential_flow}, upward
from $\Lambda'$ to $\Lambda$,
\begin{align}
  \label{eq:rgc_initial_condition1}
  &\Omega_{\text{rgc},\Lambda}^{\text{eff}} (\sigma,\Delta;\mu) 
\nonumber
\\  
  &\quad =\Omega_{\Lambda'}^\text{eff}(\sigma,\Delta)
      - 4Z_\Delta\mu^2\Delta^2
  + \mathcal{F}_{\Lambda'\to\Lambda}(\sigma,\Delta; 0, 0) \; ,
\end{align}
with the $\mu$-dependent but
scale independent $Z_\Delta$ term inherited from \eqref{eq:Ominimu}
and the flow integral 
\begin{align}
\label{eq:25}
  & \mathcal{F}_{\Lambda' \to \Lambda}  (\sigma,\Delta; T, \mu) =
\int\limits_{\Lambda'}^{\Lambda}dk \; \partial_k f_k (\sigma,\Delta; T,\mu)                                                                 \\
 & \quad =   L_{\Lambda'}(m_q,\Delgap; T, \mu) -
  L_{\Lambda}(m_q,\Delgap; T, \mu) \; . 
\end{align}
As indicated, the flow and loop contributions defined in
\cref{eq:definition_fk,eq:f_equal_L} and consequently the flow
integral $\mathcal{F}$ generally depend on $T$ and $\mu$.  However, in
\eqref{eq:rgc_initial_condition1} the UV effective potential at
$k=\Lambda$ is connected to the UV effective potential at $k=\Lambda'$
via a flow contribution in vacuum ($T=\mu=0$), in accordance with the
general idea outlined above.

The infrared effective potential at arbitrary $T$ and $\mu$ is then found by
integrating the flow from $k=\Lambda$ to $k=0$,
\begin{equation}
\label{eq:rgc_effective_potential}
    \Omega^{\text{eff}}_{\text{rgc}}(\sigma,\Delta; T, \mu)   
    =
    \Omega^{\text{eff}}_{\text{rgc},\Lambda}(\sigma,\Delta;\mu)
     + \mathcal{F}_{\Lambda \to 0}(\sigma, \Delta; T, \mu) \; .
\end{equation}

As we will discuss next, the initial conditions need to be modified
because of medium divergences. However,
\eqref{eq:rgc_effective_potential} remains valid for all RGC schemes
introduced in this context.

\subsection{Treatment of medium divergences}
\label{sec:schemes}

For the QMD model, the procedure outlined above removes possible
cutoff artifacts at finite $T$, but not at finite $\mu$. This is due
to the presence of the medium divergence in
\cref{eq:vacuum_divergences}. Since the modified initial condition in
\cref{eq:rgc_initial_condition1} contains only vacuum flow
contributions, it is insensitive to divergences that vanish when
$\mu=0$.\footnote{Here, we treat the diquark wave function
  renormalization $Z_\Delta$ as independent of the cutoff scale. In
  the renormalized approach discussed in \cref{sec:renormalization},
  it is exactly the running of this term that cancels the medium
  divergence.}

To address this issue, the solution originally proposed in
Ref.~\cite{Braun:2018svj} was to perform a Taylor expansion of the
upward flow in powers of $\mu$ and add this term to the right-hand
side of \cref{eq:rgc_initial_condition1}:
\begin{align} \label{eq:rgc_initial_condition2}
         & \Omega_{\sigma\Delta,\Lambda}^{\text{eff}}(\sigma,\Delta;\mu)
        = \Omega^\text{eff}_{\Lambda'}(\sigma,\Delta)       - 4Z_\Delta\mu^2\Delta^2
          \\ & \quad \nonumber
        + \mathcal{F}_{\Lambda'\to\Lambda}(\sigma,\Delta; 0, 0)      
        + \frac{\mu^2}{2} \left. \Big(
        \partial_\mu^2 \mathcal{F}_{\Lambda'\to\Lambda}(\sigma,\Delta; 0, \mu)
        \Big) \right|_{\mu=0} \; .
\end{align}
The RG-consistent approximation  to the effective potential is then again
obtained by integrating the flow from $k=\Lambda$ to $k=0$, as given in \cref{eq:rgc_effective_potential}.

In \cref{eq:rgc_initial_condition2}, we implicitly allowed the last
term to depend on the background fields \(\sigma\) and \(\Delta\).
For this reason we refer to this scheme as the $\sigma\!\Delta$
scheme.  Note however that this dependence is \emph{not} determined by
the requirement of RG consistency; any counterterm that removes the
logarithmic medium divergence
\(\propto\mu^{2}\Delta^{2}\,\ln(\Lambda)\) is admissible.
Consequently, there exists an intrinsic {scheme ambiguity} in the
construction of an RG-consistent mean-field potential.
 
An illustration of this ambiguity was given in
Ref.~\cite{Gholami:2024diy}.  In addition to the \(\sigma\Delta\)
scheme (which was called the \textit{massive scheme} in
Ref.~\cite{Gholami:2024diy}) the authors proposed a \textit{minimal
  scheme} that takes advantage of the fact that the divergent term
depends only on \(\mu^{2}\Delta^{2}\).  By setting \(\sigma=0\) and
expanding only to second order in \(\mu^{2}\) and \(\Delta^{2}\), one
obtains the modified initial condition
\begin{equation}
  \begin{aligned}\label{eq:RG_min}
    &\Omega^{\text{eff}}_{\min,\Lambda}(\sigma,\Delta; \mu) \\
     &\;= \Omega^\text{eff}_{\Lambda'}(\sigma,\Delta)       - 4Z_\Delta\mu^2\Delta^2
     +\mathcal{F}_{\Lambda'\!\to\!\Lambda}(\sigma,\Delta;0,0) \\
    &\quad+ \frac{1}{4}\mu^{2}\Delta^{2}
      \left.\Bigl(\partial_{\mu}^{2}\partial_{\Delta}^{2}
      \mathcal{F}_{\Lambda'\!\to\!\Lambda}(0,\Delta;0,\mu)
      \Bigr)\right|_{\mu=\Delta=0} \; ,
  \end{aligned}
\end{equation}
which removes the logarithmic divergence without introducing any
additional field-dependent counterterms.  Again, the RG-consistent
effective potential is obtained by integrating the flow from
$k=\Lambda$ to $k=0$ as in \cref{eq:rgc_effective_potential}.

We note that the last term in \cref{eq:RG_min} can be combined with
the $Z_\Delta$-term to define a scale-dependent wave-function
renormalization constant
\( Z_\Delta(k) \!\propto\!\partial_{\mu}^{2}\partial_{\Delta}^{2}
\Omega_{k} \).  In fact, the minimal prescription reproduces exactly
the running of \(Z_\Delta\) that appears in the fully renormalized
model, see \cref{sec:renormalization}.  In this sense it is the
\emph{closest} RG-consistent analogue of the renormalized mean-field
approximation known in the literature.

\subsubsection*{Vacuum matching scheme}

The fact that the subtraction term of the \emph{minimal scheme}
defined in \cref{eq:RG_min} can be reinterpreted as a
scale--dependent wave function renormalization of the diquark field,
makes this scheme particularly engaging.
However, there is a drawback once one wishes to confront the
RG--consistent framework with the standard (cutoff--)regularized
mean--field model:
Starting from the same effective potential at the UV scale, i.e.,
identifying $\Omega_{\Lambda'}^\text{eff}(\sigma,\Delta;\mu)$ in
\eqref{eq:Ominimu} with $\Omega_{\text{reg}}^{\text{eff}}$ in
\eqref{eq:regularized_potential} without the loop contribution, the
minimal scheme generates a \emph{different} vacuum value of the
diquark renormalization constant \(\vcZDelta\).
To make this point explicit, using the definition of $\vcZDelta$ in
\cref{tab:observables_description}, and using
\cref{eq:rgc_effective_potential} for the minimal scheme we find
\begin{align}\label{eq:zdeltamin}
    \vcZDelta^{(\text{min})}=  Z_\Delta
       &- \frac{1}{16} \left.\Big(
       \partial_\mu^2 \partial_\Delta^2
        \mathcal{F}_{\Lambda'\to\Lambda}(0,\Delta; 0, \mu)
        \Big)
         \right|_{\mu=\Delta=0}\nonumber\\  \quad
       &- \frac{1}{16} \left.\Big(
       \partial_\mu^2 \partial_\Delta^2
        \mathcal{F}_{\Lambda\to 0}(\vcfpi,\Delta; 0, \mu)
        \Big)\right|_{\mu=\Delta=0} \; ,
\end{align}
whereas for the regularized model \( \vcZDelta \) is obtained as
\begin{align}
  \label{eq:ozdelta}
  \vcZDelta^{(\text{reg})}&=  -\frac{1}{16} \partial_\mu^2 \partial_\Delta^2\Omega_\text{reg}^{\text{eff}}\Big|_\text{vac}\nonumber\\
 & =  
 Z_\Delta
       - \frac{1}{16} \left.\Big(
       \partial_\mu^2 \partial_\Delta^2
        \mathcal{F}_{\Lambda'\to 0}(\vcfpi,\Delta; 0, \mu)
        \Big)
         \right|_{\mu=\Delta=0} \; ,
\end{align}
where we used \cref{eq:regularized_potential,eq:25}, remembering that
the initial cutoff is now $\Lambda'$ and $L_{k=0}=0$.  Hence, taking
the same $Z_\Delta$, the two frameworks yield different results for
$\vcZDelta$.  Since we want to preserve $\vcZDelta$ ---as one of the vacuum parameters listed in \cref{tab:observables_description}--- across all approximations in order to enable a meaningful, parameter-synchronised comparison, this is a problem. In principle,
this could be cured by readjusting the bare wave function
renormalization constant $Z_\Delta$ in the minimal scheme, such that
$\vcZDelta^{(\text{min})}$ becomes equal to
$\vcZDelta^{(\text{reg})}$.  On the other hand, it is one of the
advantages of the RG-consistent approximation that cutoff artifacts
present in the regularized model can be removed \emph{without}
changing the bare parameters (in contrast to full renormalization).

To resolve this issue, we introduce an alternative approach—the RGC
vacuum matching scheme—which yields the same value for \( \vcZDelta \)
as for the regularized model while keeping $Z_\Delta$ (and all other
bare parameters of the model) identical to the ones in the regularized
model.  In this scheme, the initial condition is fixed in vacuum, that
is, evaluated at \(\sigma=\sigvac\), where \(\sigvac\) denotes the
vacuum chiral condensate, and at \(\Delta=0\). This leads to the
following modified initial conditions
\begin{equation}
    \begin{aligned}\label{eq:RG_vacmat}
        & \Omega^{\text{eff}}_{\text{vm},\Lambda}(\sigma,\Delta)
        \\
        &\;= \Omega_{\Lambda'}^{\text{eff}}(\sigma, \Delta)  - 4Z_\Delta\mu^2\Delta^2
        + \mathcal{F}_{\Lambda'\to\Lambda}(\sigma,\Delta;0,0) \\ &\quad 
        + \frac{1}{4}\mu^2\Delta^2 \left. \Big(
        \partial_\mu^2\partial_\Delta^2
        \mathcal{F}_{\Lambda'\to\Lambda}(\sigvac,\Delta; 0, \mu)
        \Big) \right|_{\mu=\Delta=0} \; .
    \end{aligned}
\end{equation}
This RGC scheme yields the same expression for \( \vcZDelta \) as the
regularized model:
\begin{align}
    \vcZDelta^{(\text{vm})} =
    Z_\Delta
    &  
       - \frac{1}{16} \left.\Big(
       \partial_\mu^2 \partial_\Delta^2
        \mathcal{F}_{\Lambda'\to\Lambda}(\vcfpi,\Delta; 0, \mu)
        \Big)
         \right|_{\mu=\Delta=0}\nonumber\\ & 
       - \frac{1}{16} \left.\Big(
       \partial_\mu^2 \partial_\Delta^2
        \mathcal{F}_{\Lambda\to 0}(\vcfpi,\Delta; 0, \mu)
        \Big)
         \right|_{\mu=\Delta=0}\nonumber\\= 
    Z_\Delta
    &          
       - \frac{1}{16} \left.\Big(
       \partial_\mu^2 \partial_\Delta^2
        \mathcal{F}_{\Lambda'\to0}(\vcfpi,\Delta; 0, \mu)
        \Big)
         \right|_{\mu=\Delta=0}\nonumber\\= 
    \vcZDelta&\!\!^{(\text{reg})} \; .
    \label{eq:ZDelta_vm}
\end{align}
We note that \eqref{eq:RG_vacmat} is equivalent to \eqref{eq:RG_min}
with a modified bare coupling $Z_\Delta$. However, the vacuum matching
scheme has the advantage that no explicit refit of $Z_\Delta$ is
necessary in order to preserve the vacuum parameters of the
regularized model.

From now on we will always assume that all bare parameters, including
$Z_\Delta$, in the RGC schemes are the same as the regularized
model. For the minimal scheme this implies that $\vcZDelta$ will take
a different vacuum value. A comparison of the results for the minimal
and the vacuum matching schemes will therefore give us a hint on the
importance of this parameter.

Note that the results obtained in our
RG‑consistent treatment differ in general from those of a renormalized
mean‑field model.  This can be explained as follows.  In the RGC
treatment, the effective potential is initialized at a finite scale
$\Lambda'$ as a polynomial of fourth order in $\sigma$ and $\Delta$,
see \eqref{eq:Ominimu}.  At any other scale $k\neq \Lambda'$, however,
the quark loop generates infinitely many higher-order contributions to
the effective potential beyond the initial fourth-order polynomial.
This includes RG scales $k > \Lambda'$, and in particular the limit
$k\to\infty$. In contrast, the renormalized model is defined by a
potential with the same fourth-order polynomial form at
$k \to \infty$.
In this case, the RGC treatment and the renormalized model correspond to
different UV effective actions. As a consequence, even if both models
are fitted to the same set $\mathcal{P}_{\text{vac}}$ of vacuum
parameters listed in \eqref{eq:obs} and defined in
\cref{tab:observables_description}, all {\it other} correlators --
such as the 6th-order derivatives of the effective potential --
generally differ.  In fact, even the fourth-order derivatives
$\tilde\lambda_\phi = \frac{1}{6}
\lrvac{\partial_\sigma^4\,\Omega^\text{eff}}$ are not identical in
both approaches because $\tilde\lambda_\phi$ does not belong to
$\mathcal{P}_{\text{vac}}$.\footnote{Alternatively, we could have
  chosen $\tilde\lambda_\phi$ to be an element of
  $\mathcal{P}_{\text{vac}}$ instead of \(\vcfpi\). However, we prefer
  to fit all models to the pion decay constant, which has a more
  direct physical interpretation.}

From the arguments above, it follows that the discrepancy increases as
the UV cutoff~\(\Lambda'\) is lowered.  On the other hand, in the
limit $\Lambda' \rightarrow\infty$, the UV effective action of the RGC
treatment converges to that of the renormalized model. Accordingly,
for sufficiently large $\Lambda'$, the differences between the two
approaches become negligible.  However, as we will see in
\cref{sec:results}, even for a relatively low cutoff
$\Lambda'= 600$~MeV, the prediction of the vacuum matching scheme and
the renormalized approach are mostly similar. In the next
\cref{sec:analytical}, we get back to this issue and analytically
study the differences between these schemes in more details.

\section{Analytical studies}
\label{sec:analytical}

Before presenting the numerical results of our calculations, it is
worthwhile to analytically examine certain aspects of the
approximations employed in this work.  Since the integrand in the gap
equation or thermodynamic potential is typically sharply peaked at
$p = \mu$, a common approximation, often found in the color
superconductivity literature \cite{Bailin:1983bm}, is to restrict the
momentum integration to a narrow window around the Fermi surface.
This approximation simplifies the calculation while retaining the
dominant physics of Cooper pairing -- particularly in the
weak-coupling regime, where the density of state is largest.  For
instance, in Ref.~\cite{Bowers:2002xr}, a sharp regulator was imposed
on the Fermi surface, yielding a finite and physically meaningful
expression for the diquark gap
\begin{align}\label{eq:deltaT0reg}
    \bDelgapZero = 2\,\omega \exp\left(-\frac{\pi^2}{2\lambda \mu^2}\right),
\end{align}
where \(\bDelgapZero\) denotes the diquark gap evaluated at $\sigma=0$
and vanishing temperature. \(\lambda\) is the coupling constant,
\(\mu\) the chemical potential, and \(\omega\) is the half of the size
of the chosen cutoff around the Fermi surface.  Another commonly
employed approximation, used to obtain finite and cutoff-independent
results, replaces parts of the integration measure in the
three-momentum integral by the chemical potential and extracts it from
the integral \cite{Schafer:1999jg, Hong:1998tn, Rajagopal:2000wf,
  Huang:2004am}
\begin{align}\label{eq:replace}
    \int_0^\Lambda dp \, p^2 f(p)\quad \longrightarrow \quad \mu^2\int_0^\infty dp f(p)\, .
\end{align}

Historically, such approximations were indispensable: without a proper
renormalization procedure, loop integrals diverge as the ultraviolet
integration bounds are taken to infinity. Moreover, a key benchmark in
any superconducting framework is the BCS relation, which connects the
critical temperature $T_c$ to the zero‑temperature pairing gap,
\begin{equation}
\label{eq:originalbcs}
   T_c \;=\; \frac{\mathrm e^{\gamma}}{\pi}\,\bDelgap(T=0)\simeq0.567 \, \bDelgap(T=0) \, ,
\end{equation}
where $\gamma\simeq0.5772$ is the Euler--Mascheroni constant.  In
weak‑coupling QCD, an analogous relation has been derived for color
superconductivity using similar approximations (see, e.g.,
Ref.~\cite{Pisarski:1999tv}).

In contrast, our renormalized QMD model, as well as its RG-consistent
treatment, allows us to evaluate the relevant integrals without such
assumptions or encountering divergences.
For massless quarks, i.e., $\sigma = 0$, at \(T=0\), we can solve the
gap equation for the diquark condensate \(\bar\Delta\) analytically,
obtaining an explicit relation between \(\bar\Delta\) and \(\mu\).
Moreover, we derive a closed-form expression of the gap equation at
the critical temperature \(T_c\), corresponding to the vanishing gap,
in terms of \(\mu\) and \(\bDelgap(T=0)\).  With both quantities
accessible analytically for arbitrary chemical potentials, we can
explicitly test the BCS relation \eqref{eq:originalbcs}. We note that
the case $\sigma=0$ considered in the following section can
alternatively be viewed as the mean-field solution of a quark-diquark
model \textit{without} mesons, see for example
Refs.~\cite{Braun:2018svj,Braun:2020bhy,Braun:2022olp} for discussions
on similar models.

In the remainder of this section we derive compact analytic formulas
for the pressure, the asymptotic diquark gap, the speed of sound and
for the BCS ratio \(T_c/\bDelgapZero\).  In addition to the
calculations for the renormalized model, we also study the asymptotic
diquark gap for different RGC schemes, and identify which of the
proposed approximations respects the BCS limit \eqref{eq:originalbcs}.

We note that some of our results for the renormalized model,
\cref{eq:weak_coupling_pressure,eq:delatasymp,eq:analytical_sos2},
have also been derived in Ref.~\cite{Andersen:2024qus} using
dimensional regularization, demonstrating the consistency between the
two approaches.  Moreover, the BCS relation was independently derived
in a model similar to the QMD model in Ref.~\cite{Deng:2006ed}.

As before, and throughout the following, we use the overbar
  notation to distinguish the solution of the gap equation,
  $\bar{\Delta}$, from the general diquark field $\Delta$.  The
  subscript \emph{gap} denotes the corresponding gap, obtained by
  multiplying $\Delta$ or $\bar{\Delta}$ with the diquark coupling
  $g_{\Delta}$. Finally, the subscript '\(0\)' indicates that the
  quantity, $\bar{\Delta}$ or $\bDelgap$, is evaluated at \(\sigma=0\)
  and \(T=0\).

\subsection{Pressure at $T=0$}
\label{sec:pressure_at_T0}

We start with the derivation of an analytical expression for the
pressure in the renormalized model, which we compare with the
Stefan-Boltzmann limit\footnote{Here we consider only quark degrees of
  freedom since mesons and diquarks do not contribute to the thermal
  pressure in mean-field approximation. In QCD there would be an
  additional gluonic contribution, which could be taken into account
  by coupling the present model to a Polyakov-loop potential.}
\begin{equation} \label{eq:pSB}
      p_{\mathrm{SB}} = \frac{N_f N_c}{6}\!\left(
        \frac{7\pi^{2}}{30}\,T^{4} \;+\; \mu^{2}T^{2} \;+\; \frac{\mu^{4}}{2\pi^{2}}
      \right) \; .
\end{equation}
For $\sigma =0$ and $T=0$, the loop and counterterm contribution
\eqref{eq:Lren} can be computed analytically
\begin{equation}
\begin{aligned}
    & L_{\text{ren}}(0,\Delgap; 0,\mu)
    = \frac{1}{8\pi^2} \bigg\{ 
    3 \Delgap^4 - 12 \Delgap^2 \mqvac^2  \\ & \quad
    - 4 \mu^4
    + 4\Delgap^2 (\Delgap^2 - 4 \mu^2) \ln \frac{\mqvac}{\Delgap}
    \bigg\} \; ,
\end{aligned}
\end{equation}
which yields for the effective potential
\begin{equation}
\begin{aligned}
    \label{eq:Omega_asympt}
    & \Omega^{\text{eff}}_{\text{ren}}(0,\Delta; 0, \mu) = 
    \left(
    \vcmDelta
    - \frac{1}{2} \vclambdamix\sigvac^2
    \right)\Delta^2 \\ & \quad
    -4 \vcZDelta \Delta^2 \mu^2 
    + \vclambdaDelta \Delta^4
    + L_{\text{ren}}(0,\Delgap; 0, \mu) \; .
\end{aligned}
\end{equation}

Following the procedure in Ref.~\cite{Shovkovy:2002kv} we use the
gap equation, \cref{eq:gap_eq_Delta}, to replace $\vcmDelta$.  Then,
using the definition of the pressure density
\begin{equation}
    p_{\text{ren}}(T,\mu) = 
    \Omega^{\text{eff}}_{\text{ren}}(\bar\sigma,\bar\Delta; 0, 0) -
    \Omega^{\text{eff}}_{\text{ren}}(\bar\sigma,\bar\Delta; T, \mu) \; ,
\end{equation}
this yields the analytical form
\begin{equation}\label{eq:poverpsb}
\begin{aligned}
    & \frac{p_{\text{ren}}}{p_{\text{SB}}} = 
    1 + 2 \frac{\bDelgapZero^2}{\mu^2} \\ & \quad
    + \frac{1}{4} \frac{\bDelgapZero^4}{\mu^4} \left( 
    1 
    + \frac{8\pi^2}{g_\Delta^4}\vclambdaDelta
    + 4 \ln \frac{\mqvac}{\bDelgapZero}
    \right) \; .
\end{aligned}
\end{equation}

We see that we recover the well-known first correction from the
superconducting gap to the Stefan-Boltzmann pressure proportional to
$\bDelgapZero^2 / \mu^2$
\cite{Rajagopal:2000ff,Shovkovy:2002kv}. 
Moreover, we observe from \cref{eq:poverpsb} that the quartic coupling
only affects sub-leading contributions proportional to
$\bDelgapZero^4$, which are negligible at high densities,
\begin{equation}\label{eq:weak_coupling_pressure}
    \frac{p}{p_{\text{SB}}} = 1 
    + \frac{2\bDelgapZero^2}{\mu^2} 
    + \mathcal{O}\Big(\frac{\bDelgapZero^4}{\mu^4}\Big) \; .
\end{equation}
We note that this expansion is a general feature of the model and
remains valid, even when additional terms beyond the quartic coupling
are included in the classical effective action.

\subsection{Diquark condensate at $T=0$}
\label{sec:delta_at_T0}

Next, we derive the diquark gap at vanishing temperature, which is
obtained as the solution of the gap equation
\cref{eq:gap_eq_Delta}. At zero temperature, the gap equation can be
directly computed from \cref{eq:Omega_asympt} and yields

\begin{align}\label{eq:T0_Delta_gap}
\begin{split}
  &     \left.
    \frac{\partial \Omega^\text{eff}_\text{ren}(\sigma, \Delta; 0, \mu)}{\partial\Delta}
    \right|_{\substack{\sigma=0\\\Delta=\bDeltaZero}} = \\
  &\bDeltaZero\bigg[
  4\vclambdaDelta\,\bDeltaZero^2 
  + 2\vcmDelta 
  - \vclambdamix \,\sigvac^2 
  - 8 \vcZDelta\mu^2 \\ &
+ \frac{g_\Delta^2}{\pi^2} \Bigl( \bDelgapZero^2 + 2\mu^2 - 3\mqvac^2 \\
&\qquad\qquad+ 2 \left( \bDelgapZero^2 - 2 \mu^2 \right) \ln \frac{\mqvac}{\bDelgapZero}
\Bigr)\bigg]=0 \; .
    \end{split}
\end{align}

The solution $\bDeltaZero(\mu)$ to this equation is the inverse
function of
\begin{equation}\label{eq:mu_of_delta}
    \mu^2(\bDeltaZero)= \frac{ \mathcal{C}(\bDeltaZero,\mathcal{P}_{\text{vac}})
    }{
    2\,g_\Delta^2 
    - 8 \pi^2 \vcZDelta
    - 4\,g_\Delta^2\,\ln\frac{\mqvac}{\bDelgapZero}
    } \; ,
\end{equation}
where the numerator is defined as
\begin{equation}
\begin{aligned} \label{app:numerator}
    & \mathcal{C}(\bDeltaZero,\mathcal{P}_{\text{vac}}) \equiv
    - 2 \pi^2 \vcmDelta
    + 3 g_\Delta^2\mqvac^2 + \pi^2 {\tilde f_\pi}^2 \vclambdamix    \\ & \quad 
    - 4 \pi^2 \bDeltaZero^2 \vclambdaDelta
    -  \bDelgapZero^2 \,g_\Delta^2 \Bigl(1 + 2\,\ln\frac{\mqvac}{\bDelgapZero}\Bigr)  \; ,
\end{aligned}
\end{equation}
and depends on the diquark condensate $\bDeltaZero$ and the
different vacuum parameters $\mathcal{P}_{\text{vac}}$ defined in
\cref{eq:obs}.

Because the denominator in \cref{eq:mu_of_delta} may vanish for
specific values of \(\bDeltaZero\), a pole can appear in this
expression. At this pole, \(\mu\) diverges while \(\bDeltaZero\)
remains finite. This defines the asymptotic value of the diquark
condensate,
\begin{equation}\label{eq:delatasymp}
    \lim_{\mu\to\infty} \bar\Delta^{(\mathrm{ren})}_{\text{gap}} = 
    \mqvac \exp{\left(
    - \frac{1}{2} 
    + \frac{2\pi^2}{g_\Delta^2}\vcZDelta
    \right)} \; .
\end{equation}
which coincides with the result of Ref.~\cite{Andersen:2024qus} if we
adopt their choice \(\vcZDelta = 1\).  The explicit dependence on
$\vcZDelta$ in our expression is a consequence of our treatment of the
diquark wave function renormalization. If we instead worked with the
canonically normalized diquark field,
$\Delta \to \vcZDelta^{-1/2} \Delta$, then the quantity
$g^2_\Delta / \vcZDelta$ would correspond to the infrared Yukawa
coupling of the canonically normalized field.  See the last paragraph
of \cref{sec:parameter_fixing_and_renormalization} for more details.
Contrary to expectations from QCD \cite{Alford:2008, Son:1998uk,
  Schafer:1999jg, Fukushima:2024gmp}, the QMD model predicts a finite
asymptotic limit for $\Delta$ as $\mu\to\infty$ in MFA.  Among the
parameters of the model only the infrared diquark coupling
$g_\Delta / \vcZDelta$ and the vacuum quark mass $\mqvac$ determine
the asymptotic value of the diquark gap.

Note that the numerator of \cref{eq:mu_of_delta} is nonzero at the
pole solution for the parameters chosen here. However, one could
select parameters such that \(\mu^2\) becomes negative, making \(\mu\)
imaginary and hence non-physical for some
\(\Delta < \lim_{\mu\to\infty}
\bar\Delta^{(\mathrm{ren})}_{\text{gap}} \). In this case, diquark
condensation might start at a large value already when \(\mu = 0\) and
then decrease to the asymptotic value as \(\mu\) grows. 
While this behavior is mathematically admissible for certain
  parameter choices, we do not consider it to be physically
  meaningful.

A similar analysis for the RGC minimal scheme gives
\begin{align}\label{eq:deltasymp_minimal}
  \lim_{\mu\to\infty} \bar\Delta^{(\mathrm{min})}_\text{gap} =
  2\Lambda' \exp{\left(
  - \frac{3}{2} 
  + \frac{2\pi^2}{g_\Delta^2} Z_\Delta
  \right)}
  \; ,
\end{align}
which resembles the $\mu\to\infty$ limit of \cref{eq:deltaT0reg}, as
it is proportional to the cutoff scale $\Lambda'$.  In contrast, the
RGC vacuum matching scheme yields (see \cref{app:RGCexpressions} for
details)
\begin{align} \label{eq:delasympvm}
    &  \lim_{\mu\to\infty} \bar\Delta^{(\mathrm{vm})}_\text{gap} = \mqvac \exp
    \bigg\{
        - \frac{1}{2} +\frac{2\pi^2}{g_\Delta^2} Z_\Delta
    \\& \nonumber
    \qquad- \frac{\Lambda'}{\sqrt{\Lambda'^2 + \mqvac^2}} 
        + \arctanh\frac{\Lambda'}{\sqrt{\Lambda'^2 + \mqvac^2}}
    \bigg\} \; ,
\end{align}
which then by using the relation for $\vcZDelta$ from \cref{eq:ZDelta_vm},
\begin{align}
    \vcZDelta =Z_\Delta
   - \frac{g_\Delta^2}{2\pi^2} 
   \bigg\{ &
   \frac{\Lambda'}{\sqrt{\Lambda'^2 + \mqvac^2}}\nonumber\\
   &- \arctanh
   \frac{\Lambda'}{\sqrt{\Lambda'^2 + \mqvac^2}}\bigg\} \; ,
\end{align}
reproduces the expression given in \cref{eq:delatasymp},
\begin{align}
\label{eq:deltasymp_vm}
      \lim_{\mu\to\infty} \bar\Delta^{(\mathrm{vm})}_\text{gap} =  \lim_{\mu\to\infty} \bar\Delta^{(\mathrm{ren})}_{\text{gap}} \; .
\end{align}

\subsection{Speed of sound at $T=0$}\label{sec:cs2}

From the pressure, the squared speed of sound at constant specific
entropy \(s/n\) can be computed directly as
\begin{equation}
    c_s^2 = \frac{d p}{d\epsilon} \bigg|_{s/n}
    = \frac{dp}{d\mu} \left(\mu\frac{d^2p}{d\mu^2}\right)^{-1} \; ,
\end{equation}
where the last equality holds only at zero temperature, $T=0$. Using
the fact that the diquark condensate saturates to a constant value as
$\mu\to\infty$, and employing \cref{eq:poverpsb}, we derive the exact
expression
\begin{equation} \label{eq:analytical_sos}
    c_s^2 = \frac{\mu^2+\bDelgapZero^2}{3\mu^2+\bDelgapZero^2} \; ,
\end{equation}
where we have assumed that the gap has reached its asymptotic value
and neglected its $\mu$ dependency. Expanding in powers of
$\bDelgapZero^2 / \mu^2$, we find
\begin{equation}
\label{eq:analytical_sos2}
    c_s^2 = \frac{1}{3} 
    + \frac{2}{9} \frac{\bDelgapZero^2}{\mu^2} 
    + \mathcal{O}\Big(\frac{\bDelgapZero^4}{\mu^4}\Big) \; ,
\end{equation}
which shows that the squared speed of sound will always approach the
conformal limit $c_s^2 = 1 / 3$ from above. This stands in contrast to
predictions from perturbative QCD \cite{Freedman:1976ub,
  Kurkela:2009gj, Kurkela:2016was, Gorda:2018gpy}. However, once
improvements to the perturbative expansion are implemented, either
through resummation techniques \cite{Fujimoto:2020tjc} or by
incorporating the pairing gap
\cite{Geissel:2024nmx,Fukushima:2024gmp,Geissel:2025vnp}, the
situation becomes less clear.

\subsection{The critical temperature}\label{subsec:bcs}

Next we derive the critical temperature. For this purpose we follow
the procedure that leads to the derivation of the BCS relation.  With
\(\sigma=0\), we consider the gap equations for the diquark field at
\(T=0\) and at \(T=T_c\) (with \(\Delta=0\)),
\begin{subequations}\label{eq:bcsgapeqs}
\begin{align}
    \left.
    \frac{\partial \Omega^\text{eff}_\text{ren}(\sigma, \Delta; 0, \mu)}{\partial\Delta}
    \right|_{\substack{\sigma=0\\\Delta=\bDeltaZero}} = 0 \; , \\
    \left.
    \frac{\partial \Omega^\text{eff}_\text{ren}(\sigma, \Delta; T_c, \mu)}{\partial\Delta}
    \right|_{\substack{\sigma=0\\\Delta=0}} = 0 \; .
\end{align}
\end{subequations}

Taking the difference of these equations leads to
\begin{align}
    &  
    \left.
    \frac{\partial \Omega^\text{eff}_\text{ren}(\sigma, \Delta; 0, \mu)}{\partial\Delta}
    \right|_{\substack{\sigma=0\\\Delta=\bDeltaZero}}-
    \left.
    \frac{\partial \Omega^\text{eff}_\text{ren}(\sigma, \Delta; T_c, \mu)}{\partial\Delta}
    \right|_{\substack{\sigma=0\\\Delta=0}}= \nonumber \\ &
    4 \vclambdaDelta\,\bDeltaZero^2 
    - \frac{g_\Delta^2}{3\pi^2} \bigg[
    2\pi^2 T_c^2 
    + 3 \bDelgapZero^2 \left( 
    2 \ln\frac{\bDelgapZero}{\mqvac}
    - 1
    \right)
     \nonumber \\ &
     + 12\mu^2 \left( 
        \ln\frac{\pi\; T_c}{\bDelgapZero} - \gamma
    \right)
    \bigg]
    = 0 \; ,
\end{align}
which can be solved to obtain \(T_c(\bDeltaZero,\mu)\) as
\begin{equation}\label{eq:Tc_of_Delta0}
\begin{split}
    & T_c^2(\bDeltaZero,\mu) = \frac{3\mu^2}{\pi^2}\,W\!\Biggl(
    \frac{\bDelgapZero^2}{3\mu^2}\,\exp\!\Biggl[
    2\gamma 
    \\ & \qquad + \frac{\bDelgapZero^2}{\mu^2}\Biggl(
    \frac{1}{2} 
    + \frac{2\pi^2}{g_\Delta^4}\vclambdaDelta
    - \ln\!\frac{\bDelgapZero}{\mqvac}
    \Biggr)
    \Biggr]
    \Biggr) \; ,
\end{split} 
\end{equation}
where \(W\) denotes the Lambert \(W\)-function (for more details see
\cref{app:TcBCS}). 
Furthermore, we obtain an explicit analytical relation between
  \(T_c\) and \(\mu\) by following the procedure detailed in
  \cref{app:TcBCS}. We get
\begin{align}\label{eq:Tcasmu}
 T_c^2(\mu)&=
  \frac{3\mu^{2}}{\pi^2}\,
  W\!\Bigl(
    \frac{\mqvac^2}{3\mu^{2}}\,
    \exp\Bigl[
    -1 + 2 \gamma + 
    \frac{4\pi^{2}\vcZDelta}{g_{\Delta}^{2}} \nonumber \\ &
    + \frac{1}{2\mu^{2}} \Bigl(
      (3 + \frac{\pi^{2}\vclambdamix}{g_{\Delta}^{2}\,g_\sigma^2}) \mqvac^{2}
      - \frac{2\pi^{2}}{g_{\Delta}^{2}} \vcmDelta
    \Bigr)
    \Bigr]
  \Bigr) \; .
\end{align}
The weak‑coupling regime is characterized by a vanishing ratio
\(\bDelgapZero/\mu\), which can be reached in two distinct ways:
either by taking \(\bDelgapZero\!\to 0\) ,
or by sending \(\mu\!\to\!\infty\). Using the relation \cref{eq:Tc_of_Delta0} and taking the
\(\bDelgapZero\!\to 0\) limit of \(T_c/\bDelgapZero\), we recover the
standard BCS ratio,
\begin{align}\label{eq:bcsweak}
   \lim_{\bDelgapZero\to0}\frac{T_c}{\bDelgapZero} = 
   \frac{e^\gamma}{\pi} \; ,
\end{align}
where we used $\lim_{x\to0} W(x) = x$.
Taking the limit \(\mu\to\infty\) of \cref{eq:Tcasmu} and using \cref{eq:delatasymp} yields
\begin{equation}\label{eq:BCSUS}
   \lim_{\mu\to\infty}  T_c = 
   \frac{\mathrm{e}^{\gamma}}{\pi}\lim_{\mu\to\infty} \bDelgapZero \; ,
\end{equation}
confirming the BCS relation as well.
Note, however, that the limits in \cref{eq:BCSUS,eq:bcsweak} are reached
along different trajectories in parameter space and therefore do not
coincide in general.
We will come back to this point in \cref{sec:testBCS}.

Similar expressions to \cref{eq:Tc_of_Delta0,eq:Tcasmu} can be derived
for all RGC schemes. These results are provided in
\cref{app:RGCexpressions}. Eventually, in the limit of
$\bDelgapZero\to0$, the BCS relation \cref{eq:bcsweak} is obtained for
all the RGC schemes studied in this work. However, in the limit of
$\mu\to\infty$ \cref{eq:BCSUS}, the BCS relation is only realized for
the RGC minimal scheme and the vacuum matching scheme but not in the
$\sigma\!\Delta$ scheme. Particularly, due to the $\Delta$-dependence
of the wave function renormalization factor in this scheme, one cannot
derive a closed-form expression for the diquark gap in the limit
$\mu\to\infty$ (see \cref{app:RGCexpressions} for a detailed
analysis.). Nevertheless, in the $\sigma\!\Delta$ scheme the critical
temperature can be expressed explicitly as a function of $\mu$
(see~\eqref{app:tcasmusigdel}), and this expression coincides with the
result obtained in the RGC‑minimal scheme.  Hence, for $\sigma=0$ the
two schemes yield identical 2SC phase boundaries.  With the choice of
$Z_\Delta=0$, the asymptotic value for the critical temperature for
these two schemes is shown to be directly connected to the scale
$\Lambda'$ via
\begin{align}\label{eq:bcssigdel}
   \lim_{\mu\to\infty}  T_c=\frac{e^{\gamma}\,}{\pi}\frac{2\Lambda'}{e^{\frac{3}{2}}} \; .
\end{align}
This holds even though \cref{eq:deltasymp_minimal}
\emph{is not} a solution of the $\sigma\!\Delta$ scheme's gap equation
at $T=0$.  Hence, while for the $\sigma\!\Delta$ scheme, the
standard BCS relation $T_{c}/\bDelgapZero=e^{\gamma}/\pi$ holds in the limit of $\bDelgapZero\to0$, it does
\emph{not} hold as $\mu\to\infty$. Instead, the ratio $\lim_{\mu\to\infty}T_{c}/\bDelgapZero$ converges to a value that
must be evaluated numerically and depends on the model parameters
specific to this scheme.

\subsection{Finite quark masses}

In the analysis above, we have assumed vanishing quark masses, i.e.,
$\sigma = 0$. This is correct in the chiral limit $c=0$ above the
chiral phase transition. In the following we argue that the results
remain valid as $\mu \to \infty$, even for $c\neq0$, i.e., away from
the chiral limit.  This assertion is supported by the following
observations:

\noindent
In the absence of diquark pairing ($\Delta = 0$), the chiral
condensate $\sigma$ tends to zero as $\mu$ approaches infinity.  This
aligns with the expectation that chiral symmetry is restored at high
densities and can be readily verified by solving the gap equation
\cref{eq:gap_eq_sigma}.
Conversely, when the chiral condensate $\sigma$ vanishes, the diquark
gap $\Delta$ asymptotically approaches a finite constant. This
behavior can be rigorously demonstrated away from the chiral limit as a
simultaneous solution to the gap equations \cref{eq:gap_eq} in the limit
$\mu \to \infty$.
To establish this result, we applied the dominated convergence theorem
to the gap equations, which justifies interchanging the limit and the
integral in the gap equations. This rigorous treatment confirms that
$\sigma \to 0$ and $\Delta \to \Delta_{\text{asymp}}$ constitute the
mutual solution to the gap equations at $T=0$ as $\mu \to
\infty$. Given the technical nature of this analysis, we refrain from
presenting the detailed calculations here.

\section{Numerical Results}\label{sec:results}

\subsection{Fixing the vacuum parameters}
\label{sec:parameter_fixing}

In this section, we compare numerical results obtained from the
different mean-field approximations to the QMD model introduced above.
To ensure a meaningful comparison, we fix the parameters such that all
versions reproduce the same vacuum physics, ideally based on a common
set of observables or quantities derived from QCD.  Due to the limited
availability of such data from either observations or first-principles
calculations, we adopt values obtained within the regularized
mean-field approximation (regMFA).

Our strategy proceeds as follows.  As detailed below, we first fix the
UV parameters of the regMFA model at a given cutoff scale $\Lambda'$.
These parameters determine the vacuum quantities listed in
\cref{tab:observables_description}, which then serve to define the
renormalized model, as outlined in
\cref{sec:parameter_fixing_and_renormalization}.  For the RGC schemes,
by contrast, we directly adopt the UV parameters of the regMFA as
input for the effective potential at the initial scale $\Lambda'$. In
the \emph{vacuum matching scheme}, the IR vacuum parameters are, by
construction, identical to those obtained in the regMFA approach and,
consequently, also to those of the renormalized model. In the minimal
scheme, however, the resulting value of $\vcZDelta$ differs (see
\cref{sec:schemes}), while the remaining vacuum parameters still
coincide with the other schemes. In the regMFA and likewise in the RGC
variants that retain an explicit wave function renormalization, the
vacuum value of the diquark renormalization constant is
\(\vcZDelta \neq 1\).  Because our renormalization conditions are
imposed on \emph{off-shell} correlators, the phenomenological input
for the diquark sector is therefore not fixed by the physical pole
mass.  For that reason we keep \(\vcZDelta\) explicit throughout, so
every scheme can be compared on an equal footing.

Let us now explain in detail how the parameters of the regMFA model
are fixed.  Starting from the action \eqref{eq:minimal_qmdm}, and
recalling that the quark and meson wave function renormalizations do
not enter directly in our equations, we are left with nine UV
parameters: the bare meson and diquark masses $m_\phi$ and $m_\Delta$,
the Yukawa couplings $g_\phi$ and $g_\Delta$, the quartic couplings
$\lambda_\phi$, $\lambda_\Delta$ and $\lambda_\text{mix}$, the
explicit symmetry breaking parameter $c$, and the diquark wave
function renormalization constant $Z_\Delta$.

Among these, $m_\phi$, $g_\phi$, $\lambda_\phi$ and $c$ belong to
  the quark-meson sector and can at least be partially constrained by
  vacuum observables such as the pion decay constant and the pion
  mass.  In contrast, since diquarks do not exist as asymptotic states
  in vacuum, the diquark sector cannot directly be fixed from
  experimental observables.  Here, we adopt values commonly used in
  the literature and compatible with typical theoretical expectations,
  see, e.g., Refs.~\cite{Hess:1998sd, Oettel:2000jj, Maris:2002yu} for
  estimates of the vacuum diquark mass.

In the absence of further constraints, we simplify 
  the bosonic potential, \cref{eq:potential_min}, by setting
  $\lambda_\Delta=\lambda_\text{mix}=0$, yielding the reduced form
\begin{equation}\label{eq:treelevel}
    U(\phi^2, |\Delta|^2) =
    \frac{1}{2} m^2_\phi \phi^2
    + \frac{1}{4} \lambda_\phi \phi^4
    + m^2_\Delta |\Delta|^2 \; .
\end{equation}
Additionally,
we set $Z_\Delta=0$.\footnote{This choice is particularly convenient
  for the RGC schemes, as it renders the UV initial conditions
  $\mu$-independent at the scale $\Lambda'$ in line with the NJL model
  spirit, see \eqref{eq:Ominimu}.}

We are thus left with six nonvanishing UV parameters: $m_\phi$,
$m_\Delta$, $g_\phi$, $g_\Delta$, $\lambda_\phi$, and $c$.  The
diquark Yukawa coupling is set to \(g_\Delta=4.5\), which is not
strongly constrained but yields a physically reasonable diquark gap at
the onset of the color-superconducting phase (see
\ref{sec:phase-struct-therm}).  To fix the remaining five parameters,
we choose a UV cutoff \(\Lambda'=600\,\mathrm{MeV}\) and fit the model
to the following vacuum properties:
\begin{enumerate}
    \item \textit{chiral condensate (pion decay constant)} 
    \begin{equation}
    \label{eq:fpi}
        \vcfpi= 92.4\,\mathrm{MeV} \; ,        
    \end{equation}
    
  \item \textit{pion mass}
    \begin{equation}
    \label{eq:mpi}
        m_\pi^2 = 2\lrvac{\partial_{\sigma^2}\Omega} = (137 \,\mathrm{MeV})^2 \; ,
    \end{equation}
    
    \item \textit{sigma mass}
    \begin{equation}
    \label{eq:msigma}
        \vcmsigma = (560 \, \mathrm{MeV})^2 \; ,
    \end{equation}
    
    \item \textit{quark mass}
    \begin{equation}
    \label{eq:mq}
        \mqvac = g_\phi \sigvac = 300\,\mathrm{MeV} \; ,
    \end{equation}

    \item \textit{diquark mass}
    \begin{equation}
        \label{eq:mDelta}
        \vcmDelta = (600 \MeV)^2 \; .
    \end{equation}
\end{enumerate}
The choice of the mesonic properties follows Ref.~\cite{Otto:2019zjy},
while for the diquark mass we just made the reasonable assumption that
it is twice the quark mass. This value is also consistent with
Refs.~\cite{Hess:1998sd,Oettel:2000jj,Maris:2002yu}.

The UV effective action parameters that reproduce these vacuum
properties are then determined as
\begin{align*}
m^2_\phi &= (951.5\,\mathrm{MeV})^2 \; , \\
m^2_\Delta &= (959.2\,\mathrm{MeV})^2 \; , \\
\lambda_\phi &= -1.34\, \; , \\
c &= (120.7\,\mathrm{MeV})^3 \; , \\
g_\phi &= 3.25 \; ,
\end{align*}
together with our choice \(g_\Delta=4.5\) and
$\lambda_\Delta=\lambda_\text{mix} = Z_\Delta=0$.  With this set of
bare parameters we can now calculate the six vacuum parameters
\cref{eq:obs} defined in \cref{tab:observables_description}.  The
numerical values are listed in \cref{tab:observables}.  Note that this
set of vacuum parameters has partial overlap with but is not identical
to the vacuum quantities in
\cref{eq:fpi,eq:mpi,eq:msigma,eq:mq,eq:mDelta}, which we used to fix
the UV parameters.  This is related to the fact that some of the UV
parameters, like $g_\phi$ and $c$ do not receive loop corrections in
mean-field approximation but, of course, need to be fixed nonetheless
(e.g., $c$ via the pion mass).  On the other hand, the vacuum
parameters listed in \cref{tab:observables_description} all receive
loop corrections, so that $\lambda_\Delta$, $\lambda_\text{mix}$ and
$Z_\Delta$, which we set equal to $0$ in the UV have nonzero vacuum
values in the IR.

\begin{table}[ht]
    \centering
    \def\arraystretch{1.3}
    \begin{tabular}{c c}
        \textbf{Vacuum parameters} & \textbf{Value} \\ \hline\hline 
         \(\vcfpi\)         & \(92.4\,\mathrm{MeV}\) \\
        \(\vcmsigma\)    & \((560\,\mathrm{MeV})^2\) \\
        \(\vcmDelta\)    & \((600\,\mathrm{MeV})^2\) \\
        \(\vclambdamix\) & {$-7.24$}  \\
        \(\vclambdaDelta\) & \(11.4\) \\
        \(\vcZDelta\)    & \(0.56\) \\
    \end{tabular}
    \caption{Values for vacuum parameters}
    \label{tab:observables}
\end{table}

With these values at hand, we now have access to the renormalized
effective potential.  Notably, if the integration in the renormalized
effective potential is carried out only up to a finite cutoff
\(\Lambda'\) rather than to infinity, the resulting expression remains
identical to that of the regularized potential. This property ensures
consistency between the regularized and renormalized models.

For the RGC schemes, as explained above, we adopt the same UV
parameters at the initial scale $\Lambda'$. This yields to identical
IR vacuum parameters as those listed in \cref{tab:observables}, with
the exception of $\vcZDelta$ in the minimal scheme, which takes the
value $\vcZDelta^{(\text{min})}=0.39$.  As we will see below, this
deviation has consequences for some of the results.

In the following, we present numerical results for the different
approximations discussed in this work. Details regarding the
computational procedures and numerical implementation are provided in
\cref{app:numerical_implementation}.
  
\subsection{Phase structure and thermodynamics}
\label{sec:phase-struct-therm}

\begin{figure}[thb]
    \centering
    \begin{subfigure}{0.48\textwidth}
        \centering
        \includegraphics[width=\textwidth]{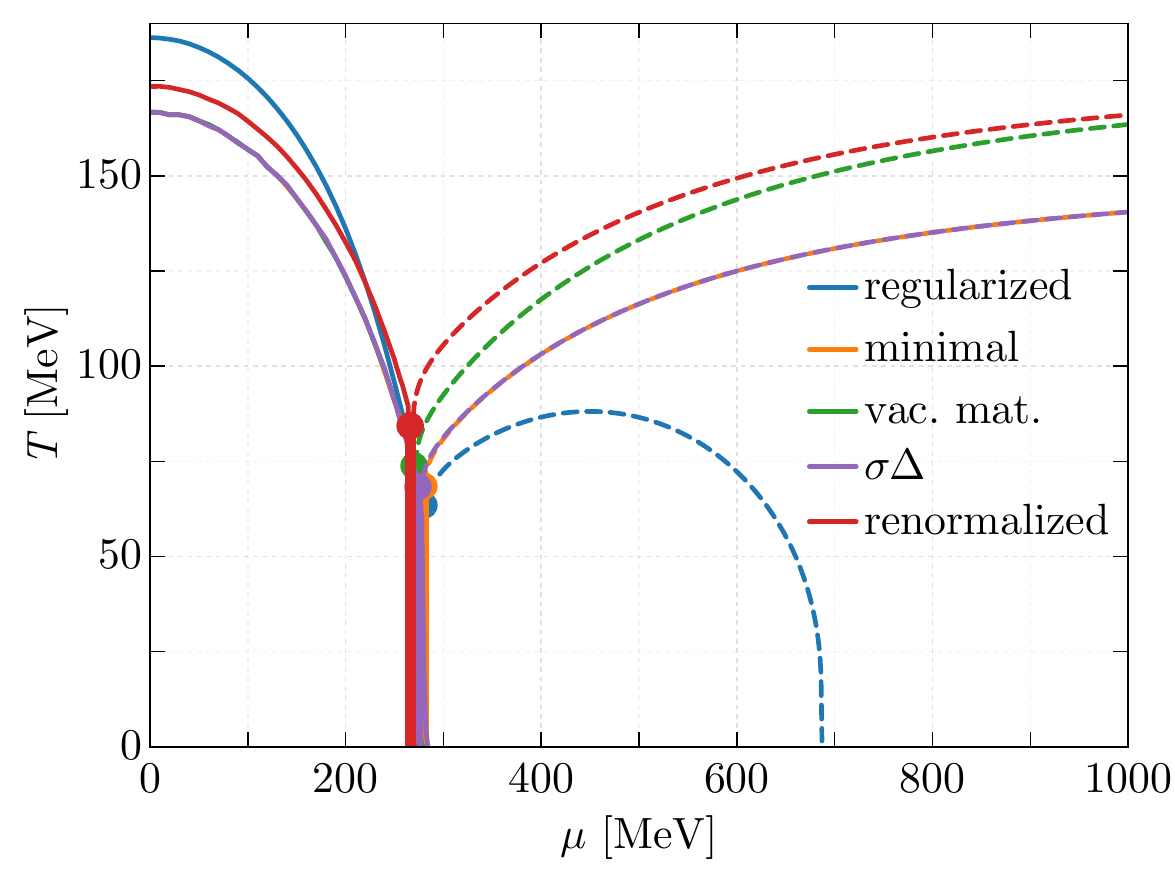}
    \end{subfigure}
    \caption{Phase diagram corresponding to the different
      approximation schemes considered in this work. Thin solid lines
      denote crossover transitions of the chiral condensate, thick
      solid lines indicate first-order transitions for both the chiral
      and diquark condensates, and dashed lines represent second-order
      transitions for the diquark condensate. Dots mark the location
      of critical endpoints.  Note that the critical temperatures in the minimal and
      \(\sigma\!\Delta\) schemes are identical.
      This is shown analytically for $\sigma =0$, see \cref{app:tcasmusigdel}, and is confirmed here numerically, even away from the chiral limit.
      \label{fig:phase_diagram}}
\end{figure}
\begin{figure*}[thb]
    \centering
    \begin{subfigure}{0.49\textwidth}
        \centering
                \includegraphics[width=\textwidth]{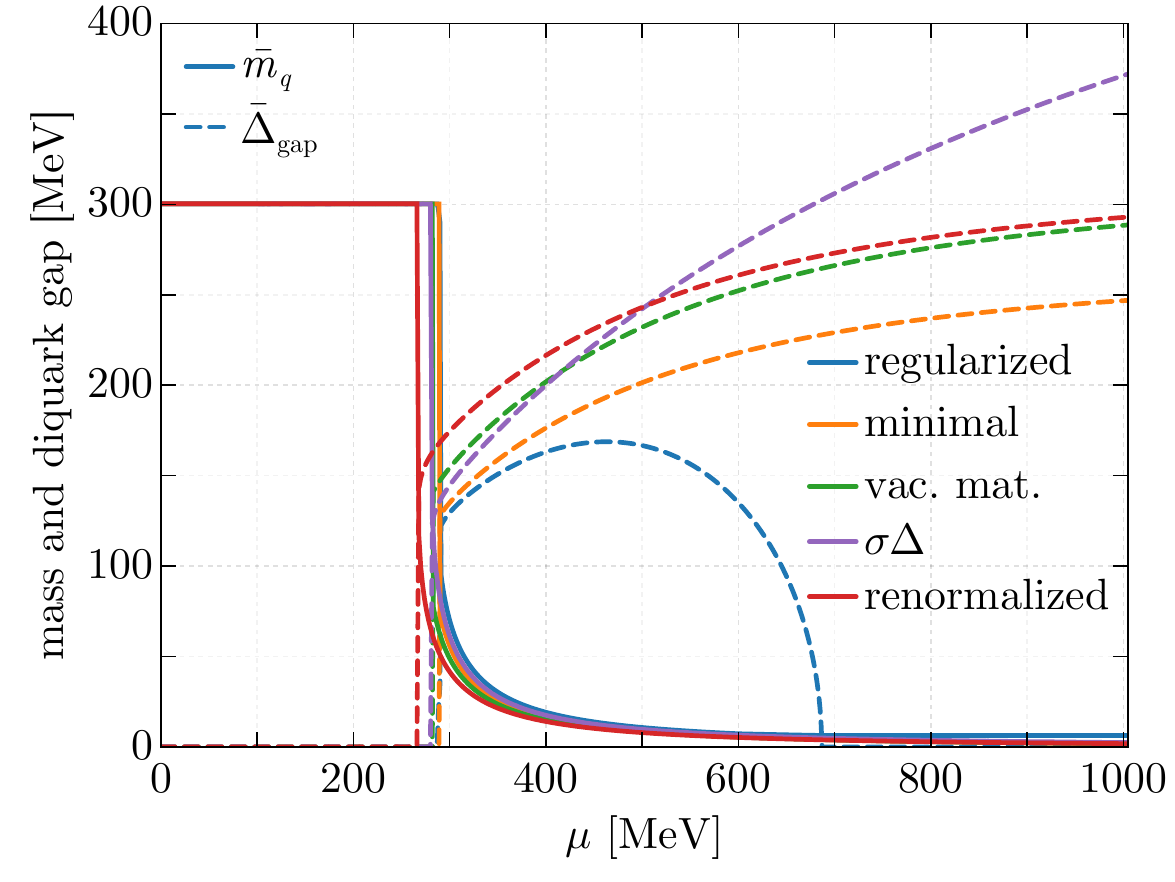}
    \end{subfigure}
    \hfill
    \begin{subfigure}{0.48\textwidth}
        \centering
        \includegraphics[width=\textwidth]{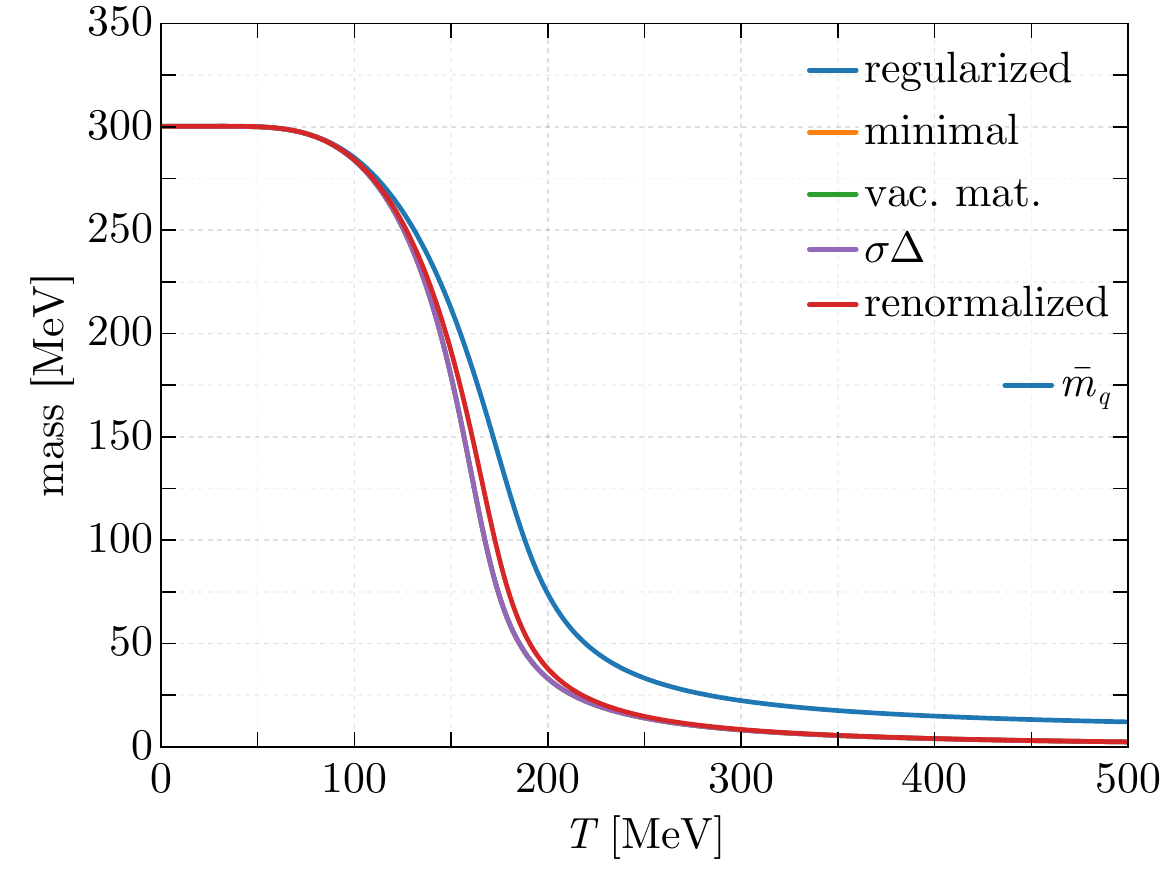}
    \end{subfigure}
    \caption{Quark mass $\bar m_q=g_\phi \bar \sigma$ and diquark gap
      $\bar\Delta_{\text{gap}}=g_\Delta \bar \Delta$ at $T=0\MeV$ (left), and quark mass at $\mu=0$ (right) for the different approximations
      considered in this work. For $\mu=0$ (right) the diquark gap always vanishes, $\bDelgap=0$, and is not shown. Also note that for $\mu=0$ the the minimal and the vacuum matching RGC schemes are identical, while the $\sigma\!\Delta$ scheme yields slightly different results.
      \label{fig:phase_diagram_and_T0_vs_mu}}
\end{figure*}

In \cref{fig:phase_diagram}, we present the QMD phase diagram obtained
within the regularized and renormalized mean‐field approximations,
including all RGC schemes ($\sigma\!\Delta$, vacuum matching and
minimal). Thin solid lines denote the chiral crossover, defined by the
minimum of \(\ \partial_\sigma^2\Omega\) at fixed \(\mu\), while thick
solid lines indicate first‐order phase transitions, where both chiral
and diquark condensate are discontinuous. Dashed lines correspond to
second‐order diquark transitions, and critical endpoints are marked by
dots. All approximations share the same qualitative structure: at low
\(T\) and \(\mu\), the system resides in a chirally broken phase
without diquark pairing; at low \(T\) and high \(\mu\), it enters the
chirally restored, color‐superconducting (2SC) phase; and at high
\(T\), chiral symmetry is restored with no diquark
condensate. Quantitatively, however, significant differences
emerge. In particular, the regMFA fails at large \(\mu\): its diquark
boundary bends downward and eventually disappears, reflecting the
absence of high‐momentum contributions in the medium integrals. In
contrast, all RGC schemes and the renormalized treatment maintain a
2SC phase up to arbitrarily large \(\mu\). Moreover, the regMFA
overestimates the chiral crossover temperature at low \(\mu\),
shifting the crossover line to higher \(T\), and delaying the onset of
the 2SC region to larger \(\mu\) at low $T$.

To understand the origin of these differences,
\cref{fig:phase_diagram_and_T0_vs_mu} shows the behavior of the
condensates as functions of \(\mu\) at \(T=0\) (left panel) and of
\(T\) at \(\mu=0\) (right panel). At zero temperature, the diquark gap
\(\bDelgap\) in the regMFA initially increases but
subsequently decreases and eventually vanishes at \(\mu\simeq 686.5\MeV\). In contrast, the
curves in the RGC schemes and the renormalized model exhibit the
expected monotonic rise of \(\bDelgap\), followed by
saturation to constant values.

Similarly, the quark mass \(\bar m_q=g_\phi\bar\sigma\) remains
relatively large in the 2SC phase and tends toward a nonzero constant
as \(\mu\to\infty\) in the regMFA, while it asymptotically vanishes in
the RGC and renormalized schemes.  These differences can be traced
back to the limited integration domain in the regMFA, which suppresses
diquark pairing and consequently delays the 2SC onset. A detailed
analysis for this behavior is provided in \cref{app:mfasos}.

Similarly, at \(\mu=0\), the regularized MFA restores chiral symmetry
only at higher \(T\), because the truncated thermal integrals slow
down the decrease of \(\bar m_q\) and leave it nonzero even as
\(T\to\infty\), similar to what happens at $T=0$ in the \(\mu\)
direction. In contrast, both the RGC and renormalized treatments drive
\(\bar\sigma\to0\) at large \(T\), giving a lower crossover
temperature. Thus, in both the \(\mu\)- and \(T\)-directions, the
limited integration domains in the regMFA favor chiral condensate
persistence, shifting the chiral‐restoration crossover as well as the
2SC onset to higher external parameters when compared to the properly
renormalized approaches.

\begin{figure*}[thb]
        \centering
        \includegraphics[width=\onefig]{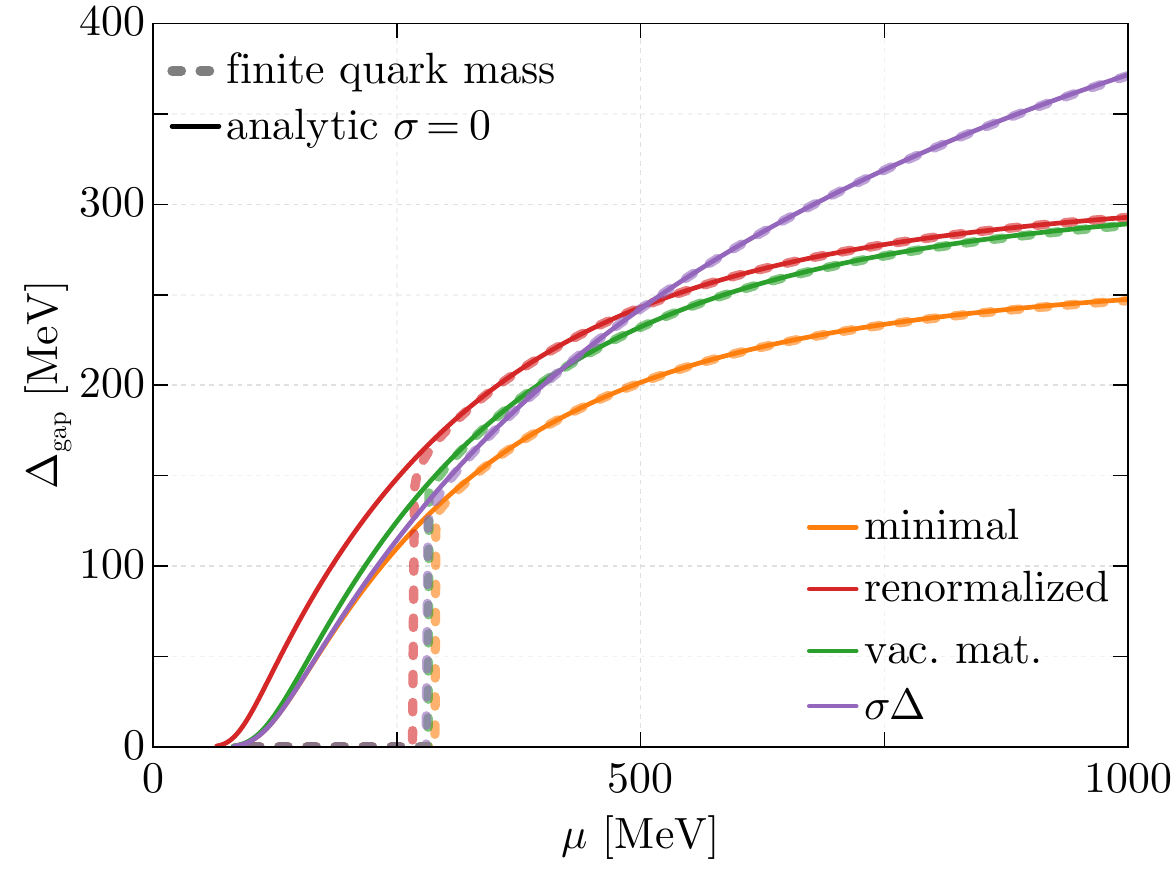}
        \hfill
        \includegraphics[width=\onefig]{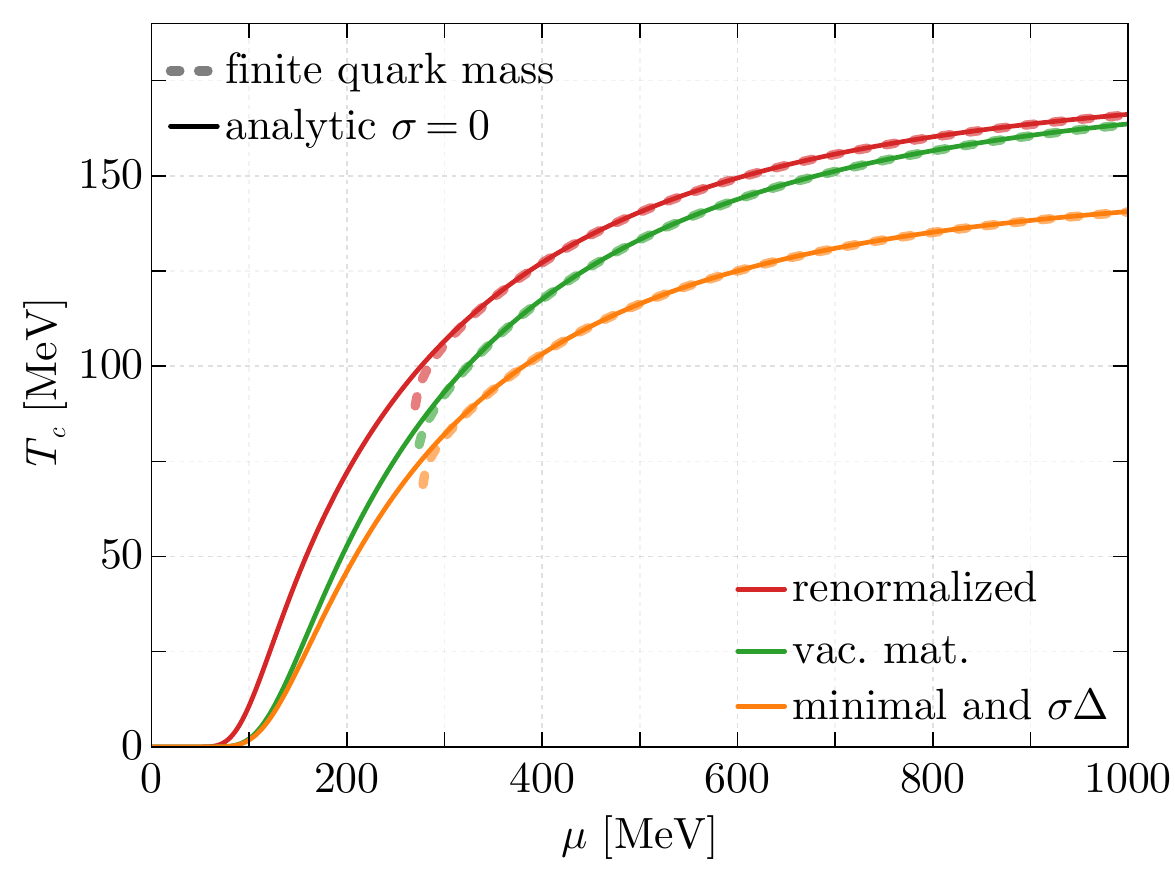}
    \caption{
Diquark gap $\Delgap$ (left) and critical temperature $T_c$ (right) against the chemical potential $\mu$. Solid lines indicate closed-form expressions derived in the $\sigma=0$ limit, while the dashed lines indicate results obtained numerically for the full solutions of the gap equation, including explicit chiral symmetry breaking, see also \cref{fig:phase_diagram_and_T0_vs_mu,fig:phase_diagram}.
}
    \label{fig:pd_analytic}
\end{figure*}

In \cref{sec:analytical,app:TcBCS,app:RGCexpressions}, we derived
closed‑form expressions for \(\mu(\bDelgapZero)\) and \(T_c(\mu)\) in
the limit \(\sigma= 0\).  To illustrate their accuracy,
\cref{fig:pd_analytic} compares these analytic results (solid curves)
with the corresponding numerical results at physical quark masses
(dashed lines taken from
\cref{fig:phase_diagram_and_T0_vs_mu,fig:phase_diagram}) for the
renormalized model and for RGC schemes.  In the left‑hand panel,
\(\bar\Delta_{\text{gap}}(T=0)\) is plotted against~\(\mu\). In every
case the numerical results merge with the chiral‑limit curves almost
immediately after diquark condensation sets in, indicating that at
high density the dynamics is governed by quark--diquark properties
while the chiral condensate (mesonic part) plays only a minor
role. The same rapid convergence is visible in the right‑hand panel
for the 2SC phase boundaries \(T_c(\mu)\).

The RGC vacuum matching and renormalized schemes differ only in their
UV completion of the vacuum sector. Their small residual discrepancies
arise from the choice of the RG matching scale \(\Lambda'\) and the
specific set of vacuum operators retained at that scale.  Both
approaches nevertheless agree closely with each other, showing that
once vacuum matching is imposed, the detailed structure of the UV
effective action has only a minor impact on the medium physics. The
RGC minimal scheme, however, deviates further due to its different
value of \(\vcZDelta\), and accordingly its diquark gap approaches a
different constant at large \(\mu\): in the minimal RGC scheme one
finds (see \cref{eq:deltasymp_minimal})
\[
\lim_{\mu\to\infty}\bar\Delta^{(\mathrm{min})}_{\mathrm{gap}}
=267.8 \MeV \; ,
\]
whereas both the RGC vacuum matching and the renormalized schemes
converge to (see \cref{eq:delatasymp,eq:deltasymp_vm})
\[
\lim_{\mu\to\infty}\bar\Delta^{(\mathrm{vm})}_{\mathrm{gap}}
= \lim_{\mu\to\infty}\bar\Delta^{(\mathrm{ren})}_{\mathrm{gap}}
=315.2 \MeV \; ,
\]
consistent with the numerical results at \(\mu=1000\MeV\) in
\cref{fig:phase_diagram_and_T0_vs_mu} (left). Finally, the
$\sigma\!\Delta$-scheme shows the most different behavior compared to
the other schemes as the diquark gap grows much faster with
$\mu$. This is confirmed by the asymptotic value, which can be
computed {from the procedure outlined in \cref{app:RGCexpressions}}
(see \cref{eq:asympt_delta_sigmadel} and also \cref{fig:bcs}) to yield
\begin{equation*}
    \lim_{\mu \to \infty} \bDelgap^{(\sigma\Delta)} =
    584.2 \MeV \; .
\end{equation*}

Finally, as discussed in \cref{subsec:bcs}, the renormalized, as well
as the minimal and the vacuum matching RGC schemes exhibit BCS scaling
for the critical temperature \(T_c\) associated with the melting of
the 2SC phase in the limit $\mu \rightarrow \infty$ as given by
\cref{eq:BCSUS}.  Inserting the gaps at \(\mu=1000\MeV\), one finds
\begin{align*}
    T_c^\text{(min)}&\;\approx \; 0.567\times 246.7 \MeV \; \approx 139.9 \MeV \; ,\\
    T_c^\text{(vm)} \; \approx \; T_c^\text{(ren)}&\;\approx \; 0.567 \times 292.8 \MeV \; \approx 166.0 \MeV \; ,
\end{align*}
in very good agreement with the 2SC boundaries in \cref{fig:phase_diagram}
already at $\mu=1000$ MeV:
\begin{align*}
    T_c^{(\text{min})} & = 140.5 \MeV \; , 
    \quad \text{and}\quad \\
    T_c^{(\text{ren})} & = 166.0 \MeV \; .
\end{align*}
For the $\sigma\!\Delta$ scheme, we do not expect the BCS relation to
hold.  As shown analytically, the 2SC phase boundaries of the minimal
scheme and the $\sigma\!\Delta$ scheme coincide for $\sigma=0$, and
our results in~\cref{fig:phase_diagram} indicate that this agreement
persists even away from the chiral limit.  Despite the identical
critical temperature curves in ~\cref{fig:phase_diagram},
~\cref{fig:phase_diagram_and_T0_vs_mu} reveals a pronounced difference
in the diquark gap, confirming that the system is far from the BCS
limit.  Using the asymptotic critical temperature from
\cref{eq:tcasympsigdel},
\( \lim_{\mu\to\infty}T_c^{(\sigma\Delta)} = 151.8~\text{MeV}, \)
together with the asymptotic gap
\( \lim_{\mu\to\infty}\bDelgapZero^{(\sigma\Delta)} =
584.2~\text{MeV}, \) we obtain the ratio
\( \lim_{\mu\to\infty}\frac{T_c}{\bar\Delta} = 0.260 \) for the chosen
parameter set. Using this ratio, at \(\mu=1000\MeV\), we find
\begin{align*}
    T_c^{(\sigma\Delta)}&\;\approx \; 0.260\times 373.6 \MeV \; \approx 97.14 \MeV \; .
\end{align*}
This value does not match the computed 2SC boundary at
\(T_c^{(\sigma\Delta)}(\mu=1000\;\text{MeV}) = 140.5\;\text{MeV}\).
The discrepancy arises because, in the $\sigma\!\Delta$ scheme,
both \(T_c\) and the diquark gap approach their asymptotic limits much
more slowly; the gap, in particular, saturates only at considerably
larger~\(\mu\) (see also~\cref{sec:testBCS}).

\begin{figure*}
    \centering
    \begin{subfigure}[t]{0.49\textwidth}
        \centering
        \includegraphics[width=\textwidth]{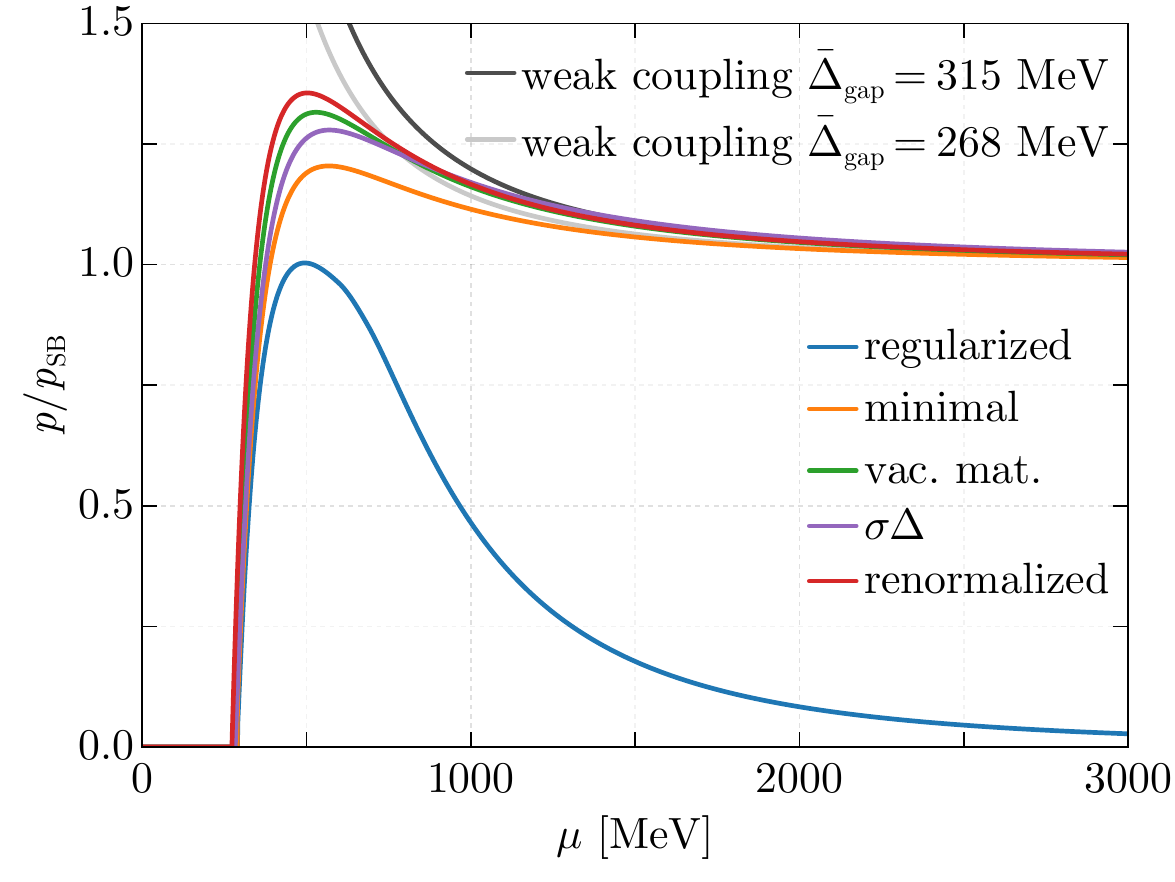}
    \end{subfigure}
    \hfill
    \begin{subfigure}[t]{0.49\textwidth}
        \centering
        \includegraphics[width=\textwidth]{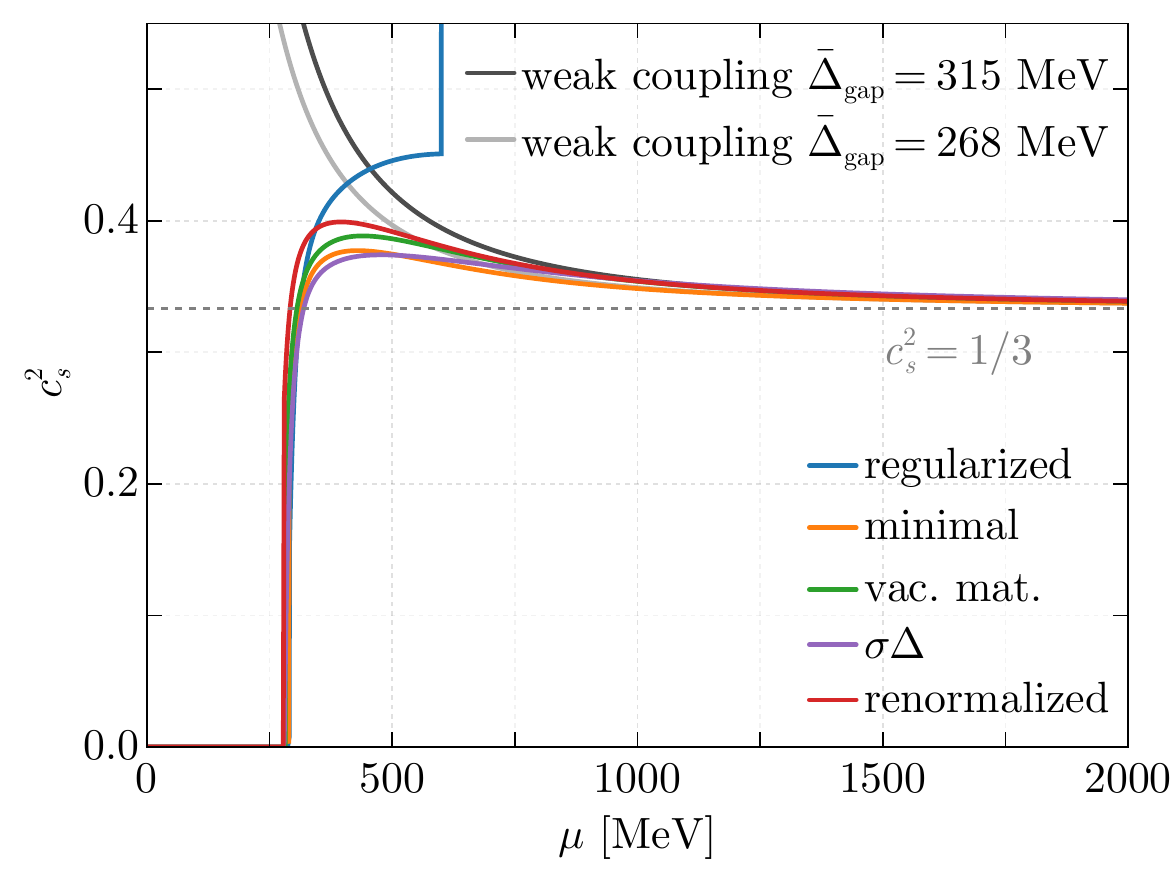}
    \end{subfigure}
    \caption{Pressure $p$ normalized to the Stefan-Boltzmann pressure
      $p_{\text{SB}}$ (left) and squared  speed of sound (right) as
      functions of $\mu$ at $T=0\MeV$. 
      The black and gray solid lines indicate
      the asymptotic behavior of the pressure, 
      \eqref{eq:weak_coupling_pressure}, and of the speed of sound, \eqref{eq:analytical_sos2},
      expected from a
      weak-coupling expansion.           
      The gray dashed line marks the
      conformal limit, $c_s^2 = 1/3$ (right).}
    \label{fig:T0_pressure_sos}
\end{figure*}

In \cref{fig:T0_pressure_sos}, we show the pressure $p$ and the
squared speed of sound $c_s^2$ at $T=0$ as functions of the chemical
potential.  As observed, the regMFA fails to approach the
Stefan-Boltzmann limit for $\mu\to\infty$. Again, this deficiency is
absent in the RGC and renormalized schemes, where the Stefan-Boltzmann
limit is gradually attained in the high-density regime. For the speed
of sound (right panel) we observe a peak exceeding the conformal limit
$c_s^2 = 1/3$ , which is a characteristic feature of models
incorporating diquark condensation~\cite{Leonhardt:2019fua,
  Braun:2021uua, Andersen:2024qus, Andersen:2025ezj,
  Gholami:2024ety}. As discussed in \cref{sec:cs2}, this quantity
approaches the conformal value from above at large $\mu$.

Additionally, in the regMFA the speed of sound exhibits an unphysical
divergence at higher chemical potentials. This can be traced back to
the behavior of the condensates in this approximation: as shown in
\cref{fig:phase_diagram_and_T0_vs_mu}, the diquark condensate vanishes
and the chiral condensate $\bar\sigma$ tends toward a constant at
large $\mu$. Under these conditions, the regulated medium integrals
lead to a linear increase of the pressure $p$ with $\mu$, while the
energy density $\epsilon$ asymptotically saturates to a constant value
(see \cref{app:mfasos} for details).  Since the squared speed of sound
is defined as $ c_s^2 = \frac{dp}{d\epsilon}, $ this leads to
$\frac{dp}{d\epsilon} \to \infty\ ,$ implying a divergent $c_s^2$ at
high chemical potentials in the regMFA. This unphysical behavior of
the speed of sound was discussed for the NJL model in
Ref.~\cite{Pasqualotto:2023hho}.

We also compare the pressure and speed of sound to the weak coupling
expansions, \cref{eq:weak_coupling_pressure,eq:analytical_sos2}
respectively, for two constant diquark gap values corresponding to the
asymptotic values for the minimal and renormalized schemes. At large
densities $\mu \gtrsim 1.5 \GeV$, we find a very good agreement
between the weak coupling expansion and the numerical results as
already noted in Ref.~\cite{Braun:2018svj}.

\begin{figure*}
    \centering
    \begin{subfigure}[t]{0.49\textwidth}
        \centering
        \includegraphics[width=\textwidth]{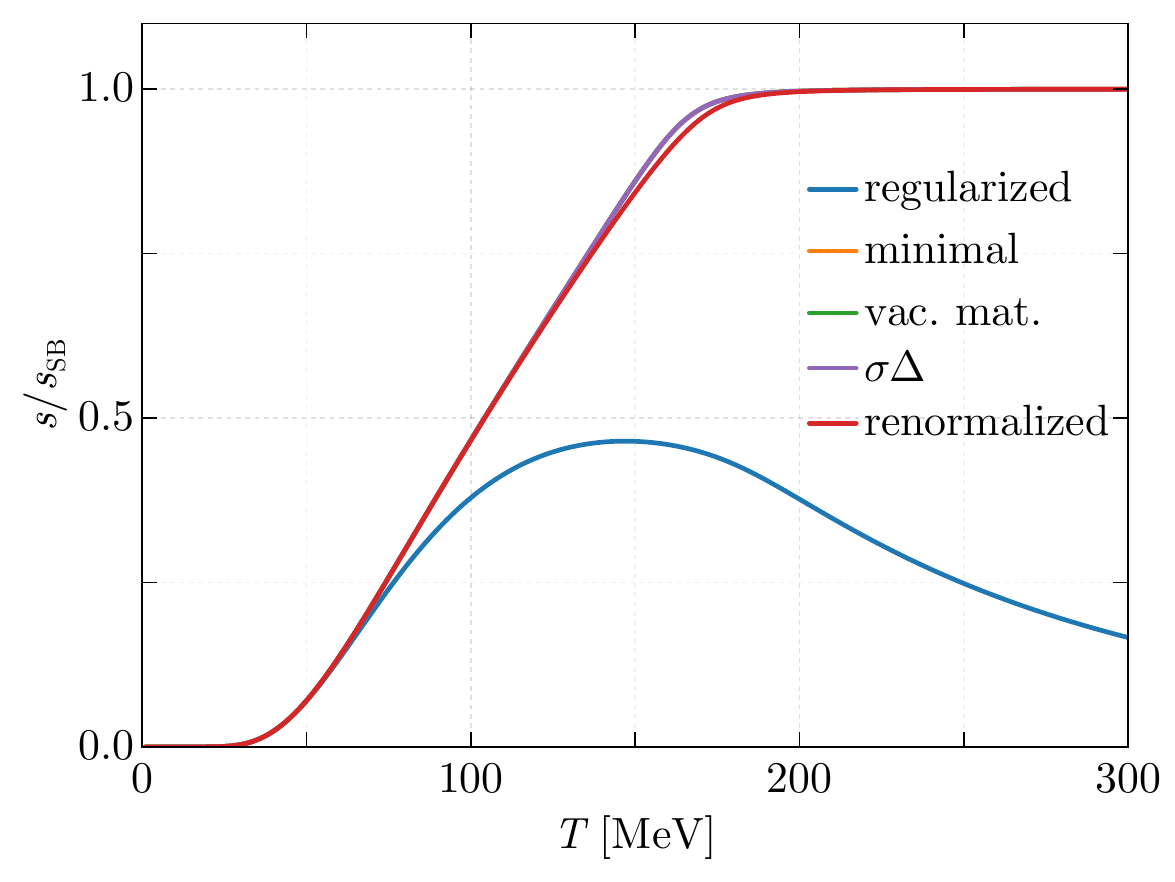}
        \caption{$\mu=0\MeV$}
    \end{subfigure}
    \hfill
    \begin{subfigure}[t]{0.49\textwidth}
        \centering
        \includegraphics[width=\textwidth]{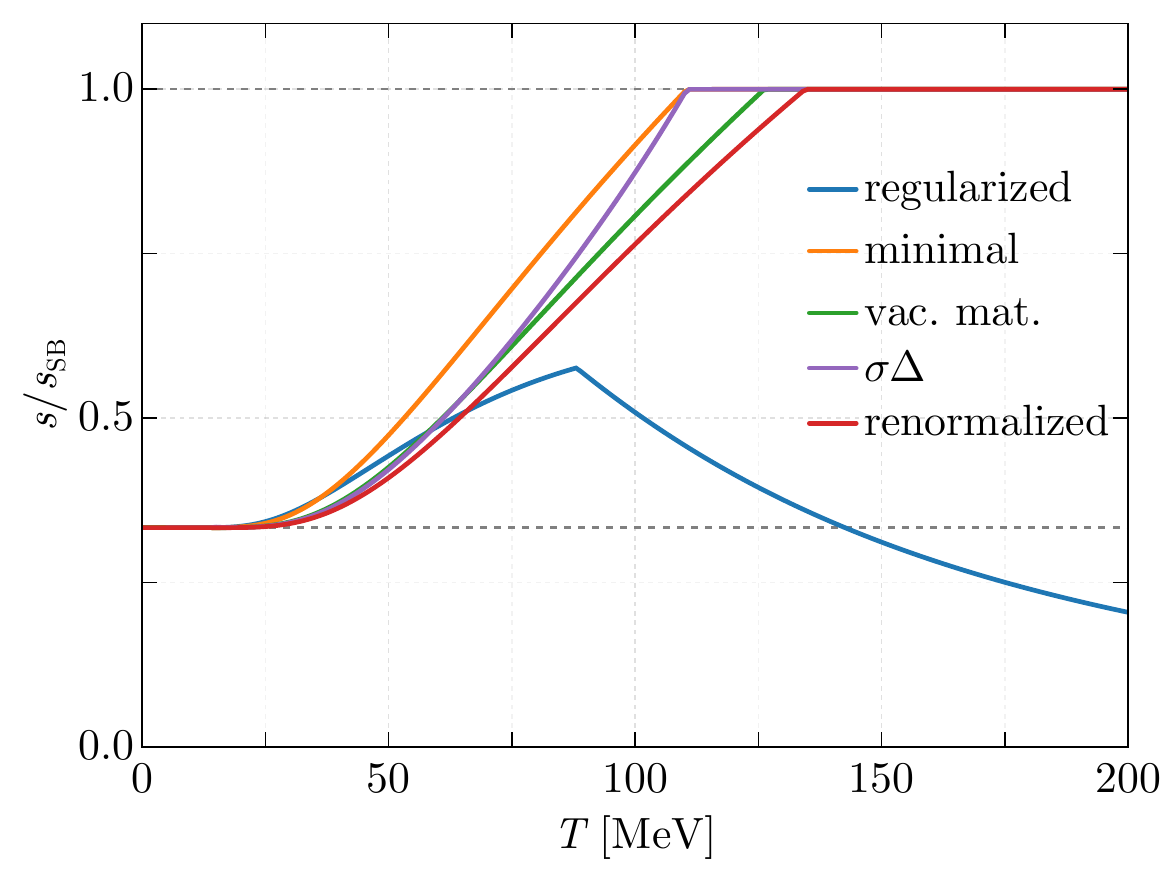}
        \caption{$\mu=450\MeV$}
    \end{subfigure}
    \caption{Entropy density $s$, normalized to the Stefan-Boltzmann
      entropy $s_{\text{SB}}$ (\eqref{eq:sSB}), as a function of $T$
      at $\mu=0\MeV$ (left) and $\mu=450\MeV$ (right). The horizontal
      dashed lines indicate $s/s_{\text{SB}}=1$ and
      $s/s_{\text{SB}}=1/3$. }
    \label{fig:entropy_fixed_mu}
\end{figure*}

The same cutoff artifacts that distort the phase diagram and the speed
of sound at higher values of $\mu$, are also visible in thermal
observables.  Because the regularized MFA omits high-momentum modes,
its thermal pressure and entropy fall short of the Stefan--Boltzmann
limits.  \Cref{fig:entropy_fixed_mu} compares the entropy density of
the five approximations with the SB limit
\begin{equation} \label{eq:sSB}
      s_{\mathrm{SB}} = \frac{N_f N_c}{3}\!\left(
        \frac{7\pi^{2}}{15}\,T^{3} \;+\; \mu^{2}T
      \right) \; .
\end{equation}
For \(\mu=0\) (left panel) no diquark condensate is present, and after
the chiral crossover the quark mass drops to nearly zero.  One
therefore expects \(s/s_{\mathrm{SB}}\!\to\!1\) at high temperature.
All curves with RG consistent treatment and the renormalized curve
meet this expectation, whereas the regMFA never reaches unity because
the truncated momenta omit an increasing fraction of thermal modes as
\(T\) grows.

The right panel shows the same ratio at \(\mu=450\;\text{MeV}\), where
a 2SC condensate persists up to the critical temperature.  Below
\(T_c\) the red--green quarks are gapped and exponentially suppressed,
so only the ungapped blue quasi-quarks contribute; consequently
\(s/s_{\mathrm{SB}}\simeq1/3\), a limit reached by all five
approximations.  Above \(T_c\) the diquark gap vanishes and the quark
mass is small; the RG-consistent and renormalized schemes therefore
approach \(s/s_{\mathrm{SB}}\!\to\!1\).  For temperatures near but
below $T_c$, a linear increase in this ratio is observed in both the
renormalized and RGC vacuum matching schemes, as well as in the
minimal scheme, while this linear behavior is absent in the
$\sigma\!\Delta$ scheme.  The regularized MFA, however, breaks down
once again, with the entropy ratio turning over and falling because
the missing high-momentum thermal modes cannot be recovered by any
further increase in \(T\).

We emphasize that \emph{all} thermodynamic observables tell the same
story.  Whether one probes along the temperature axis or along the
chemical potential axis, the truncated integrals in the regularized
MFA restrict the available phase space and, as a consequence, distort
chiral and superconducting transitions as well as bulk quantities such
as the entropy.  By contrast, the RG-consistent and renormalized
calculations, which integrate over the full momentum range,
consistently recover the Stefan--Boltzmann limits and preserve the
expected pattern of phase transitions throughout the entire
\((T,\mu)\) plane.

\subsection{Testing the BCS relation}
\label{sec:testBCS}

\begin{figure*}
    \centering
    \begin{subfigure}[t]{0.49\textwidth}
        \centering
       \includegraphics[width=\textwidth]{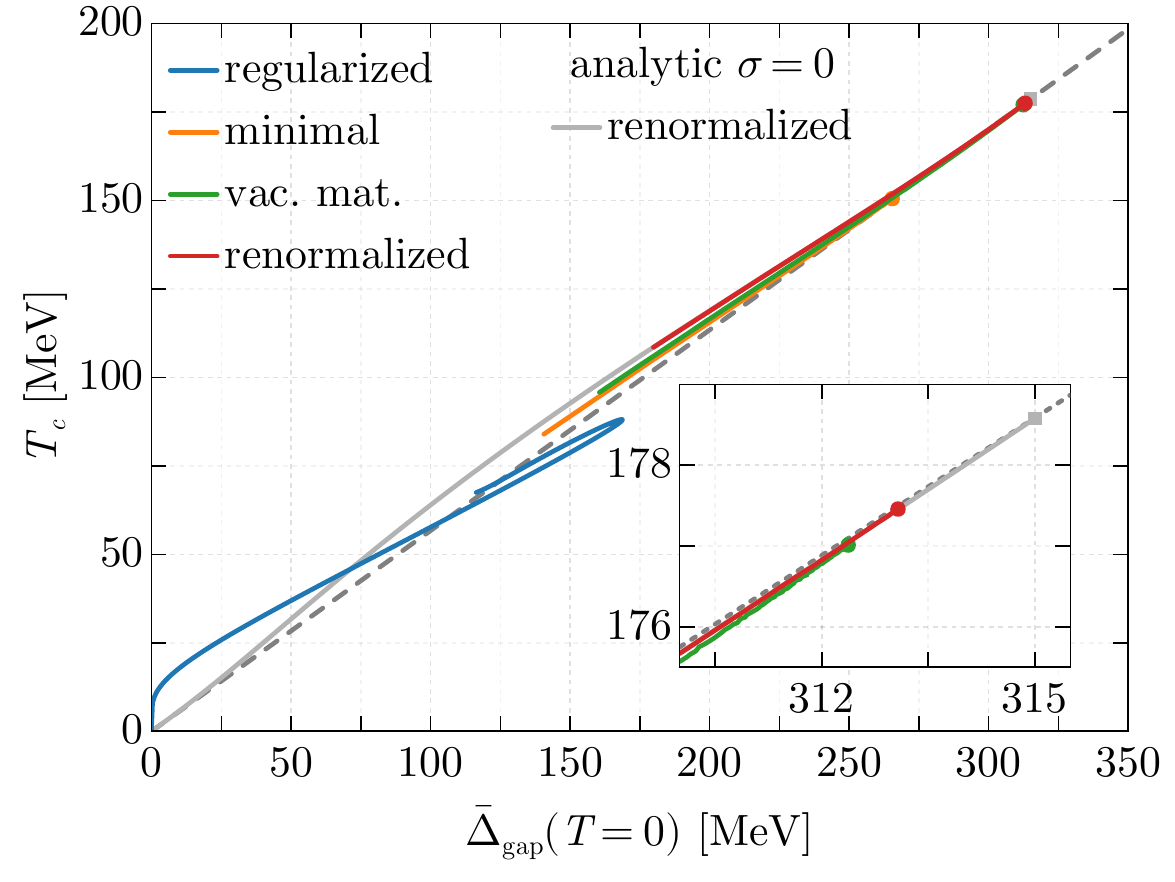}
    \end{subfigure}
    \hfill
    \begin{subfigure}[t]{0.49\textwidth}
        \centering
          \includegraphics[width=\textwidth]{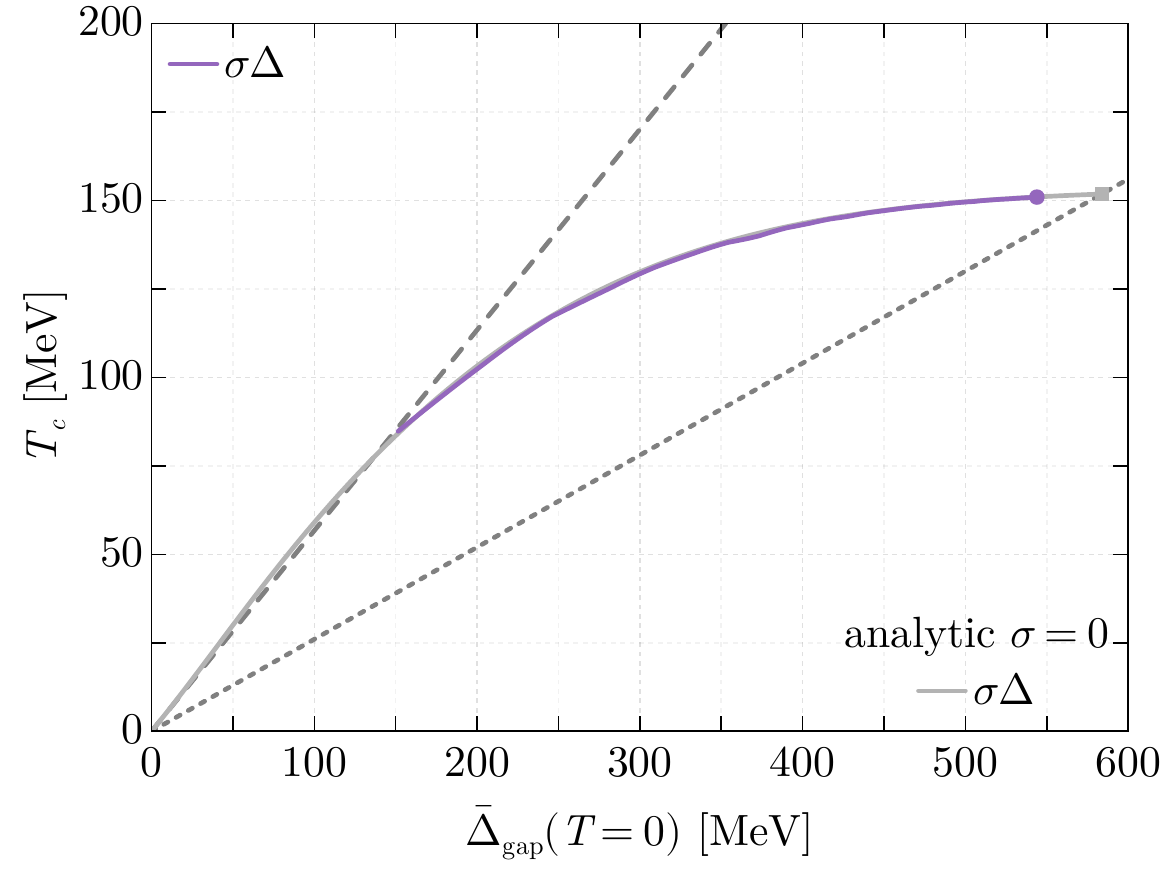}
    \end{subfigure}
    \caption{
Critical temperature $T_c$ against the diquark gap at vanishing temperature $\bDelgap(T=0)$. 
The gray solid lines indicate results obtained analytically for $\sigma=0$, see \cref{eq:Tc_of_Delta0,app:tcasmuanddeltamin},
and solid squares mark their limits at $\mu\to\infty$. Solid dots are obtained numerically at $\mu=4\GeV$. The dashed gray lines indicate the BCS scaling relation $T_c=0.567\bDelgap(T=0)$, while the dotted gray line indicates the asymptotic ratio of the $\sigma\!\Delta$ scheme obtained for our parameter choice, $T_c=0.260\bDelgap(T=0)$.
}
        \label{fig:bcs2}  
        \label{fig:bcs}
\end{figure*}

With $\bar\Delta(T=0)$ and $T_c$ at our hands, we can compare their
ratio with the BCS result and also with the numerical results obtained
away from the chiral limit.  The results are shown in \cref{fig:bcs},
where \(T_c\) is plotted as a function of the zero-temperature diquark
gap, \(\bar\Delta_{\text{gap}}(T=0)\): the left panel of
\cref{fig:bcs} shows the regularized model together with the
RGC-minimal and vacuum matching schemes in comparison with the
renormalized results , while the right panel displays the
\(\sigma\!\Delta\) scheme.

First, recall that for the renormalized QMD, the BCS ratio
\(T_c/\bDelgapZero = e^{\gamma}/\pi\) is recovered both in the limit
\(\mu \to \infty\) and \(\bDelgapZero \to 0\), see
\cref{eq:bcsweak,eq:BCSUS}.  The solid gray curve in the left panel of
\cref{fig:bcs} shows the analytic result for \(T_c/\bDelgapZero\)
obtained from the closed‑form analytic expressions in
\cref{eq:Tc_of_Delta0,app:tcasmuanddeltamin} for the renormalized
model, whereas the dashed line indicates the BCS ratio
\(T_c/\bDelgapZero= e^{\gamma}/\pi\).  Accordingly, the analytic curve
approaches the dashed BCS line smoothly at both ends: once as
\(\bDelgapZero \to 0\) and once as \(\mu \to \infty\), the latter
using the asymptotic gap \(\bDelgapZero\) from \cref{eq:delatasymp}.
The lower endpoint \(\bDelgapZero \to 0\) is, however, never realized
in the physical calculation with finite quark masses (colored curves):
In all approximations, the 2SC phase is reached through a first‑order
transition at a finite gap (see
\cref{fig:phase_diagram_and_T0_vs_mu}), so only the high‑density limit
\(\mu \to \infty\) is able to approach the BCS ratio in practice.  At
large chemical potential, the quark mass vanishes, and all three
curves satisfy the BCS relation~\cref{eq:BCSUS}.  The calculation with
physical quark masses in the renormalized model (red curve) lies
almost exactly on top of the analytic curve obtained for massless
quarks, demonstrating that explicit chiral‑symmetry breaking has only
a minor effect.  Finally, the regularized model (blue curve) never
shows the BCS scaling.  Its multi‑valued behavior is a cutoff
artifact: at large~\(\mu\) the finite three--momentum cutoff forces
\(\bDelgap\) to decrease (see \cref{fig:phase_diagram_and_T0_vs_mu}),
which causes the curve to bend back on itself.  Even within the
displayed range, the slope of \(T_c/\bDelgapZero\) differs
substantially from the BCS value, and no systematic BCS behavior is
visible. The values in the lower‑left corner correspond to the point
in the phase diagram where both the diquark condensate and the 2SC
phase boundary drop to zero, at \(\mu\simeq 686.5\MeV\)(see
\cref{fig:phase_diagram_and_T0_vs_mu,fig:phase_diagram}).

In the right panel of \cref{fig:bcs}, the solid gray curve depicts the
analytic result for the \(\sigma\!\Delta\) scheme for $\sigma=0$.  In
this scheme, the BCS ratio (dashed gray line) is reached only in the
\(\bDelgapZero\!\to 0\) limit, which explains why the curve touches
the dashed BCS line in the lower‑left corner.  At the opposite end,
\(\mu\!\to\!\infty\), the scaling deviates from the BCS value toward
the computed value of $T_c/\bDelgapZero=0.260$, shown in the gray
dotted curve. So the analytic curve bends away from the dashed line
and toward the dotted line at the end, as expected.  As in the other
schemes, the lower‑left corner is never realized once the chiral
dynamics is included. The purple curve in the right panel of
\cref{fig:bcs}, which incorporates finite quark masses in the
\(\sigma\!\Delta\) scheme, follows the analytic chiral‑limit curve
almost perfectly but starts at a finite lower value of the gap. This
starting point is almost exactly on the BCS line, but we consider this
to be accidental.

As discussed in \cref{subsec:bcs}, the \(\sigma\!\Delta\) scheme
yields a different asymptotic value for the diquark gap, yet, the
critical temperature approaches the same limit as in the minimal
scheme, consistent with \cref{eq:bcssigdel}.  This similarity and
distinction is clearly visible in comparing the two panels of
\cref{fig:bcs}. Therefore among the RGC schemes discussed in this
paper, only the \(\sigma\!\Delta\) scheme fails to follow the BCS
relation at $\mu\to\infty$. Finally, we note that in the
\(\sigma\!\Delta\) scheme the asymptotic regime is reached only at
much larger chemical potentials than in the other schemes.  Near the
asymptotic point (gray square), \(T_c\) remains almost constant while
\(\bar\Delta_{\text{gap}}\) continues to grow, so no extended linear
window develops around the dotted line in the right panel of
\cref{fig:bcs}.  Even at \(\mu = 4\;\text{GeV}\) (solid purple dot)
the system is still far from its asymptotic point (gray square).

We conclude that although both the \(\sigma\!\Delta\) and the vacuum
matching schemes employ the same set of vacuum parameters, the
additional field dependent subtractions in the \(\sigma\!\Delta\)
scheme leads to a distinct behavior of the diquark gap and entropy
density
(cf. \cref{fig:phase_diagram_and_T0_vs_mu,fig:entropy_fixed_mu}).  As
noted in Ref.~\cite{Gholami:2024diy}, such subtractions in RGC schemes
can yield unexpected consequences, not anticipated from a physical
standpoint, for example regarding the melting pattern of the different
diquark condensates. The discrepancy observed here regarding the BCS
relation is of the same nature; the additional subtraction in all
\(\Delta\)-dependent RGC schemes (including schemes used in
Ref.~\cite{Braun:2018svj} and mass-dependent and massless schemes in
Ref.~\cite{Gholami:2024diy}) results in a deviation from the expected
BCS relation \footnote{We note that those subtractions are motivated
  by considerations \emph{beyond} MFA. At the mean‑field level they
  are not required, but once bosonic fluctuations are taken into
  account they may become essential.}.  It should be stressed that, we
employ the BCS relation only as a convenient diagnostic.  It is
derived for conventional (phonon-mediated) superconductors and
therefore should \emph{not} be viewed as a rigorous benchmark for
color-superconducting models, where the pairing mechanism and
underlying dynamics are qualitatively different and no experimental
confirmation of the ratio exists.

It is also worth noting that the extent to which the BCS relation is
realized and how the asymptotic limit is approached are sensitive to
the choice of parameters. With different parameter choices, the BCS
relation might become a good approximation only at substantially
larger values of \(\mu\). Thus, the linear scaling along a large range
of $\bDelgap$ values observed in \cref{fig:bcs,fig:bcs2} (left) may be
coincidental.

\section{Summary, Discussion and Conclusions}
\label{sec:conc}

In this work, we have presented a systematic comparison between two
conceptually distinct approaches for rendering the two--flavor
Quark--Meson--Diquark (QMD) model ultraviolet complete: a
\emph{renormalized} formulation, and an \emph{RG-consistent}
mean-field treatment that incorporates ideas from the functional
renormalization group.  In order to make the comparison meaningful, we
fix the couplings in both approaches to a common set of vacuum
parameters, related to low-energy properties of the model. With this
vacuum fit, they yield descriptions of cold and dense quark matter
that are not only qualitatively but also, to a large extent,
quantitatively consistent.  Both frameworks generate identical phase
structures, approach the Stefan--Boltzmann limit at large chemical
potential, respect the BCS relation
\(T_{\mathrm c}=e^{\gamma}\bDelgapZero/\pi\) (with the exception of
the $\sigma\!\Delta$ scheme) and avoid the cutoff artifacts that
typically affect naïvely regularized mean-field calculations.

Our analysis reveals that the regularized mean-field approximation
with explicit diquark degrees of freedom leads to \emph{unphysical}
thermodynamics at high densities and temperatures. In particular, the
pressure grows only linearly with~$\mu$, the Stefan--Boltzmann limit
is never reached, and the speed of sound diverges, signaling a
breakdown of the approximation.

Both renormalization and RG consistent treatment cure these artifacts.
The two approaches nevertheless differ in important details.  In the
renormalized model, all vacuum and medium divergences are absorbed
into scale-dependent bare couplings, rendering the effective action
completely independent of the regulator.  The price for this regulator
independence is an enlarged parameter space: additional couplings
proportional to \(\phi^2|\Delta|^2\), \(|\Delta|^4\) and, via the
diquark wave function renormalization, to \(\mu^2|\Delta|^2\) must be
allowed to run with the cutoff.  In contrast, RG-consistent mean-field
schemes retain a fixed microscopic action and compensate for the
residual cutoff dependence by imposing renormalization group
consistency through suitable matching conditions.

Although an RG-consistent (RGC) mean-field construction ensures that
physical observables become independent of the ultraviolet cutoff in
the limit $\Lambda \to \infty$, it does not uniquely determine how to
eliminate medium-induced divergences.  The corresponding counterterms
are therefore scheme-dependent, and different prescriptions can affect
finite-density observables even after the vacuum parameters have been
fixed.  In Ref.~\cite{Gholami:2024diy}, it was shown how these schemes
can be related to the wave function renormalization of the diquark
field, $Z_{\Delta}$.  Motivated by this insight, we introduced a
vacuum-matching RGC scheme in which $Z_{\Delta}$ is fixed in the
vacuum to a value that is, in principle, accessible via
first-principles QCD calculations.

With this choice, the in-medium flow reproduces the diquark gap
obtained in the fully renormalized QMD model up to sub-leading
$1/\mu^2$ corrections, while preserving the cutoff independence
characteristic of all RGC constructions. In this sense, the
vacuum-matching scheme retains the conceptual clarity of the
prescription in Ref.~\cite{Braun:2018svj} and the minimal prescription
in Ref.~\cite{Gholami:2024diy}. By anchoring the construction to a
physical vacuum observable, it largely eliminates the residual scheme
ambiguity that would otherwise persist among different RGC
implementations.

A final crisp benchmark is the BCS relation,
$T_c=e^{\gamma}\bDelgap(T=0)/\pi$.  We find that this ratio is
\emph{exactly} reproduced in the $\mu\to\infty$ limit by the
renormalized QMD model and by RG-consistent schemes whose medium
subtraction can be reinterpreted as a diquark wave function
renormalization, specifically, the minimal and vacuum-matching
variants.  In contrast, schemes in which the subtraction acquires an
additional \(\Delta\)-dependent structure (such as the
\(\sigma\!\Delta\) scheme of Ref.~\cite{Braun:2018svj}) violate the
BCS ratio. This underscores, once again, that seemingly innocuous
choices for removing medium divergences can significantly alter key
in-medium observables. We emphasize, however, that this should
\emph{not} be taken as a criterion for discarding specific
RG-consistent schemes.  Rather, it highlights the intrinsic scheme
dependence in the subtraction of medium divergences, an ambiguity that
persists even after matching to vacuum parameters.

Our analytic solution of the $T=0$ gap equation reveals that, in the
limit $\mu\to\infty$, the diquark gap approaches a \emph{finite}
constant --- see \cref{eq:delatasymp,eq:deltasymp_minimal}.
Remarkably, in both the renormalized model and RGC vacuum-matching
scheme, this asymptotic value depends \emph{only} on the vacuum quark
mass, the diquark Yukawa coupling, and the vacuum wave function
renormalization of the diquark field, with every explicit trace of the
UV regulator eliminated.  Renormalization and RG consistency thus
remove the cutoff from the formalism without severing the link between
vacuum physics and dense-matter observables: once the vacuum
parameters are fixed, the entire finite-$T$ / finite-$\mu$ sector of
the QMD model becomes a genuine prediction of mean-field dynamics.

The most compelling next step is therefore to determine off-shell
correlators directly from QCD --- for example, via lattice calculations
or by matching to continuum FRG/DSE computations.  Such an
\emph{ab-initio} calibration would render the model a parameter-free
tool for exploring the phase structure and equation of state of dense
matter. This procedure would provide a decisive test of whether a
mean‑field QMD model, once anchored to realistic vacuum observables,
can simultaneously satisfy astrophysical constraints without the need
to incorporate beyond‑mean‑field fluctuations.

With the vacuum parameter fixing provided by first-principles QCD in
place, several phenomenologically important extensions become both
natural and reliable. For instance, the present two‑flavor framework
can be generalized to the three‑flavor case. In
Ref.~\cite{Gholami:2024diy}, the melting pattern of the
color--flavor‑locked (CFL) phase was analyzed within an RG‑consistent
NJL framework, successfully reproducing earlier Ginzburg--Landau
predictions \cite{Iida_2004}, in contrast to the failure of
conventional cutoff‑regularized NJL models. It would be instructive to
extend this analysis to a three-flavor QMD model.

For astrophysical applications, a repulsive vector interaction,
together with electric charge, isospin and color-neutrality
constraints, must be incorporated to obtain an equation of state
  compatible with $\sim2\,M_\odot$ neutron stars and their
tidal-deformability bounds \cite{Baym:2017whm, Bonanno:2011ch,
  Klahn:2013kga, Otto:2020hoz, Gholami:2024ety}.
In addition, the regulator-independent framework developed here
provides a clean setting to revisit spatially modulated chiral or
pairing phases, whose very existence crucially depends on careful
control of cutoff artifacts \cite{Buballa:2020nsi, Pannullo:2022eqh,
  Pannullo:2024sov}.  Work along these lines is ongoing and will be
reported elsewhere.

Since many earlier hybrid-star studies rely on regMFA diquark
equations of state at high densities, they inevitably inherit
unphysical artifacts, potentially leading to erroneous conclusions
regarding stellar properties. A properly renormalized or RG-consistent
treatment is therefore essential to ensure reliable astrophysical
predictions. In this work, we have demonstrated that both
renormalization and RG consistency eliminate the regulator ambiguities
in the QMD model, yielding mutually consistent and analytically
tractable descriptions of dense quark matter. Together, these
approaches provide a powerful and flexible framework for future
studies aiming to bridge microscopic QCD dynamics with astrophysical
observations of compact stars.

\section*{Acknowledgment}

\noindent
We thank Marco Hofmann and Shreedhar Rajesh for collaboration on
related topics, and Jan M. Pawlowski and Lorenz von Smekal for
valuable discussions.  H.G. further acknowledges insightful
conversations with Jens Andersen, Jens Braun, Tomáš Brauner, Andreas
Geissel, Mathias Nødtvedt and Dirk Rischke, and U.M. thanks Janos Polonyi for discussions. The authors gratefully
acknowledge support from the Helmholtz Graduate School for Hadron and
Ion Research (HGS-HIRe) for FAIR, the GSI Helmholtzzentrum für
Schwerionenforschung, and the Deutsche Forschungsgemeinschaft (DFG,
German Research Foundation) through the CRC-TR211 'Strong-interaction
matter under extreme conditions' project number 315477589 – TRR 211.

\section*{Data Availability}
\noindent
The numerical data presented in all figures in this work are openly
available in the ancillary files of the corresponding arXiv
submission.

\onecolumngrid

\appendix
\renewcommand{\thesubsection}{\Roman{subsection}}

\section{Derivation of the renormalized potential}
\label{app:ren_pot_derivation}
\noindent
Fixing $\vcfpi$, the remaining vacuum parameters in
\cref{tab:observables_description} are found by evaluating derivatives
of the effective potential \eqref{eq:regularized_potential} in vacuum,
which yields \begingroup \allowdisplaybreaks
\begin{align}
    \vcmsigma =           & m_\phi^2(\Lambda)
    + 3\lambda_\phi(\Lambda) {\sigvac}^2 + \lrvac{\partial_\sigma^2L_\Lambda} \; , \label{eq:44} \\
    \vcmDelta=           &  m_\Delta^2(\Lambda) 
    + \frac{1}{2}\lambda_\text{mix}(\Lambda) {\sigvac}^2 + \frac{1}{2}\lrvac{\partial_\Delta^2L_\Lambda}\; , \\
    \vclambdamix= & \lambda_\text{mix}(\Lambda)
    + \frac{1}{2} \lrvac{\partial_\sigma^2\partial_\Delta^2L_\Lambda} \; , \\
    \vclambdaDelta=     & \lambda_\Delta(\Lambda)
    + \frac{1}{24} \lrvac{\partial_\Delta^4L_\Lambda} \; , \\
    \vcZDelta=           & Z_\Delta(\Lambda) 
    - \frac{1}{16} \lrvac{\partial_\mu^2\partial_\Delta^2L_\Lambda} \; . \label{eq:48}
\end{align}
\endgroup
\cref{eq:sigma_gap_ren} and \crefrange{eq:44}{eq:48} form a simple
linear system of equations that can be solved for an expression of the
6 UV scale-dependent bare parameters in terms of the correlators. The
solution of this linear system yields
\begingroup
\allowdisplaybreaks
\begin{align}
    & m_\phi^2(\Lambda) = -\frac{1}{2{\sigvac}} \Big\{
        \vcmsigma {\sigvac} - 3c
        - {\sigvac} \lrvac{\partial_\sigma^2L_\Lambda}
        + 3 \lrvac{\partial_\sigma L_\Lambda}
    \Big\} \; , \\
    & m_\Delta^2(\Lambda) = 
         \vcmDelta
        - \frac{1}{2} \vclambdamix {\sigvac}^2
        - \frac{1}{2} \lrvac{\partial_\Delta^2L_\Lambda}
        + \frac{{\sigvac}^2}{4}  \lrvac{\partial_\sigma^2\partial_\Delta^2L_\Lambda} \; , \\
    & \lambda_\phi(\Lambda) =
        \frac{1}{2{\sigvac}^3}
        \Big\{ 
        \vcmsigma {\sigvac} - c
        - \sigvac \lrvac{\partial_\sigma^2L_\Lambda}
        + \lrvac{\partial_\sigma L_\Lambda}
    \Big\} \; , \\
    & \lambda_\text{mix}(\Lambda) = 
    \vclambdamix
    - \frac{1}{2} \lrvac{\partial_\sigma^2\partial_\Delta^2L_\Lambda} \; , \\
    & \lambda_\Delta(\Lambda) =     
     \vclambdaDelta 
     - \frac{1}{24} \lrvac{\partial_\Delta^4L_\Lambda} \; , \\
    & Z_\Delta(\Lambda) =          
    \vcZDelta 
    + \frac{1}{16} \lrvac{\partial_\mu^2\partial_\Delta^2L_\Lambda} \; .
\end{align}
\endgroup

Using the expression for the loop contribution
\cref{eq:Omega_reg_therm}, we can write the bare parameters as
\begingroup \allowdisplaybreaks
\begin{align}
    & m_\phi^2(\Lambda) = 
        - \frac{1}{2} \vcmsigma + \frac{3c}{2\sigvac}
        + 2N_f \int_{|\vec{p}|<\Lambda} 
        \frac{3 g_\phi^2}{2\epsilon_{q,\text{vac}}^3} 
        \left( 2 p^2 + 3 m_{q,\text{vac}}^2 \right) \; , \\
    & m_\Delta^2(\Lambda) = 
        \vcmDelta
        - \frac{1}{2} \vclambdamix {\sigvac}^2 
        + 2 N_f \int_{|\vec{p}|<\Lambda} 
        \frac{g_{\Delta}^2 p^2}{2\epsilon_{q,\text{vac}}^5} 
        \left( 2 p^2 + 5 m_{q,\text{vac}}^2 \right) \; , \\
     & \lambda_\phi(\Lambda) = 
        \frac{1}{2{\sigvac}^2} \vcmsigma
        - \frac{c}{2{\sigvac}^3} 
        - 2 N_f \int_{|\vec{p}|<\Lambda} 
        \frac{3 g_{\phi}^4}{2 \epsilon_{q,\text{vac}}^3} \; , \\
    & \lambda_\text{mix}(\Lambda) = 
        \vclambdamix 
            - 2N_f \int_{|\vec{p}|<\Lambda} 
            \frac{g_{\Delta}^2 g_{\phi}^2 }{\epsilon_{q,\text{vac}}^5} 
            \left(p^2 - 2 m_{q,\text{vac}}^2\right) \; , \\
    & \lambda_\Delta(\Lambda) =     
    \vclambdaDelta 
    - 2 N_f \int_{|\vec{p}|<\Lambda} 
    \frac{g_{\Delta}^4}{4 \epsilon_{q,\text{vac}}^3} \; , \\
    & Z_\Delta(\Lambda) =          
    \vcZDelta 
    - 2N_f \int_{|\vec{p}|<\Lambda} 
     \frac{g_{\Delta}^2}{4 \epsilon_{q,\text{vac}}^3} \; ,
\end{align}
\endgroup
where we introduced $m_{q,\text{vac}}=g_\phi\sigvac$ and $\epsilon_{q,\text{vac}}=\sqrt{p^2 + m_{q,\text{vac}}^2}$. Finally, we obtain the  renormalized and UV scale independent potential by sending $\Lambda\to\infty$ in the momentum integration of the effective potential, which results in \cref{eq:ren_potential}.

\section{$\beta$-functions of the model couplings}
\label{app:coupling_flow}

\noindent
From the mean-field flow \cref{eq:potential_flow}, we can evaluate the
$\beta$-functions of the relevant couplings of the model. At vanishing
temperature and for the sharp regulator, the mean-field flow reduces
to
\begin{equation}
  \label{eq:vacuum_flow}
    \begin{aligned}
      & \partial_k \Omega_k = \frac{N_f k^2}{\pi^2} \bigg\{
      \epsilon_q(k) + E_q^-(k) + E_q^+(k) - \epsilon_q^-(k)
      \theta\left(\epsilon_q^-(k)\right) \bigg\} \; .
    \end{aligned}
\end{equation}
In the following, we neglect the last term
$- \epsilon_q^-(k) \theta\left(\epsilon_q^-(k)\right)$ arising from
the thermal part of the potential, as it yields an irrelevant
contribution in the limit $k \to \infty$. We then perform a Taylor
expansion of the effective potential
\begin{equation}
    \begin{aligned}
         \Omega_k = \frac{1}{2} m^2_{\phi,k} \phi^2
        + \frac{1}{4} \lambda_{\phi,k} \sigma^4
        + \frac{1}{2} \lambda_{\text{mix},k} \sigma^2 \Delta^2
        + m^2_{\Delta,k} \Delta^2
        + \lambda_{\Delta,k} \Delta^4
        - 4 Z_{\Delta,k} \mu^2 \Delta^2
        + \cdots \; ,
    \end{aligned}
\end{equation}
and insert this expansion in the mean-field flow
\cref{eq:vacuum_flow}.  The $\beta$-functions of the couplings are
obtained by comparing coefficients, yielding
\begin{equation}
\begin{aligned}
    & \partial_k m^2_{\phi,k} = \frac{6g_\phi^2}{\pi^2} k \; ,\quad 
    \partial_k \lambda_{\phi,k} = -\frac{3g_\phi^4}{\pi^2} \frac{1}{k} \; ,\quad 
    \partial_k m^2_{\Delta,k} = \frac{2g_\Delta^2}{\pi^2} k \; ,\quad \\
    & \partial_k \lambda_{\Delta,k} = -\frac{g_\Delta^4}{2\pi^2} \frac{1}{k} \; ,\quad 
    \partial_k \lambda_{\text{mix},k} = -\frac{2g_\Delta^2 g_\phi^2}{\pi^2} \frac{1}{k} \; ,\quad 
    \partial_k Z_{\Delta,k} = -\frac{g_\Delta^2}{2\pi^2} \frac{1}{k} \; .
\end{aligned}
\end{equation}
This directly leads to the following solutions for the scale
dependence of the couplings:
\begin{align}
    m^2_{\phi,k}           & = \frac{3g_\phi^2}{\pi^2}( k^2 - {k_{0}}^2 )
    + m^2_{\phi,k_{0}} \; ,                                                            \\
    \lambda_{\phi,k}       & = - \frac{3g_\phi^4}{\pi^2} \ln \frac{k}{k_{0}}
    + \lambda_{\phi,k_{0}} \; ,                                                        \\
    m^2_{\Delta,k}         & = \frac{g_\Delta^2}{\pi^2} ( k^2 - {k_{0}}^2 )
    + m^2_{\Delta,k_{0}}   \; ,                                                        \\
    \lambda_{\Delta,k}     & = - \frac{g_\Delta^4}{2\pi^2} \ln \frac{k}{k_{0}}
    + \lambda_{\Delta,k_{0}} \; ,                                                      \\
    \lambda_{\text{mix},k} & = - \frac{2g_\Delta^2g_\phi^2}{\pi^2} \ln \frac{k}{k_{0}}
    + \lambda_{\text{mix},k_{0}} \; ,                                                  \\
    Z_{\Delta,k}           & = - \frac{g_\Delta^2}{2\pi^2} \ln \frac{k}{k_{0}}
    + Z_{\Delta,k_{0}} \; .
\end{align}
Here, $k_0$ represents an arbitrary reference scale. Notably, if we
choose $k=\Lambda$, the RG scale dependence of the couplings exactly
matches the $\Lambda$-dependence of the renormalized model couplings,
as given in \crefrange{eq:34}{eq:39}. This explicitly demonstrates
that, as expected, the RG-consistency generates the appropriate
counterterms introduced in the renormalized model.

\section{BCS analysis at $T_c$}\label{app:TcBCS}
\noindent
Subtraction of the gap equations in \cref{eq:bcsgapeqs} yields the 
relation
\begin{align}
    \begin{split}\label{app:bcssubtr}
            &
    \left.
    \frac{\partial \Omega^\text{eff}_\text{ren}(\sigma, \Delta; 0, \mu)}{\partial\Delta}
    \right|_{\substack{\sigma=0\\\Delta=\bDeltaZero}}-
     \left.
    \frac{\partial \Omega^\text{eff}_\text{ren}(\sigma, \Delta; T_c, \mu)}{\partial\Delta}
    \right|_{\substack{\sigma=0\\\Delta=0}} =
            4\vclambdaDelta\bDeltaZero^2 \\ & \quad
            - 2 N_f g_\Delta^2 \int_{\vec{p}} \left\{ \frac{1}{E_{q,0}^-} + \frac{1}{E_{q,0}^+} + \frac{\bDelgapZero^2}{\epsvac^3} - \frac{1}{\epsilon_q^-}\tanh\frac{\epsilon_q^-}{2T_c} - \frac{1}{\epsilon_q^+}\tanh\frac{\epsilon_q^+}{2T_c} \right\} \; ,
    \end{split}
\end{align}
where, assuming a vanishing chiral condensate, we have 
\[
E_{q,0}^{\pm} \;=\; \sqrt{(\epsilon_q^{\pm})^{2} + \bDelgapZero^{2}} \; ,
\qquad
\epsilon_q^{\pm} \;=\; p \pm \mu \; ,
\]
with $p=|\vec{p}|$. Since both gap equations vanish, their difference
also vanishes.  The resulting condition implicitly defines the
location of the second-order phase boundary of the 2SC phase at a
given chemical potential~$\mu$ is determined by the corresponding
zero-temperature gap $\bar\Delta_{0}(\mu)\equiv\bar\Delta(T=0,\mu)$.
Solving this condition for~$T$ yields the critical
temperature~$T_{c}$.

To proceed, we split the remaining momentum integral into two
convergent parts, one for each $\tanh$ term.  Because each part
converges separately, we can shift and unify the integration limits.
Rewriting \cref{app:bcssubtr} in this way, we obtain
\begin{align}
    0 = {}& 4\vclambdaDelta\bDeltaZero^2
    - g_\Delta^2 2N_f \left\{ \int_{\vec{p}} \bigg( 
    \frac{1}{2}\frac{\bDelgapZero^2}{\epsvac^3}
    + \frac{1}{E_{q,0}^-}  
    - \frac{1}{\epsilon_q^-} \tanh \frac{\epsilon_q^-}{2T_c} \bigg) 
    + \int_{\vec{p}} \bigg( 
    \frac{1}{2}\frac{\bDelgapZero^2}{\epsvac^3}
    + \frac{1}{E_{q,0}^+}  
    - \frac{1}{\epsilon_q^+} \tanh \frac{\epsilon_q^+}{2T_c} \bigg)
    \right\} \; .
\end{align}
By shifting the integration variable in the first integral,
\(p \to p - \mu\), and in the second integral, \(p \to p + \mu\), and
then separating the even and odd parts of the integrand, the two
integrals can be combined into a single one,
\begin{equation}\label{eq:appc3}
    \begin{split}
    0={}& 4 \vclambdaDelta\bDeltaZero^2 - g_\Delta^2 2 N_f \int_{\vec{p}} \bigg\{
    \frac{\bDelgapZero^2\,(p-\mu)^2}{2\left( (p-\mu)^2+ \mqvac^2 \right)^{3/2}} 
    + \frac{\bDelgapZero^2\,(p+\mu)^2}{2\left((p+\mu)^2+ \mqvac^2 \right)^{3/2}}
    \\ & 
    + \frac{(p-\mu)^2}{\sqrt{p^2+\bDelgapZero^2}} 
    + \frac{(p+\mu)^2}{\sqrt{p^2+\bDelgapZero^2}}
    - 2(p^2+\mu^2) \frac{1}{p} \tanh\frac{p}{2T_c}
    \bigg\} \; .
    \end{split}
\end{equation}
The integral can be evaluated analytically by inserting the series representation
\begin{equation}
\tanh x \;=\; 2\,\sinh x \sum_{k=1}^{\infty} (-1)^{\,k+1} e^{\,(1-2k)x} \; .
\end{equation}
For all $k\ge 2$ the resulting terms can be resummed into the infinite product
\begin{equation}
\prod_{k=1}^{\infty}\!\Bigl(1-\frac{1}{4k^{2}}\Bigr)\;=\;\frac{2}{\pi} \; ,
\end{equation}
while the remaining contribution from the $k=1$ term is obtained by an
integral,
\begin{equation}
    \int_{0}^{\infty} dx \left\{
    -\frac{1}{1+x} + \frac{2}{x} e^{-x} \sinh(x)
    \right\}
    = \ln 2 + \gamma \; ,
\end{equation}
where $\gamma \simeq 0.5772$ is the Euler--Mascheroni constant.
\noindent
Summing all contributions for the integral in~\eqref{eq:appc3}, this whole expression can be evaluated as
\begin{align}
    0 = 4 \vclambdaDelta\bDeltaZero^2 
    - \frac{g_\Delta^2}{{3\pi^2}} \left\{
    2\pi^2 T_c^2 
    + 12 \mu^2 \Big( 
    \ln\frac{\pi T_c}{\bDelgapZero} - \gamma
    \Big)
    + 3 \DelgapZero^2 \Big(  
    2\ln\frac{\bDelgapZero}{\mqvac} - 1
    \Big)
    \right\} \; .
\end{align}
The expression above can be re-written as
\begin{align}\label{app:transformation}
   \Big(\frac{\pi^2}{3\,\mu^2} T_c^2\Big)
   \;\exp{\Big(\frac{\pi^2}{3\,\mu^2} T_c^2\Big)}=
\frac{\bDelgapZero^2}{3\mu^2}\,\exp\!\Biggl[
2\gamma 
+ \frac{\bDelgapZero^2}{\mu^2}\Biggl(
\frac{1}{2} 
+ \frac{2\pi^2}{g_\Delta^4} \vclambdaDelta
- \ln\!\frac{\bDelgapZero}{\mqvac}
\Biggr)
\Biggr] \; .
\end{align}
Now, starting from the transcendental equation
\begin{equation}\label{eq:ze^z}
  z\,e^{z}=C \; ,
\end{equation}
one introduces the (multi-valued) \emph{Lambert W-function} \cite{Corless1996},
defined implicitly by
\begin{equation}
  W(x)\,e^{W(x)} \; = \; x \; .
\end{equation}
Applying this definition to \eqref{eq:ze^z} gives the formal
solution\footnote{For real arguments \(C\in[-e^{-1},\infty)\) the
  Lambert \(W\) function has exactly two real branches: the principal
  branch \(W_{0}(C)\ge -1\) and the secondary branch
  \(W_{-1}(C)\le -1\).  Consequently, \eqref{eq:z=W} can supply up to
  two distinct real roots, one from each branch. For \(C<-e^{-1}\) the
  two real branches combine and become complex. Detailed discussions
  of these branches may be found in Refs.~\cite{Corless1996,DLMF}.}
\begin{equation} \label{eq:z=W}
  z \;=\; W(C) \; .
\end{equation}
Employing \eqref{eq:z=W}, the critical temperature \(T_{c}\) in
\eqref{app:transformation} can be written compactly as

\begin{equation}\label{app:Tc_of_Delta0}
T_c^2(\bDeltaZero,\mu)=\frac{3\mu^2}{\pi^2}\,W\!\Biggl(
\frac{\bDelgapZero^2}{3\mu^2}\,\exp\!\Biggl[
2\gamma 
+ \frac{\bDelgapZero^2}{\mu^2}\Biggl(
\frac{1}{2} 
+ \frac{2\pi^2}{g_\Delta^4} \vclambdaDelta
- \ln\!\frac{\bDelgapZero}{\mqvac}
\Biggr)
\Biggr]
\Biggr) \; ,
\end{equation}
where the branch of the \(W\)-function must be chosen such that the
second root is positive and the resulting \(T_{c}\) is real and
positive. Furthermore, \(T_c\) can also be expressed purely in terms
of the chemical potential \(\mu\).  To derive this relation, we start
from the gap equation \eqref{eq:T0_Delta_gap} and establish the
following identity
\begin{align}\label{app:identityforreno}
  \ln\frac{\bDelgapZero}{\mqvac}
\;=\;
\frac{
    \vcmDelta
    + 2\vclambdaDelta \bDeltaZero^2
    - 4\vcZDelta \mu^2
    - \frac{1}{2} \vclambdamix \sigvac^2
    + \frac{g_\Delta^2}{2\pi^2} \left(
        \bDelgapZero^2 + 2\mu^2 - 3 \mqvac^2
    \right)
}{
      \frac{g_{\Delta}^{2}}{\pi^{2}}\,
      \left(\bDelgapZero^{2}-2\mu^{2}\right)
}.
\end{align}
Plugging the above identity into Eq.~\eqref{app:Tc_of_Delta0}, all
\(\Delta\)-dependence cancels out, and we obtain an explicit expression
for \(T_c\) that depends \emph{only} on the chemical potential~\(\mu\)
\begin{equation}
T_c^2(\mu)=
  \frac{3\mu^{2}}{\pi^2}\,
  W\!\bigg(
    \frac{\mqvac^2}{3\mu^{2}}\,
    \exp\!\Bigl[
    - 1 + 2\gamma + \frac{4\pi^{2}\vcZDelta}{g_{\Delta}^{2}}
    + \frac{1}{2\mu^{2}}
\Bigl(
        3\mqvac^2 
        + \frac{\pi^{2}}{g_{\Delta}^{2}} \Big(
        \vclambdamix \sigvac^2
        - 2 \vcmDelta
        \Big)
\Bigr)
    \Bigr]
  \bigg).
\end{equation}
Taking the limit \(\mu \to \infty\) of the above equation yields
\begin{align}
   \lim_{\mu\to\infty}T_c(\mu)= \frac{e^\gamma}{\pi}\,
   \mqvac
    \exp\Bigl(
      -\frac12
      + \frac{2\pi^{2} \vcZDelta}{\,g_{\Delta}^{2}}
\Bigr)= \frac{e^\gamma}{\pi}\lim_{\mu\to\infty} \bDelgapZero,
\end{align}
which is the BCS relation \eqref{eq:originalbcs}. Additionally, taking
the weak‑coupling limit \(\bDelgapZero\!\to 0\) of $T_c/\bDelgapZero$,
from \cref{app:Tc_of_Delta0} we find
\begin{equation}
\lim_{\bDelgapZero\to 0}\frac{T_c}{\bDelgapZero}
      = \frac{e^{\gamma}}{\pi},
\end{equation}
which agrees with the standard BCS ratio.
This explains why the analytic curve in~\cref{fig:bcs} approaches
the BCS line at both ends: once for \(\mu\!\to\!\infty\) and once for
\(\bDelgapZero\!\to\!0\).

\section{Analytic RGC schemes expression at {\boldmath$\sigma=0$}}
\label{app:RGCexpressions}

In this appendix, we provide the analysis for the case $\sigma=0$,
leading to expressions for the diquark gap at \( T = 0 \) and the
dependence of the critical temperature on the chemical potential
\( \mu \) for all RGC schemes presented in this work. To obtain the
expression for critical temperature, we follow the same procedure as
in \cref{app:TcBCS}.

\subsubsection*{Minimal scheme}
Starting from the effective potential given in
\cref{eq:rgc_effective_potential} and the minimal scheme ansatz in
\cref{eq:RG_min}, and taking the limit \( \Lambda \to \infty \), the
gap equation for the diquark field becomes
\begin{align}\label{app:mingapeq}
  & \nonumber \frac{\partial \Omega^\text{eff}_\text{min}(\sigma, \Delta; 0, \mu)}{\partial\Delta}
\bigg|_{\substack{\sigma=0\\\Delta=\bDeltaZero}} = 2\bDeltaZero \bigg[
  m_{\Delta}^2 - 4Z_{\Delta}\mu^{2} \\ & \qquad
  + \frac{g_\Delta^2}{\pi^2} \Bigl(
  - \Lambda' \sqrt{\Lambda'^2 + \bDelgapZero^2}
    + \bDelgapZero^{2} \operatorname{artanh}\frac{\Lambda'}{\sqrt{\Lambda'^2 + \bDelgapZero^2}}
  + \mu^{2}\Bigl( 3 + \ln \frac{\bDelgapZero^2}{4\Lambda'^2} \Bigr)
  \Bigr)
\bigg] = 0 \; .
\end{align}
Solving this equation for $\mu^2$ and identifying for which value of
the diquark gap does $\mu^2$ diverges, similarly to
\cref{sec:delta_at_T0}, one finds the asymptotic diquark gap to be
\begin{align}\label{app:delasympmin}
    \lim_{\mu\to\infty} \bDelgapZero^{(\text{min})} =
2\Lambda' \exp\left(
-\frac{3}{2} + \frac{2\pi^{2}}{g_{\Delta}^{2}}Z_{\Delta}
\right) \; ,
\end{align}
as given by \cref{eq:deltasymp_minimal}.
Following the procedure outlined in \cref{app:TcBCS}, we obtain the equation for \( T_c \) as
\begin{align}\label{app:tcasmuanddeltamin}
    T_c^2(\bDeltaZero,\mu) = 
  \frac{3\mu^{2}}{\pi^2}\,
  W\!\bigg(
    \frac{\bDelgapZero^2}{3\mu^{2}}
      \exp\!\Bigl[
        2\gamma
        - \frac{\Lambda'^2}{\mu^2} 
        \Bigl( -1 + \sqrt{1 + \bDelgapZero^2/\Lambda'^2} \Bigr) 
        \Bigl]\,
      \Bigl(
        \frac{-1 + \sqrt{1 + \bDelgapZero^2/\Lambda'^2}}
             {1 + \sqrt{1 + \bDelgapZero^2/\Lambda'^2}}
      \Bigr)^{-\dfrac{\bDelgapZero^2}{2\mu^{2}}}
  \bigg) \; .
\end{align}
Furthermore, making use of the identity
\begin{align}\label{app:identityforminimal}
    \frac{-\Lambda' + \sqrt{\Lambda'^2 + \bDelgapZero^2}}
     {\Lambda' + \sqrt{\Lambda'^2 + \bDelgapZero^2}} = 
     \Bigl(
            \frac{2\Lambda'}{\bDelgapZero}
    \Bigr)^{-4\mu^{2}/\bDelgapZero^2}
\exp\!\bigg[
    \frac{1}{\bDelgapZero^2} \bigg( 
      -\,2\Lambda'\sqrt{\Lambda'^2 + \bDelgapZero^2}
      + 6\mu^{2}
      + \frac{2\pi^{2}}{g^2_\Delta}\bigl(m_{\Delta}^{2} - 4Z_{\Delta}\mu^{2}\bigr)
    \bigg)
\bigg] \; ,
\end{align}
which follows from the gap equation \cref{app:mingapeq},
the critical temperature $T_c$ can be expressed as an explicit function of $\mu$ as
\begin{align}\label{app:tcasmuminimal}
    T_c^2(\mu)= \frac{3\mu^2}{\pi^2} \
  W\!\bigg(
    \frac{4\Lambda'^{2}}{3\mu^{2}}\,
    \exp\!\Bigl[
        -3 + 2\gamma
        + \frac{\pi^{2}\bigl(4Z_{\Delta} - m_{\Delta}^{2}/\mu^{2}\bigr)}{g_{\Delta}^{2}}
        + \frac{\Lambda'^{2}}{\mu^{2}}
    \Bigr]
  \bigg) \; .
\end{align}
Taking $\mu\to\infty$, we get
\begin{align}\label{app:tcasympmin}
    \lim_{\mu\to\infty} T_c(\mu) &=\frac{ 2 \Lambda'}{\pi} \exp\left(
\gamma-\frac{3}{2} + \frac{2\pi^{2}}{g_{\Delta}^{2}}Z_{\Delta}
\right)\\
&=    \lim_{\mu\to\infty}\frac{e^{\gamma}}{\pi} \bDelgapZero \; ,
\end{align}
yielding the BCS relation.

\subsubsection*{Vacuum matching scheme}

As the vacuum matching scheme is a variant of the minimal scheme with
a special value for $\vcZDelta$, both schemes are similar. This is
directly seen by computing the gap equation
\begin{equation}
    \frac{\partial \Omega^\text{eff}_\text{vm}(\sigma, \Delta; 0, \mu)}{\partial\Delta}
\bigg|_{\substack{\sigma=0\\\Delta=\bDeltaZero}} = 0 \; ,
\end{equation}
which yields the same expression as \cref{app:mingapeq} with the replacement $Z_\Delta \to Z_\Delta'$ where
\begin{equation}
    Z_\Delta' = Z_\Delta - \frac{g_\Delta^2}{2\pi^2} \bigg(
    \ln\frac{2\Lambda'}{\mqvac}
    + \frac{\Lambda'}{\sqrt{\Lambda'^2 + \mqvac^2}}
    + 1
    + \operatorname{artanh} \frac{\Lambda'}{\sqrt{\Lambda'^2 + \mqvac^2}}
    \bigg) \; .
\end{equation}
All relations derived in the previous section can be recovered for the
present scheme by the insertion of the above ansatz
into~\cref{app:delasympmin}. In particular, the resulting asymptomatic
diquark gap coincides with~\cref{eq:delasympvm}, which in turn
equals~\cref{eq:delatasymp}. Consequently, the BCS relation is
obtained for this scheme as well.

\subsubsection*{$\sigma\!\Delta$ scheme}

As discussed in the main text, the \(\sigma\!\Delta\) scheme differs
from the other two schemes because of its additional \(\Delta\)- and
\(\sigma\)-dependence.  These extra \(\Delta\) terms become
particularly important, especially in deriving the BCS relation.
Proceeding analogously to the previous cases, but starting from the
ansatz~\cref{eq:rgc_initial_condition2} and taking the limit
\( \Lambda \to \infty \), we determine the asymptotic value of
\(\Delta\) by examining the pole of the gap equation in \(\mu^{2}\).
This procedure yields
\begin{align} \label{eq:asympt_delta_sigmadel}
0 &= 
   2\,g_{\Delta}^{2}\bDelgapZero^{2}\Lambda'
   - 4\pi^{2}Z_{\Delta}\Lambda'^{2}\sqrt{\bDelgapZero^2+\Lambda'^{2}}
   + g_{\Delta}^{2}\Bigl(
       3\Lambda'^{3}
       - 4\pi^{2}Z_{\Delta}\bar\Delta^{2}_0\sqrt{\bDelgapZero^2+\Lambda'^{2}}
     \Bigr)\notag
\\[4pt]
&\quad
   - 2\,g_{\Delta}^{2}\bigl(\bDelgapZero^2+\Lambda'^{2}\bigr)^{3/2}
     \operatorname{artanh} \frac{\Lambda'}{\sqrt{\bDelgapZero^2+\Lambda'^{2}}} \; .
\end{align}
The equation above cannot be solved analytically for \(\bDelgapZero\); in its
current form only a numerical solution is feasible. Solving Eq.~\eqref{eq:asympt_delta_sigmadel} numerically with \(Z_{\Delta}=0, \Lambda'=600\MeV\)
and \(g_{\Delta}=4.5\) gives the asymptotic diquark gap in the
\(\sigma\!\Delta\) scheme as $
    \lim_{\mu \to \infty} \bDelgap^{(\sigma\Delta)} =
    584.2 \MeV $.
Lastly, to obtain the critical temperature \(T_c\) we follow the
 procedure leading to~\cref{app:bcssubtr}.
This leads to the implicit equation
\begin{align}\label{app:tcasmuanddeltamin}
   T_c^{2}(\bar\Delta_{0},\mu)
&=\frac{3\mu^{2}}{\pi^{2}}\,
  W\!\Biggl(
      \frac{4\Lambda'^{2}}{3\mu^{2}}\,
      \exp\!\Biggl[
        -3+2\gamma
        -\frac{\Lambda'}{\mu^{2}}
          \biggl(
            -\Lambda'+
            \frac{(\bDelgapZero^2+\Lambda'^{2})^{2}
                  -(2\bDelgapZero^2+3\Lambda'^{2})\mu^{2}}
                 {(\bDelgapZero^2+\Lambda'^{2})^{3/2}}
          \biggr)
      \Biggr]
\notag\\[4pt]
&\quad\;
      \times\Biggl(
        \frac{-\Lambda'+\sqrt{\Lambda'^{2} + \bDelgapZero^2}}
             {\Lambda'+\sqrt{\Lambda'^{2} + \bDelgapZero^2}}
      \Biggr)^{1-\frac{\bDelgapZero^2}{2\mu^{2}}}
   \Biggr) \; .
\end{align}
Using the following identity, derived from the gap equation 
\begin{align}
\frac{-\,\Lambda' + \sqrt{\Lambda'^{2} + \bDelgapZero^{2}}}
     {\Lambda' + \sqrt{\Lambda'^{2} + \bDelgapZero^{2}}}
&=\exp\Biggl[
   -\frac{2}{
          g_{\Delta}^{2}\bigl(\bDelgapZero^{2}+\Lambda'^{2}\bigr)^{3/2}
          \bigl(\bDelgapZero^{2}-2\mu^{2}\bigr)}
   \Bigl\{
      g_{\Delta}^{2}\Lambda'\!
      \bigl(
        \bDelgapZero^{4}
        + 2\,\bDelgapZero^{2}(\Lambda'^{2}-\mu^{2})
        + \Lambda'^{4}
        - 3\,\Lambda'^{2}\mu^{2}
      \bigr)
\notag\\[4pt]
&\qquad\quad
      -\;
      \pi^{2}\sqrt{\bDelgapZero^{2}+\Lambda'^{2}}\,
      \bigl(\Lambda'^{2}+g_{\Delta}^{2}\bDelgapZero^{2}\bigr)
      \bigl(m_{\Delta}^{2}-4Z_{\Delta}\mu^{2}\bigr)
   \Bigr\}
\Biggr],
\end{align}
 \( T_c \) can be expressed as an explicit function of \( \mu \) as
\begin{align}\label{app:tcasmusigdel}
    T_c^2(\mu)= \frac{3\mu^2}{\pi^2} \
  W\!\bigg(
    \frac{4\Lambda'^{2}}{3\mu^{2}}\,
    \exp\!\Bigl[
        -3 + 2\gamma
        + \frac{\pi^{2}\bigl(4Z_{\Delta} - m_{\Delta}^{2}/\mu^{2}\bigr)}{g_{\Delta}^{2}}
        + \frac{\Lambda'^{2}}{\mu^{2}}
    \Bigr]
  \bigg) \; .
\end{align}

\cref{app:tcasmusigdel} is the same as \cref{app:tcasmuminimal},
implying that the $\sigma\!\Delta$ scheme and the minimal scheme yield
identical 2SC phase boundary for $\sigma=0$. Taking $\mu\to\infty$ in
\cref{app:tcasmusigdel}, we get
\begin{align}\label{eq:tcasympsigdel}
    \lim_{\mu\to\infty} T_c(\mu) =\frac{ 2 \Lambda'}{\pi} \exp\left(
\gamma-\frac{3}{2} + \frac{2\pi^{2}}{g_{\Delta}^{2}}Z_{\Delta}
\right) \; ,
\end{align}
which is equal to \cref{app:tcasympmin}. However, within this scheme,
the above equation does \emph{not} reproduce the BCS relation.  The
reason is straightforward: if we insert the ansatz
\[
\bDelgapZero \;=\;
2\Lambda'\,
\exp\Bigl(
      -\frac{3}{2}
      + \frac{2\pi^{2}}{g_{\Delta}^{2}}\,Z_{\Delta}
\Bigr) \; ,
\]
into the gap equation and then take the limit \(\mu\to\infty\),
the right‑hand side does \emph{not} vanish.
Consequently, this ansatz is \emph{not} a solution of the gap equation
in the \(\sigma\!\Delta\) scheme, and the BCS relation
does not hold in general.

\section{Behavior of regMFA at high chemical potential}
\label{app:mfasos}

In the regularized QMD model, as seen in
\cref{fig:phase_diagram_and_T0_vs_mu}, the diquark condensate vanishes
at sufficiently large chemical potentials, leaving only a finite
constituent quark mass which remains at a constant value.  This
behavior can be understood analytically as follows.  For chemical
potentials satisfying $\mu > \sqrt{\Lambda^{2}+m_{q}^{2}}$ , the loop
contribution to the effective potential \eqref{eq:Omega_reg_therm} at
\(T=0\) reads
\begin{align}
  L_\Lambda\!\bigl(m_q,\Delgap;T=0\bigr)
  \;=\;
  -2N_f
  \!\int_{|\vec p|<\Lambda}\!
  \Bigl\{\,E_q^+ + E_q^- +  
   \frac{1}{2} \epsilon_q^{+}
    +\frac{1}{2} |\epsilon_q^{-}|
  \;\Bigr\} \; .
\end{align}
Expanding this expression in powers of \(\Delta/\Lambda\) gives
\begin{align}\label{eq:loop_all}
  L_\Lambda(m_q,\Delgap;T=0)
  \;=\;
  -\,\frac{N_f\,\Lambda^{3}}{\pi^{2}}\mu
  +\frac{N_f\,\,g_{\Delta}^{2}\,\Delta^{2}}{\pi^{2}}\mu  \,\Lambda
  \Biggl[
    1
     -
     \frac{\sqrt{{\mu^{2}-m_q^2}}}{\Lambda}\,
     \arctanh\!
       \frac{\Lambda}{\sqrt{\mu^{2}-m_q^2}}
   +\mathcal{O}\!\bigl(\Delta^{2}/\Lambda^{2}\bigr)\Biggr] \; .
\end{align}
Inserting this into the effective potential together with the bosonic
potential~\eqref{eq:treelevel} (with \(Z_\Delta=0\)) and differentiating
with respect to \(\Delta\) to get the gap equation yields
\begin{align}  
  \frac{\partial \Omega_\text{reg}^\text{eff}(\sigma,\Delta;T=0)}{\partial\Delta}
  \;=\;
  \Delta\,
  \Biggl\{
     2m_\Delta^{2}
     \;+\;
     \frac{2N_f g_{\Delta}^{2}}{\pi^{2}}\,\mu\,\Lambda
     \Biggl[
       1
        \;-\;
    \frac{\sqrt{{\mu^{2}-m_q^2}}}{ \Lambda}\,
     \arctanh\!
       \frac{\Lambda}{\sqrt{\mu^{2}-m_q^2}}
        +\;
     \mathcal{O}\!\bigl(\Delta^{2}/\Lambda^{2}\bigr)
     \Biggr]  \Biggr\} = 0
     \;.
\end{align}
Thus \(\bar\Delta=0\) (and $g_\phi\bar\sigma=\bar{m}_q$) is a non-trivial solution provided the bracket
vanishes.  Defining \(\mu_{c}\) by that condition,
\begin{align}\label{eq:wheremuc}
     2m_\Delta^{2}
     \;+\;
     \frac{2N_f g_{\Delta}^{2}}{\pi^{2}}\,\mu_c\, \Lambda
     \Biggl[
       1
        \;-\;
   \frac{\sqrt{{\mu_c^{2}-\bar{m}_q^2}}}{ \Lambda}\,
     \arctanh\!
       \frac{\Lambda}{\sqrt{\mu_c^{2}-\bar{m}_q^2}}
     \Biggr]=0 \; ,
\end{align}
which can be solved numerically to obtain $\mu_c$ once $\bar{m}_q$ is known (Note that $\mu_c$ should still satisfy the condition $\mu_c > \sqrt{\Lambda^{2}+\bar{m}_{q}^{2}}$\;). Accordingly, a cutoff-dependent critical chemical potential
\(\mu_{c}\) emerges, and its very appearance signals a cutoff
artifact of the regularized approximation. 
Furthermore, setting \(\Delta=0\) in \eqref{eq:loop_all} leaves only the
 term linear in $\mu$
\begin{align}
  L_\Lambda\!\bigl(m_q,\Delgap=0;T=0\bigr)
  \;=\;
  -\frac{N_f\,\Lambda^{3}}{\pi^{2}}\,
  \mu \; .
  \label{eq:loop_linear_mu}
\end{align}
Because the bosonic potential in \eqref{eq:treelevel} is
$\mu$-independent without diquarks and the loop contribution \eqref{eq:loop_linear_mu}
is $\sigma$-independent, the gap equation reduces to the tree-level
condition
\begin{align}
  -c \;+\; m_\phi^{2}\,\bar\sigma \;+\; \lambda_\phi\,\bar\sigma^{3} \;=\; 0 \; .
\end{align}
Using the vacuum parameters listed in \cref{tab:observables}, this
yields a $\mu$-independent constituent quark mass of
\(\bar{m}_q = 6.31~\text{MeV}\) at large chemical potentials in the
regMFA. This result is in perfect agreement with the values shown in
\cref{fig:phase_diagram_and_T0_vs_mu}. Inserting this value into \cref{eq:wheremuc}, we get
\(\mu_{c}=686.5\MeV\), in perfect agreement with the boundaries found in
\cref{fig:phase_diagram_and_T0_vs_mu,fig:phase_diagram}.

Furthermore, since $\bar{m}_q(\mu)$ is constant, the only
$\mu$-dependence in the pressure originates from the linear term
in~\eqref{eq:loop_linear_mu}. Consequently, the pressure scales as
$p\propto\mu$ (in contrast to the $\mu^{4}$ behavior in the
Stefan-Boltzmann limit), and the corresponding energy density
\(\epsilon = -p + \mu\,\frac{d p}{d\mu}\) becomes $\mu$-independent.
As a result, the squared speed of sound,
\( c_s^2 = \frac{d p}{d\epsilon} = \frac{dp}{d\mu}
\left(\mu\frac{d^2p}{d\mu^2}\right)^{-1} \), diverges at large $\mu$
in the regularized model, as illustrated in
\cref{fig:T0_pressure_sos}.

\section{Numerical implementation}
\label{app:numerical_implementation}

Numerical results presented in this work are obtained using the
\texttt{SciML} ecosystem offered by the \texttt{Julia} language. In
particular, \texttt{Optimization.jl}
\cite{vaibhav_kumar_dixit_2023_7738525} and solvers accessible within
\cite{NLopt,NELDERMEAD}. We also made use of automatic differentiation
through \texttt{ForwardDiff.jl} \cite{RevelsLubinPapamarkou2016} in
the numerical computation of vacuum parameters, the entropy density,
number densities and the speed of sound (by means of the Jacobian
method developed in Ref.~\cite{Gholami:2025cfq}). All figures are
produced using \texttt{Makie.jl} \cite{DanischKrumbiegel2021}.

\twocolumngrid

\bibliography{literature.bib}

\end{document}